\documentclass[aps, prd, floatfix, nofootinbib, superscriptaddress, twocolumn]{revtex4-1}

\usepackage{latexsym}
\usepackage{amsmath}
\usepackage{amssymb}
\usepackage{amsfonts}

\usepackage[mathscr,scaled=1.15]{urwchancal}
\DeclareFontFamily{OT1}{pzc}{}
\DeclareFontShape{OT1}{pzc}{m}{it}%
{<-> s * [1.15] pzcmi7t}{}
\DeclareMathAlphabet{\mathpzc}{OT1}{pzc}{m}{it}

\usepackage{CJKutf8}
\usepackage{color}
\usepackage{slashed}
\usepackage{dsfont}

\usepackage{supertabular}
\usepackage{placeins}
\usepackage{epsfig}
\usepackage{graphicx}
\usepackage{mathrsfs}

\definecolor{purple}{rgb}{0.5,0,0.5}
\definecolor{blue}{rgb}{0.0,0,0.9}
\definecolor{prdblue}{rgb}{0.133,0.118,0.498}
\usepackage[colorlinks=true, pdfstartview=FitV, linkcolor=prdblue, citecolor= prdblue, urlcolor=prdblue]{hyperref}

\allowdisplaybreaks

\begin{document}
%\begin{CJK}{UTF8}{song}

\title{$\,$\\[-6ex]\hspace*{\fill}{\normalsize{\sf\emph{Preprint no}.
USTC-ICTS/PCFT-24-49}}\\[1ex]
Screening masses of positive- and negative-parity hadron ground-states, including those with strangeness}

\author{Chen Chen}
\email[]{chenchen1031@ustc.edu.cn}
\affiliation{Interdisciplinary Center for Theoretical Study, University of Science and Technology of China, Hefei, Anhui 230026, China}
\affiliation{Peng Huanwu Center for Fundamental Theory, Hefei, Anhui 230026, China}

\author{Fei Gao}
\email[]{fei.gao@bit.edu.cn}
\affiliation{School of Physics, Beijing Institute of Technology, Beijing 100081, China}

\author{Si-xue Qin}
\email[]{sqin@cqu.edu.cn}
\affiliation{Department of Physics and Chongqing Key Laboratory for Strongly Coupled Physics, Chongqing University, Chongqing 401331, China}

\date{\today}

\begin{abstract}
Using a symmetry-preserving treatment of a vector $\times$ vector contact interaction (SCI) at nonzero temperature, we compute the screening masses of flavour-SU(3) ground-state $J^P=0^\pm$, $1^\pm$ mesons, and $J^P=1/2^\pm$, $3/2^\pm$ baryons. We find that all correlation channels allowed at $T=0$ persist when the temperature increases, even above the QCD phase transition. The results for mesons qualitatively agree with those obtained from the contemporary lattice-regularised quantum chromodynamics (lQCD) simulations. One of the most remarkable features is that each parity-partner-pair degenerates when $T>T_c$, with $T_c$ being the critical temperature. For each pair, the screening mass of the negative parity meson increases monotonously with temperature. In contrast, the screening mass of the meson with positive parity is almost invariant on the domain $T\lesssim T_c/2$; when $T$ gets close to $T_c$, it decreases but soon increases again and finally degenerates with its parity partner, which signals the restoration of chiral symmetry. We also find that the $T$-dependent behaviours of baryon screening masses are quite similar to those of the mesons. For baryons, the dynamical, nonpointlike diquark correlations play a crucial role in the screening mass evolution. We further calculate the evolution of the fraction of each kind of diquark within baryons respective to temperature. We observe that, at high temperatures, only $J=0$ scalar and pseudoscalar diquark correlations can survive within $J^P=1/2^\pm$ baryons.

\end{abstract}

\maketitle
%\end{CJK}

%%%%%%%%%%%%%%%%%%%%%%%%%%%%%%%%%%%%%%%%%%%%%%%%%%%%%%%%%%%%%%%%%%%%%%%%%%%%%%
%%%%%%%%%%%%%%%%%%%%%%%%%%%%%%%%%%%%%%%%%%%%%%%%%%%%%%%%%%%%%%%%%%%%%%%%%%%%%%

\section{Introduction}
\label{secintro}

Investigating the phase structure of quantum chromodynamics (QCD) and the physics of strongly interacting matter is a significant challenge in modern physics. This research is closely linked to understanding the early evolution of the universe and the experiments conducted in relativistic heavy ion collisions. Currently, it is widely accepted that strongly interacting matter transitions from hadronic excitations to quark-gluon plasma (QGP) at high temperatures, where quarks and gluons become deconfined and chiral symmetry is restored.

For strongly interacting matter, one key aspect to study is the behaviour of screening masses, which are defined by the exponential decay of the spatial correlation functions\,\cite{Detar:1987kae,Detar:1987hib}. The screening mass is physically connected to the response of the medium inserted into the probe hadron. On the chiral symmetry restoration domain, for any hadronic parity-partner-pair, \emph{i.e.}, a hadron and its parity partner, the associated screening masses are expected to be degenerate. Furthermore, as $T\to\infty$, the hadron bound-states dissolve and become the almost-free propagating quark gas, as a results, the screening masses of mesons and baryons approach $2\pi T$ and $3\pi T$ respectively\,\cite{Eletsky:1988an}. Besides, it may need to mention that the screening mass of any hadron does not have a simple relationship with its inertial mass obtained at $T=0$\,\cite{Wang:2013wk}. 

The masses at finite temperature have been analysed in an array of studies, utilising continuum and lattice methods, see, \emph{e.g.}, Refs.\,\cite{Florkowski:1993bq,Florkowski:1993br,Ishii:2016dln,Giusti:2024mwq,Gupta:2013vha,Datta:2012fz,Cheng:2010fe,Bazavov:2019www,DallaBrida:2021ddx,Giusti:2024ohu,Aarts:2015mma,Aarts:2017rrl,Aarts:2018glk,Aarts:2023nax}. While significant progress has been made in studying the screening masses of mesons, research on baryons, particularly those with heavier flavours, remains insufficient.
Amongst the nonperturbative methods, the Continuum Schwinger function methods (CSMs), when utilised judiciously, can provide meaningful predictions for the study of hadron physics at zero temperature, see Refs.\,\cite{Ding:2022ows,Roberts:1994dr,Eichmann:2016yit,Dupuis:2020fhh} for reviews, and recent progresses can be found in, \emph{e.g.}, Refs.\,\cite{Yao:2024ixu,Tang:2024pky,Chen:2023zhh,Xu:2023izo,Yin:2023kom,Chen:2021guo,Cui:2020rmu,Eichmann:2024glq,Hoffer:2024fgm,Hoffer:2024alv,Braun:2014ata,Alkofer:2018guy,Souza:2019ylx,Fukushima:2021ctq,Gao:2024gdj,Serna:2022yfp,daSilveira:2022pte,Serna:2024vpn,Shi:2022erw,Shi:2023oll,Shi:2024laj}. Additionally, the CSMs can be used to investigate the properties of quarks and gluons at finite temperature and chemical potential\,\cite{Lu:2023mkn,Lu:2023msn,Gao:2020fbl,Gao:2020qsj,Motta:2024agi,Bernhardt:2023hpr,Motta:2023pks,Gunkel:2021oya,Braun:2020ada}, and Refs.\,\cite{Roberts:2000aa,Fischer:2018sdj} are two comprehensive reviews. Therefore, CSMs can, in principle, be employed to calculate hadron screening masses, however, due to technical challenges, relevant publications remain scarce
\,\cite{Gao:2017gvf,Maris:2000ig,Blaschke:2000gd,Schmidt:1994di,Wang:2013wk}.

A few additional comments on Ref.\,\cite{Wang:2013wk} are necessary here, where the screening masses for ground-state light-quark mesons and baryons were computed using a symmetry-preserving treatment of a vector $\times$ vector contact interaction (SCI) at nonzero temperature. Due to its algebraic simplicity, the SCI can deliver insights for quantities with long wavelengths without using massive computing resources, \emph{e.g.}, hadron spectrum\,\cite{Chen:2012qr,Xu:2015kta,Lu:2017cln,Yin:2019bxe,Yin:2021uom,Paredes-Torres:2024mnz,Gutierrez-Guerrero:2024him,Gutierrez-Guerrero:2021rsx,Gutierrez-Guerrero:2019uwa}, and form factors on the domain where the momentum transfer is small\,\cite{Chen:2012txa,Segovia:2013rca,Segovia:2013uga,Cheng:2022jxe,Sultan:2024hep,Hernandez-Pinto:2023yin,Xu:2024frc,Wang:2022mrh,Raya:2021pyr}. A key feature of the treatment in Ref.\,\cite{Wang:2013wk} is its simultaneous consideration of scalar and axial-vector diquark correlations for the nucleon, demonstrating that the influence of the axial-vector diquark is non-negligible. However, recent studies from the SCI\,\cite{Lu:2017cln,Yin:2021uom,Raya:2021pyr} and more realistic interactions\,\cite{Chen:2017pse,Liu:2022ndb,Liu:2022nku,Eichmann:2016hgl} have shown the necessity to consider additional types of diquark correlations within baryons. Therefore, we are positioned to extend the analysis in Ref.\,\cite{Wang:2013wk} to investigate the screening masses of baryons by considering all kinds of diquarks within them. It is of great interest to see how the different diquark fractions evolve with temperature. In particular, although the negative-parity diquarks are negligible in $J^P=1/2^+$ baryons at $T=0$\,\cite{Chen:2017pse,Eichmann:2016hgl,Chen:2019fzn}, due to chiral symmetry restoration, the pseudoscalar diquarks are expected to become as important as their scalar partners at large temperatures. In this article, we will use the SCI to compute the screening masses of positive- and negative-parity hadron ground-states, including those with strangeness.

In Section\,\ref{secsci} we describe our symmetry-preserving treatment of a contact interaction at nonzero temperatures. We then solve the gap equations for different current quark masses and discuss the solutions. In Section\,\ref{secmeson} we discuss the Bethe-Salpeter equation formalism, from which we derive the screening masses of $J^P=0^\pm$, $1^\pm$ mesons.  Section\,\ref{secdiquark} presents analogous results for diquark correlations, which play a crucial role in the formation and evolution of baryons. In Section\,\ref{secbaryon} we provide a concise explanation of the quark+diquark Faddeev equation for baryons. By solving this equation, we thoroughly investigate the screening masses and the diquark contents of the $J^P=1/2^\pm$, $3/2^\pm$ baryons. Section\,\ref{secsum} provides a summary and perspective.

%%%%%%%%%%%%%%%%%%%%%%%%%%%%%%%%%%%%%%%%%%%%%%%%%%%%%%%%%%%%%%%%%%%%%%%%%%%%%%
%%%%%%%%%%%%%%%%%%%%%%%%%%%%%%%%%%%%%%%%%%%%%%%%%%%%%%%%%%%%%%%%%%%%%%%%%%%%%%

\section{Contact interaction at nonzero temperature and phase transition}
\label{secsci}

\subsection{Two-body scattering kernel}

For any practical CSMs study, it is essential to apply a sensible truncation and specify a two-body scattering kernel as the initial step. In this work, we utilise the rainbow-ladder (RL) truncation, which represents the leading order in the most commonly used truncation scheme that preserves global symmetry
\,\cite{Bender:1996bb,Munczek:1994zz}. In RL truncation, the two-body quark-antiquark(quark) scattering kernel can be expressed as
\begin{align}
\label{rlker}
{\mathcal K}_{\alpha\beta,\gamma\delta}=g^2D_{\mu\nu}(k)[i\gamma_\mu]_{\alpha\beta}[i\gamma_\nu]_{\gamma\delta}\,,
\end{align}
where $D_{\mu\nu}$ is the gluon propagator. Furthermore, in this work, we use the following SCI \textit{Ans\"atz} 
\begin{align}
\label{scianz}
g^2D_{\mu\nu}(k)=\delta_{\mu\nu}\frac{4\pi\alpha_{\rm IR}}{m_G^2}\,,
\end{align}
where $m_G = 0.8$\,GeV is the gluon mass scale\,\cite{Qin:2011xq}, and $\alpha_{\rm IR}=0.93\pi$ is a model parameter that simulates the zero-momentum strength of a running-coupling in QCD\,\cite{Aguilar:2010gm,Boucaud:2010gr}.

Inserting Eq.\,\eqref{scianz} into \eqref{rlker}, one can immediately derive the SCI-RL kernel as follows:
\begin{align}
\label{sciker}
{\mathcal K}^{\rm SCI}_{\alpha\beta,\gamma\delta}=\frac{4\pi\alpha_{\rm IR}}{m_G^2}[i\gamma_\mu]_{\alpha\beta}[i\gamma_\mu]_{\gamma\delta}\,.	
\end{align}
It is noticeable that this kernel does not support relative momentum between bound-state constituents, therefore, significant simplifications in the calculations can be expected.

%%%%%%%%%%%%%%%%%%%%%%%%%%%%%%%%%%%%%%%%%%%%%%%%%%%%%%%%%%%%%%%%%%%%%%%%%%%%%%

\subsection{Dressed-quark propagator}

One of the most important ingredients in CSMs is the dressed-quark propagator. Using Eq.\,\eqref{sciker}, the gap equation of the $T\neq0$ dressed-quark propagator is
\begin{align}
\label{gapequ}
\nonumber
S^{-1}_f(p;T)
\nonumber
&=i\vec{\gamma}\cdot\vec{p}+i\gamma_4\omega_n+m_f\\
&+\frac{16\pi\alpha_{\rm IR}}{3m_G^2}\int_{l,dq}\gamma_\mu S_f(q;T)\gamma_\mu\,,
\end{align}
where $p=(\vec{p},\omega_n)$ and $q=(\vec{p},\omega_l)$ are the quark's four momenta, $f$ represents the quark flavour, $m_f$ is the quark's current mass, $\int_{l,dq}:=T\sum^{\infty}_{\l=-\infty}\int d^3\vec{q}/(2\pi)^3$, and $\omega_n=(2n+1)\pi T$ is the fermion Matsubara frequency. The formal solution of Eq.\,\eqref{gapequ} is
\begin{align}
S^{-1}_f(p;T)=	i\vec{\gamma}\cdot\vec{p}+i\gamma_4\omega_n+M_f(T)\,,
\end{align}
where $M_f(T)$ is the dressed-quark mass, it is momentum-independent and determined by
\begin{align}
\label{gapm}
M_f(T) = m_f + \frac{16\pi\alpha_{\rm IR}}{3m_G^2}\int_{l,dq}\frac{4M_f(T)}{s_l+M_f(T)^2}\,,
\end{align}
where $s_l=\vec{q}^2+\omega_l^2$. The integral in Eq.\,\eqref{gapm} is linearly divergent, thus one should regularise it using a Pincar\'e-invariant method. To this end, we use the procedure in Ref.\,\cite{Wang:2013wk}\footnote{This procedure is equivalent to that in the vacuum ($T=0$) case, see, \emph{e.g.}, Ref.\,\cite{Ebert:1996vx}.}, \emph{viz.}, we set
\begin{subequations}
\label{regl}
\begin{align}
\frac{1}{s_l+M_f(T)^2} &=\int^\infty_0d\tau {\rm e}^{-\tau(s_l+M_f(T)^2)}\\
&\to \int^{1/\Lambda_{\rm ir}^2}_{1/\Lambda_{\rm uv}^2}d\tau {\rm e}^{-\tau(s_l+M_f(T)^2)}\\
&= \frac{{\rm e}^{-(s_l+M_f(T)^2)/\Lambda_{\rm uv}^2}-{\rm e}^{-(s_l+M_f(T)^2)/\Lambda_{\rm ir}^2}}{s_l+M_f(T)^2}\,,
\end{align}	
\end{subequations}
where $\Lambda_{\rm uv}$ and $\Lambda_{\rm ir}$ are ultraviolet and infrared regulators, respectively. However, their roles herein are different. $\Lambda_{\rm ir}$ is the infrared regularisation scale, the momenta below it are neglected; therefore, it eliminates the quark production scale and implements the confinement\,\cite{Krein:1990sf}. In this work we use the standard choice $\Lambda_{\rm ir}=0.24\,$GeV\,\cite{Wang:2013wk,Chen:2012qr}. On the other hand, since the SCI kernel \eqref{sciker} does not define a renormalisable theory, $\Lambda_{\rm uv}$ sets the scale of all dimensioned quantities. Consequently, it acts as a dynamical parameter. For the light- and strange-quarks, a typical value is $\Lambda_{\rm uv}=0.905\,$GeV\,\cite{Wang:2013wk,Lu:2017cln,Yin:2019bxe,Cheng:2022jxe}.

Using Eq.\,\eqref{regl}, the gap equation \eqref{gapm} becomes
\begin{align}
\label{gapmnum}
M_f(T) = m_f + M_f(T)\frac{4\alpha_{\rm IR}}{3\pi m_G^2}{\mathcal C}^{\rm iu}(M_f(T)^2;T)\,,
\end{align}
where
\begin{subequations}
\label{cfun}
\begin{align}
{\mathcal C}^{\rm iu}(M_f^2;T)
&= 2T\sum^{\infty}_{l=-\infty}\int^{1/\Lambda_{\rm ir}^2}_{1/\Lambda_{\rm uv}^2}d\tau {\rm e}^{-\tau(M_f^2+\omega_l^2)}\frac{\sqrt{\pi}}{\tau^{3/2}}\\
&= \int^{1/\Lambda_{\rm ir}^2}_{1/\Lambda_{\rm uv}^2}d\tau {\rm e}^{-\tau M_f^2}2T\vartheta_2({\rm e}^{-\tau4\pi^2T^2})\frac{\sqrt{\pi}}{\tau^{3/2}}\,,
\end{align}
\end{subequations}
with $\vartheta_2(x)$ the Jacobi theta function\,\cite{Gradshteyn:1943cpj}. One should also notice that the vacuum ($T=0$) results can be obtained by simply taking 
\begin{align}
\label{jacbi2}
2T\vartheta_2({\rm e}^{-\tau4\pi^2T^2})\overset{T\to0}{=}\frac{1}{\sqrt{\pi\tau}}\,.
\end{align}

\begin{table}[t]
\caption{Current- and dressed-quark masses at $T=0$. The results are obtained with $\alpha_{\rm IR}=0.93\pi$, and (in GeV) $m_G=0.8\,$, $\Lambda_{\rm ir} = 0.24\,$, $\Lambda_{\rm uv}=0.905$. In this article we assume isospin symmetry throughout.
\label{tableqmasses}
}
\begin{center}
\begin{tabular*}%{|c|c|c|c|c|c|c|}\hline
{\hsize}
{
c@{\extracolsep{0ptplus1fil}}
c@{\extracolsep{0ptplus1fil}}
c@{\extracolsep{0ptplus1fil}}
c@{\extracolsep{0ptplus1fil}}
c@{\extracolsep{0ptplus1fil}}
c@{\extracolsep{0ptplus1fil}}
c@{\extracolsep{0ptplus1fil}}
c@{\extracolsep{0ptplus1fil}}}\hline\hline
\multicolumn{4}{c}{Current masses in \rm{GeV}} & \multicolumn{4}{c}{Dressed masses  in \rm{GeV}}\\
$m_0$ & $m_u$ & $m_s$ & $m_s/m_u$ & $M_0$ &   $M_u$ & $M_s$ & $M_s/M_u$  \\\hline
0 & 0.007  & 0.17 & 24.3 & 0.36 & 0.37 & 0.53 & 1.43  \\\hline\hline
\end{tabular*}
\end{center}
\end{table}

%%%%%%%%%%%%%%%%%%%%%%%%%%%%%%%%%%%%%%%%%%%%%%%%%%%%%%%%%%%%%%%%%%%%%%%%%%%%%%

\subsection{Phase transitions}
\label{subsecpht}

The positive Nambu solutions (see below) of Eq.\,\eqref{gapmnum} are the temperature-dependent dressed-quark masses, the corresponding values at $T=0$ are presented in Table\,\ref{tableqmasses}. However, nonzero temperature introduces complexities such as chiral symmetry restoration and deconfinement transitions, requiring adjustments to the theoretical framework. Especially, the infrared regulator $\Lambda_{\rm ir}=0.24$\,GeV was introduced to implement confinement by ensuring the absence of quark production thresholds, this picture is not valid in the deconfinement phase, for which it is natural to expect that $\Lambda_{\rm ir}$ vanishes. Therefore, $\Lambda_{\rm ir}$ should be temperature-dependent and reflect the physical picture correctly.

\begin{figure}[t]
\leftline{\hspace*{0.5em}{\large{\textsf{A}}}}
\vspace*{-5ex}
\includegraphics[clip, width=0.43\textwidth]{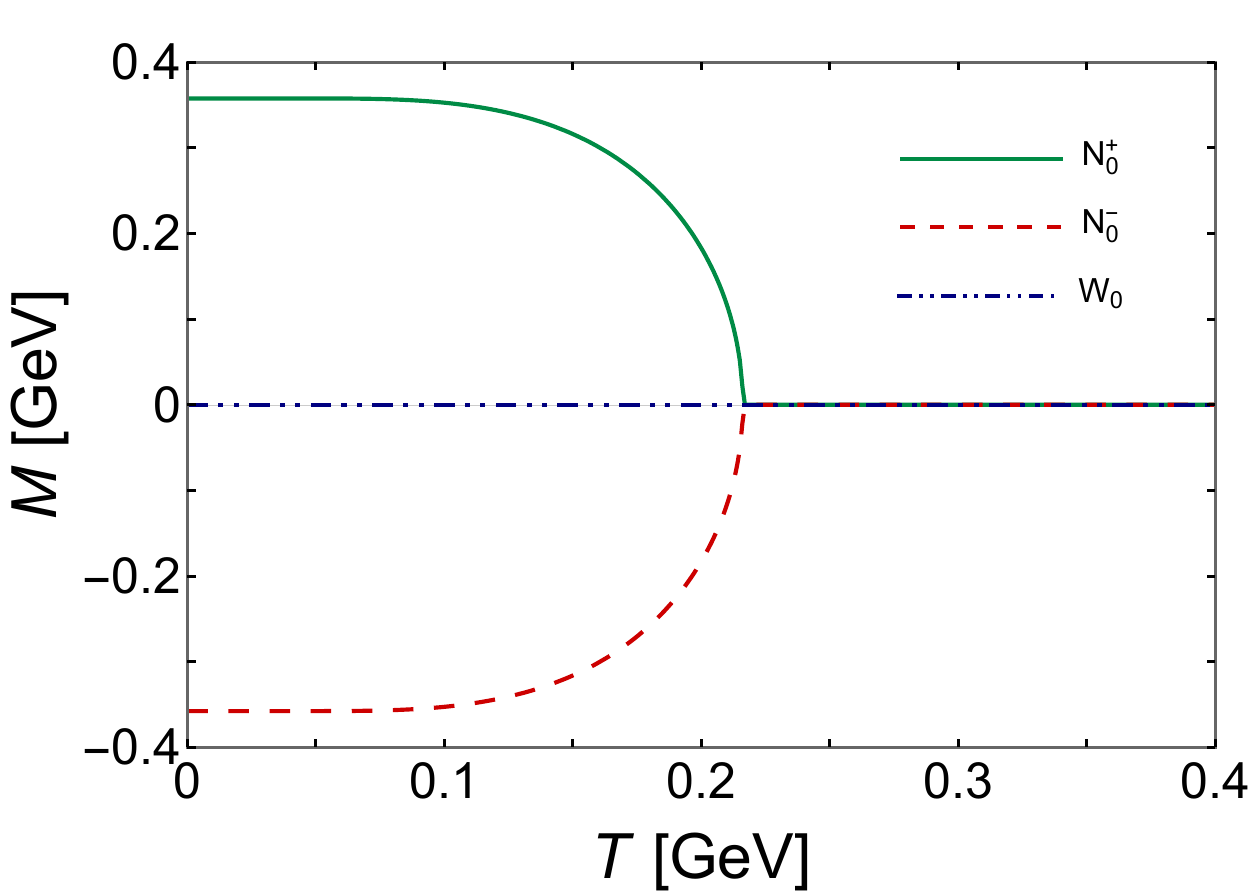}
\vspace*{1ex}
\leftline{\hspace*{0.5em}{\large{\textsf{B}}}}
\vspace*{-5ex}
\includegraphics[clip, width=0.43\textwidth]{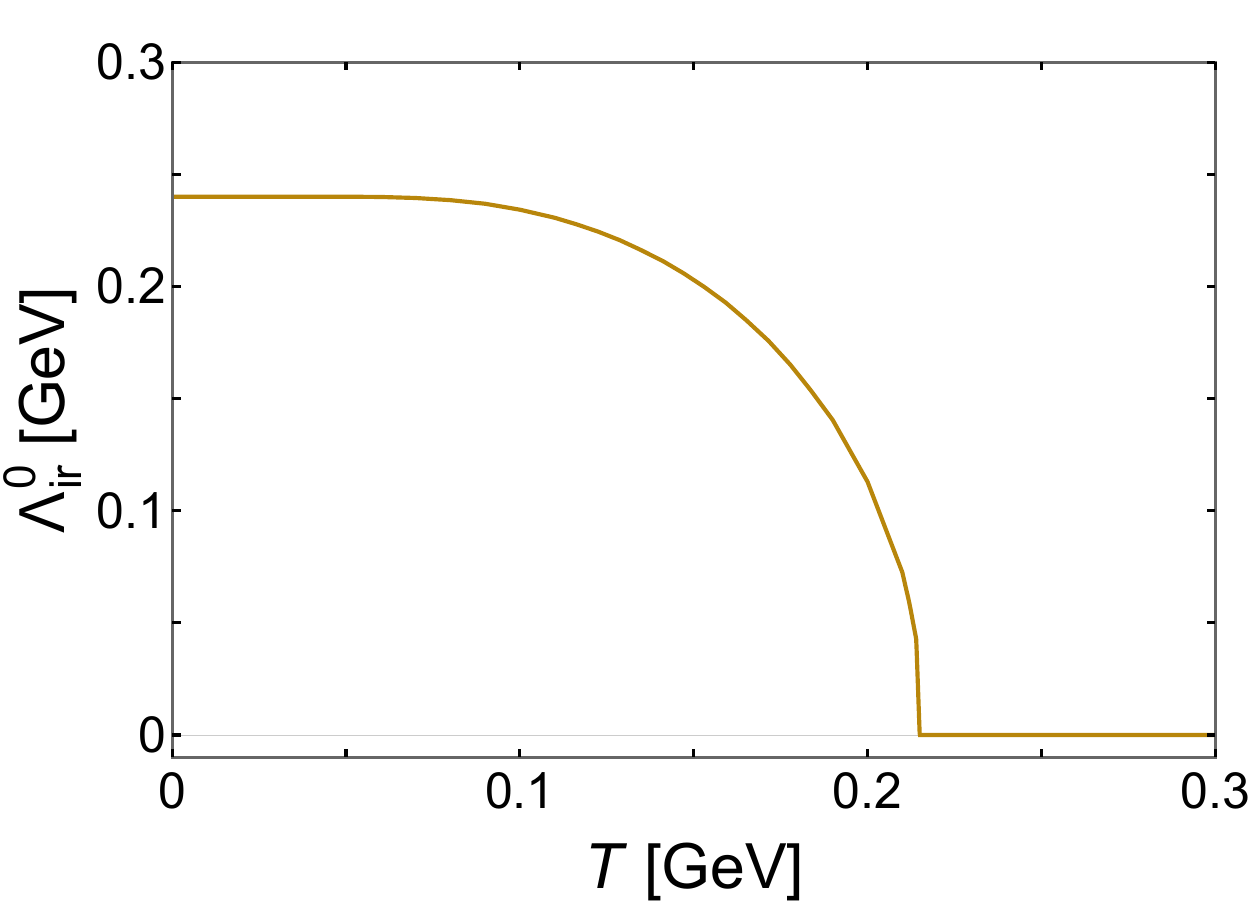}
\vspace*{4.5ex}
\caption{\label{figM0}
{\sf Panel A}.  Temperature dependence of the dressed-quark mass in the chiral limit. \emph{Solid green curve}: positive Nambu solution $N_0^+$; \emph{dashed red curve}: negative Nambu solution $N_0^-$; and \emph{dot-dot-dashed blue curve}: Wigner solution $W_0$. {\sf Panel B}. Temperature dependence of the infrared regulator. \emph{Solid gold curve}: $\Lambda^0_{\rm ir}(T)$.}
\end{figure}

First, we solve the gap equation, Eq.\,\eqref{gapmnum}, in the chiral limit, the solutions are depicted in Fig.\,\ref{figM0}. Using the SCI-RL kernel, Eq.\,\eqref{sciker}, at small $T$, the gap equation has three solutions: two Nambu solutions and one Wigner solution. The Nambu solutions are both nonzero, they are of equal size, and opposite signs; we denote them by $N_0^\pm$. The positive Nambu solution $N_0^+$ is the $T$-dependent dressed-quark mass in the chiral limit. Therefore, the appearance of the Nambu solutions is the signal of dynamical chiral symmetry breaking (DCSB). On the other hand, the Wigner solution $W_0$ is precisely zero, indicating that it is chirally symmetric.

As $T$ increases, in Fig.\,\ref{figM0}A, the Wigner solution remains unchanged, while the two Nambu solutions monotonically approach zero at the same rate. The temperature at which these solutions merge defines the critical temperature $T_c^0$, at which chiral symmetry restores, and clearly it is a second-order transition. A simple calculation yields $T_c^0=0.215$\,GeV. This value is larger than contemporary lQCD results\,\cite{Bazavov:2011nk,HotQCD:2018pds}; however, one should take into account other SCI result. For example, at $T=0$ and in the chiral limit, the SCI-RL $\rho$-meson mass is $m_\rho^0=0.919$\,GeV\,\cite{Roberts:2011wy}, which is also larger than the empirical $\rho$-meson mass. Using $m_\rho^0$ to normalise the critical temperature, one has $T_c^0/m_\rho^0=0.234$, a typical value of the RL truncation\,\cite{Wang:2013wk}.

According to the previous CSMs and lQCD studies, the chiral symmetry restoration and deconfinement transitions are highly expected to happen simultaneously at zero chemical potential\cite{Qin:2013ufa,Fischer:2013eca}. Assuming that this statement is true, one has $T_d^0=T_c^0$, where $T_d^0$ is the critical temperature of the deconfinement transition in the chiral limit. Correspondingly, the infrared regulator can be redefined via\,\cite{Wang:2013wk,Mo:2010zza}
\begin{align}
\label{lambir0}
\Lambda_{\rm ir}^0(T) = \Lambda_{\rm ir}\frac{{\mathcal B}^{1/4}_{W_0N^-_0}(T)}{{\mathcal B}^{1/4}_{W_0N^-_0}(0)}\,,
\end{align}
where
\begin{align}
\label{bchiral}
{\mathcal B}_{W_0N^-_0}(T) = \int^m_{m_0^c(T)}dt[\langle\bar{q}q\rangle^{m=t}_{W_0} - \langle\bar{q}q\rangle^{m=t}_{N^-_0}]\,,
\end{align}
with
\begin{align}
\label{qcond}
\langle\bar{q}q\rangle_{P\in\{N^-_0,W_0\}}^m = -8N_c\int_{l,dp}\frac{M_P(T)}{p_l^2+M_P(T)^2}\,.
\end{align}
At each temperature $T$, there is a critical current-quark mass $m_0^c(T)$, at which the negative Nambu ($N^-_0$) and the Wigner ($W_0$) solutions merge, therefore $\langle\bar{q}q\rangle_{N^-_0}^{m_0^c(T)}=\langle\bar{q}q\rangle_{W_0}^{m_0^c(T)}$. Using Eqs.\,\eqref{gapequ}-\eqref{gapm} and \eqref{lambir0}-\eqref{qcond}, one obtains
\begin{align}
\Lambda_{\rm ir}^0(T) = \Lambda_{\rm ir}\bigg(\frac{M_{N_0^+}(T)m_0^c(T)}{M_{N_0^+}(0)m_0^c(0)}\bigg)^{1/4}\,.
\end{align}

The $T$-dependence behaviour of $\Lambda_{\rm ir}^0(T)$ is depicted in Fig.\,\ref{figM0}B. Clearly, it decreases monotonically until it reaches $T=T^0_d$. For $T>T^0_d$, $\Lambda_{\rm ir}^0(T)\equiv0$, which physically means the quark production thresholds reappear, leading to the occurrence of the deconfinement transitions.

\begin{figure}[t]
\leftline{\hspace*{0.5em}{\large{\textsf{A}}}}
\vspace*{-5ex}
\includegraphics[clip, width=0.43\textwidth]{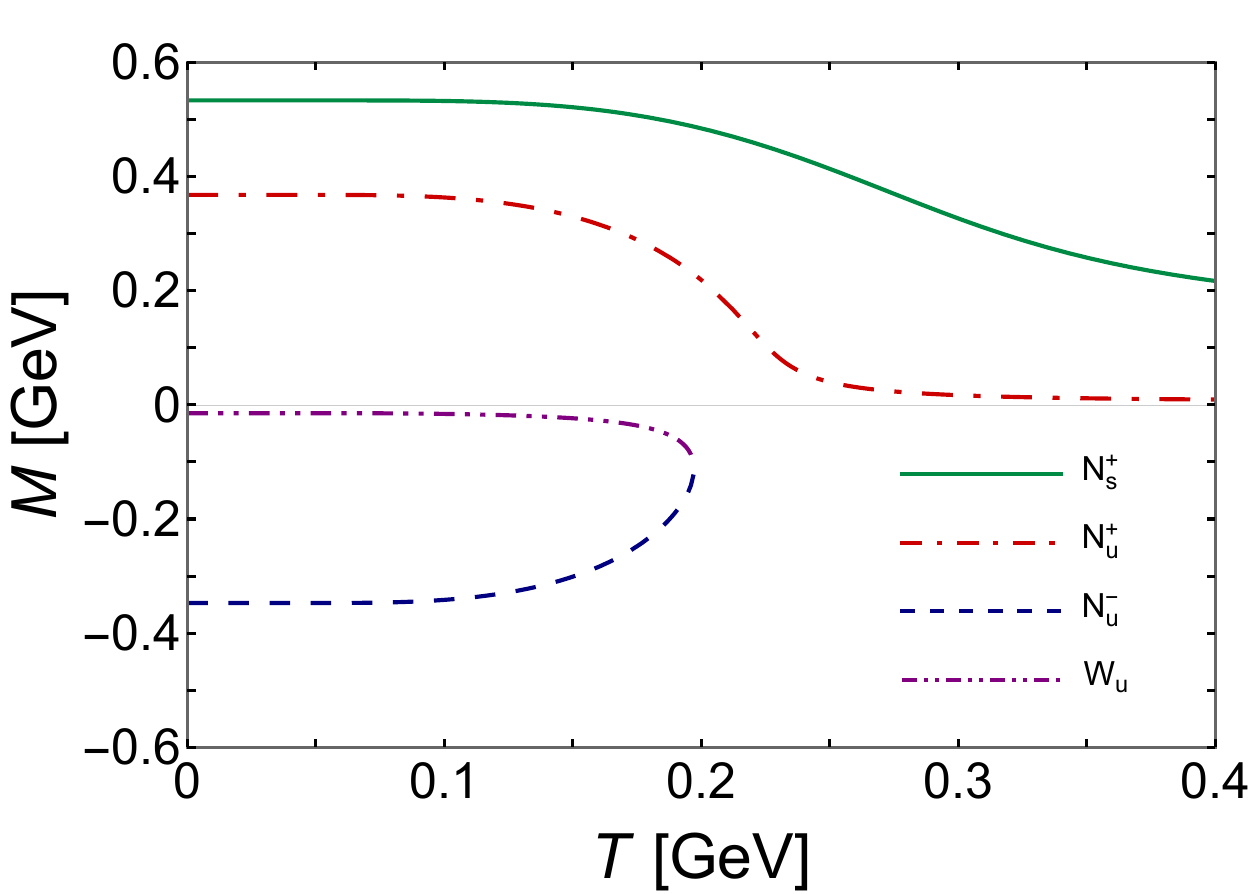}
\vspace*{1ex}
\leftline{\hspace*{0.5em}{\large{\textsf{B}}}}
\vspace*{-5ex}
\includegraphics[clip, width=0.43\textwidth]{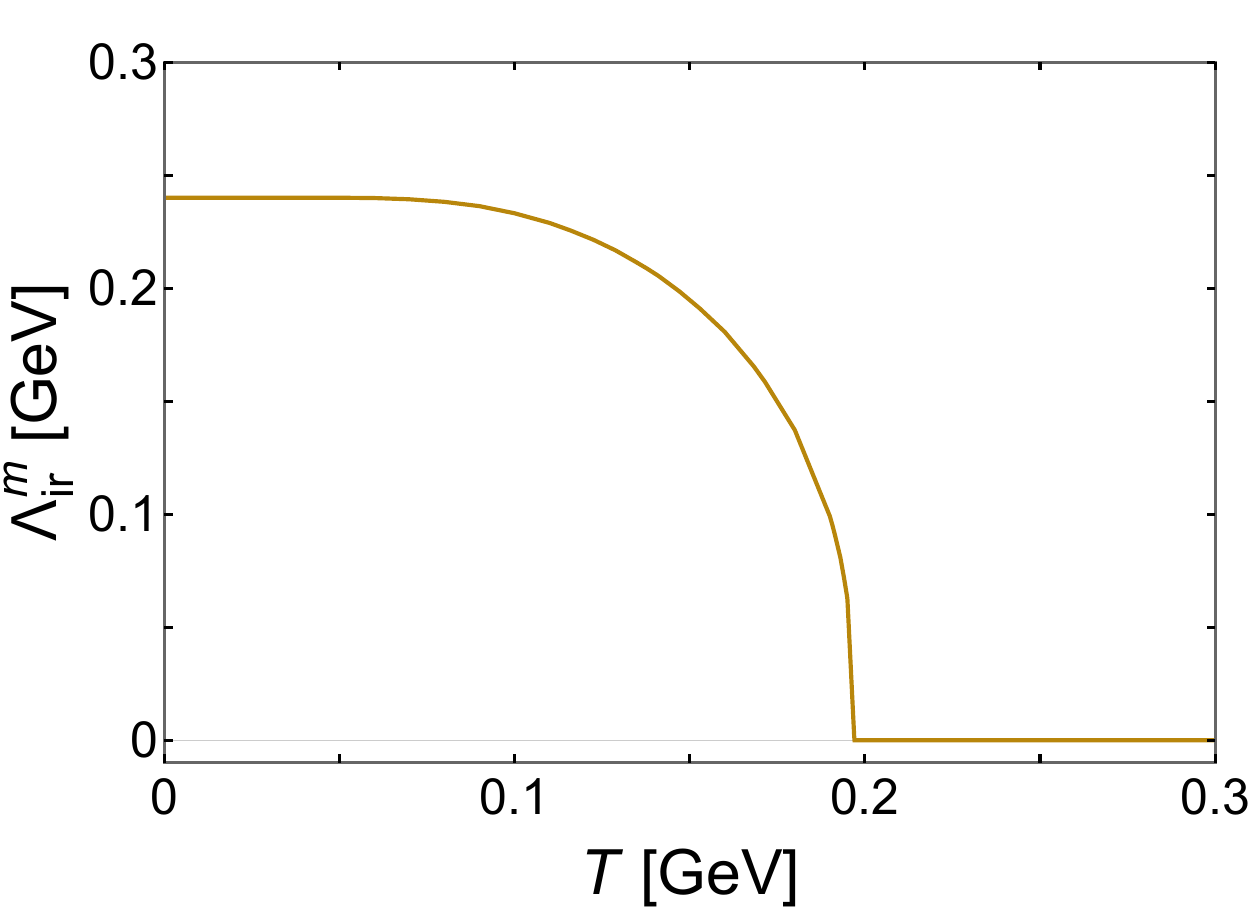}
\vspace*{4.5ex}
\caption{\label{figMq}
{\sf Panel A}.  Temperature dependence of the physical dressed-quark masses at $m=7\,$MeV ($u/d$ quark), and at $m=170\,$MeV ($s$ quark). For $u/d$ quark: \emph{dot-dashed red curve}: positive Nambu solution $N_u^+$; \emph{dashed blue curve}: negative Nambu solution $N_u^-$; and \emph{dot-dot-dashed purple curve}: Wigner solution $W_u$. The \emph{solid green curve} is the positive Nambu solution $N_s^+$ of $s$ quark. {\sf Panel B}. Temperature dependence of the infrared regulator. \emph{Solid gold curve}: $\Lambda_{\rm ir}^m(T)$.}
\end{figure}

The discussion above is only for the chiral limit case. However, to compute the screening masses of realistic hadrons, we require results from the physical current-quark masses, see Table\,\ref{tableqmasses}. For $u$-quark, its $T$-dependent dressed mass is depicted in Fig\,\ref{figMq}A. Apparently, the chiral symmetry restoration transition becomes a crossover. At large $T$, the positive Nambu solution ($N^+_u$) trends to its current mass, \emph{i.e.},
\begin{align}
\lim_{T\to\infty}M_{N^+_u}(T) = m_u=0.007\,{\rm GeV}\,.
\end{align}
While the negative Nambu ($N^-_u$) and Wigner ($W_u$) solutions get close to each other as $T$ increases, and the temperature they merge can be applied to define the critical temperature of chiral symmetry restoration of $u$-quark:
\begin{align}
T_c=0.197\,{\rm GeV}\,.
\end{align}
For $T>T_c$, only the $N^+_u$ solution exists. Assuming the deconfinement transition happens simultaneously, one has $T_d=T_c=0.197$\,GeV, with $T_d$ the critical temperature of the deconfinement transition. Using the SCI-RL $\rho$-meson mass $m_\rho=0.93$\,GeV\,\cite{Chen:2012qr}, the normalised value is $T_c/m_\rho=0.212$, quite close to the lQCD result $0.200(3)$\,\cite{HotQCD:2018pds}.

The infrared regulator for $u$-quark, $\Lambda_{\rm ir}^m(T)$, can be constructed similarly to the chiral limit situation, the expression is
\begin{align}
\label{lambirm}
\Lambda_{\rm ir}^m(T) = \Lambda_{\rm ir}\bigg(\frac{\big[M_{W_u}(T)-M_{N_u^-}(T)\big]\big[m_u-m_u^c(T)\big]}{\big[M_{W_u}(0)-M_{N_u^-}(0)\big]\big[m_u-m_u^c(0)\big]}\bigg)^{1/4}\,.
\end{align}
We depict $\Lambda_{\rm ir}^m(T)$ in Fig\,\ref{figMq}B. For $T>T_d$, $\Lambda_{\rm ir}^m(T)\equiv0$, the deconfinement transition happens.
Turning to $s$-quark, the $T$-dependence of its positive Nambu ($N^+_s$) solution is depicted in Fig.\,\ref{figMq}A. It decreases monotonically with $T$, and gradually approaches its current mass $m_s=0.17$\,GeV.  The critical temperature $T_c^s$ as the merging temperature of negative Nambu ($N^-_s$) and Wigner ($W_s$) solutions cannot be determined here as the current quark mass is too large to generate these two solutions. However, we find that the effect of changes in the value of $T_c^s$ is negligible in practical computations. So, for simplicity, we set $T_c^s=T_c=0.197$\,GeV, and also use Eq.\,\eqref{lambirm} to represent the infrared regulator for $s$-quark.

%%%%%%%%%%%%%%%%%%%%%%%%%%%%%%%%%%%%%%%%%%%%%%%%%%%%%%%%%%%%%%%%%%%%%%%%%%%%%%
%%%%%%%%%%%%%%%%%%%%%%%%%%%%%%%%%%%%%%%%%%%%%%%%%%%%%%%%%%%%%%%%%%%%%%%%%%%%%%

\section{Screening masses: mesons}
\label{secmeson}

We are now ready to study the properties of mesons at nonzero temperatures. In the CSMs framework, a meson constituted from a valence $f$-quark and valence $\bar{g}$-antiquark is represented by the Bethe-Salpeter amplitude $\Gamma^{J^P}_{[f\bar{g}]}$. Using the SCI-RL kernel \eqref{sciker}, the corresponding Bethe-Salpeter equation takes the form
\begin{align}
\label{bsem}
\nonumber
\Gamma^{J^P}_{[f\bar{g}]}(Q;T)
&= -\frac{16\pi}{3}\frac{\alpha_{\rm IR}}{m_G^2}\int_{l,dq}\gamma_\mu S_f(q+Q;T) \\
&\times\Gamma^{J^P}_{[f\bar{g}]}(Q;T)S_g(q;T)\gamma_\mu\,,
\end{align}
where $S_{f,g}$ are the positive Nambu solutions of the gap equation, Eq.\,\eqref{gapequ}; and $Q=\{\vec{Q},\Omega_m\}$ with $\Omega_m=2m\pi T$ the Boson Matusbara frequency. In this article, we only consider the zeroth Matusbara frequency $\Omega_0=0$, thus take $Q\to Q_0=\{\vec{Q},0\}$. Eq.\,\eqref{bsem} is an eigenvalue problem which has a solution $Q_0^2=-(m_{f\bar{g}}^{J^P}(T))^2$, where $m_{f\bar{g}}^{J^P}(T)$ is the meson screening mass. As we mentioned before, the screening mass in any channel does not have a simple relationship with the inertial mass, a short discussion can be found in Ref.\,\cite{Wang:2013wk}, and a tentative CSMs study of the inertial masses of the $\pi$- and $\rho$-mesons at $T>0$ is given in Ref.\,\cite{Gao:2020hwo}.

Using the SCI, the Bethe-Salpeter amplitude for a pseudoscalar ($J^P=0^-$) meson is expressed in the form
\begin{align}
\label{bsamps}
\Gamma^{0^-}_{[f\bar{g}]}(Q_0;T) &= i\gamma_5E^{0^-}_{[f\bar{g}]}(T)+\frac{1}{2M_{fg}(T)}\gamma_5\gamma\cdot Q_0F^{0^-}_{[f\bar{g}]}(T)\,,
\end{align}	
where $M_{fg}(T)=[M_fM_g/(M_f+M_g)](T)$. While for its parity partner, the scalar ($0^+$) meson, the Bethe-Salpeter amplitude is
\begin{align}
\label{bsamsc}
\Gamma^{0^+}_{[f\bar{g}]}(Q_0;T) = {\mathbb I}_D E^{0^+}_{[f\bar{g}]}(T)\,.
\end{align}	

The situation is more complicated for $J=1$ mesons. Due to the breaking of $O(4)$ symmetry, each $J=1$ Bethe-Salpeter amplitude splits into independent transverse and longitudinal components for $T>0$. For a vector ($1^-$) meson, its Bethe-Salpeter amplitude has the form
\begin{align}
\label{bsamvc}
\Gamma^{1^-}_{[f\bar{g}]}(Q_0;T) =
\left\{\begin{array}{l}
\gamma_4 \, E^{1^-,\parallel}_{[f\bar{g}]}(T) \\
\vec{\gamma}_\perp E^{1^-,\perp}_{[f\bar{g}]}(T)
\end{array}\right.\,,
\end{align}
where $\vec{\gamma}_\perp=(\delta_{ij}-Q_iQ_j/|\vec{Q}|^2)\gamma_j$, $i,j=1,2,3$. And the amplitude of its parity partner, the axial-vector  ($1^+$) meson, is 
\begin{align}
\label{bsamax}
\Gamma^{1^+}_{[f\bar{g}]}(Q_0;T) =
\left\{\begin{array}{l}
\gamma_5\gamma_4 \, E^{1^+,\parallel}_{[f\bar{g}]}(T) \\
\gamma_5\vec{\gamma}_\perp E^{1^+,\perp}_{[f\bar{g}]} (T)
\end{array}\right.\,.
\end{align}

Previous CSMs studies found that the light pseudoscalar and vector mesons can be described well by using RL truncation, however, for scalar and axial-vector mesons, the RL results are not satisfactory. The issue arises from the oversimplification of RL truncation, which fails to generate sufficient ``spin-orbit'' (SO) repulsion in the Bethe-Salpeter kernel. . This deficiency significantly impacts the scalar and axial-vector mesons. For $T=0$ inertial masses, one way to remedy this weakness is to multiply the RL kernel by a factor to simulate the SO effect, see, \emph{e.g.}, Refs.\,\cite{Chen:2012qr,Lu:2017cln,Yin:2021uom}. To calculate the screening masses, in this article we keep employing the method given in Ref.\,\cite{Wang:2013wk}, \emph{viz.}, we multiply the SCI kernel \eqref{sciker} by the $T$-dependent factors
\begin{subequations}
\label{gsom}
\begin{align}
\label{gsomsc}
{\mathfrak g}_{\rm SO}^{q\bar{q},0^+}(T) &= 1-\frac{M_u(T)}{M_u(0)}(1-{0.32}^2)\,,\\
\label{gsomax}
{\mathfrak g}_{\rm SO}^{q\bar{q},1^+}(T) &= 1-\frac{M_u(T)}{M_u(0)}(1-{0.25}^2)\,,
\end{align}
\end{subequations}
for scalar ($0^+$) and axial-vector ($1^+$) mesons separately. At $T=0$, one has
\begin{subequations}
\begin{align}
{\mathfrak g}_{\rm SO}^{q\bar{q},0^+}(0) &= {0.32}^2\,,\\
{\mathfrak g}_{\rm SO}^{q\bar{q},1^+}(0) &= {0.25}^2\,,
\end{align}
\end{subequations}
these are the values given in Refs.\,\cite{Lu:2017cln,Yin:2021uom}, which we can use to obtain the empirical values of the $a_1$-$\rho$ and $\sigma$-$\rho$ inertial mass splittings. As $T$ increases, ${\mathfrak g}_{\rm SO}^{q\bar{q},0^+}(T)$ and ${\mathfrak g}_{\rm SO}^{q\bar{q},1^+}(T)$ become smaller and trend to 1 for $T>T_c$; consequently, DCSB can be traced.

\begin{table*}[t]
\caption{\label{tab:mdsmasses}
Ground-state masses of mesons and diquark correlations, \emph{i.e.}, their screening masses at $T=0$.
\emph{Upper panel}. Mesons. Row~1: computed using the SCI-RL kernel, Eq.\,\eqref{sciker}. Row~2: empirical values from Ref.\,\cite{ParticleDataGroup:2024cfk}. ``--'' indicates no empirical results available for comparison.
\emph{Lower panel}. Diquark correlations, computed using Eq.\,\eqref{sciker}. ``--'' indicates no $[ss]_{1^-}$ correlation because of Pauli principle.
(All masses are listed in GeV.)
}
\begin{center}
\begin{tabular*}%{|c|c|c|c|c|c|c|}\hline
{\hsize}
{
l@{\extracolsep{0ptplus1fil}}
|l@{\extracolsep{0ptplus1fil}}
l@{\extracolsep{0ptplus1fil}}
l@{\extracolsep{0ptplus1fil}}
l@{\extracolsep{0ptplus1fil}}
l@{\extracolsep{0ptplus1fil}}
l@{\extracolsep{0ptplus1fil}}
l@{\extracolsep{0ptplus1fil}}
l@{\extracolsep{0ptplus1fil}}
l@{\extracolsep{0ptplus1fil}}
l@{\extracolsep{0ptplus1fil}}}\hline\hline
 & $\pi$ & $K$ &
$\rho$ & $K^\ast $ & $\phi\;\;$ &
$\sigma$ & $\kappa$ &
$a_1$ & $K_1$ & $f_1$ \\\hline
SCI & 0.14 & 0.50 &
0.93 & 1.03 & 1.13 &
1.22 & 1.32 &
1.37 & 1.48 & 1.59 \\\hline\hline
expt. & 0.14 & 0.50 &
0.78 & 0.89 & 1.02 &
-- & -- &
1.22 & 1.25 & 1.43 \\\hline\hline
\multicolumn{10}{c}{  }\\\hline\hline
 & $[ud]_{0^+}$ & $[sd]_{0^+}$ &
$\{ud\}_{1^+}$ & $\{us\}_{1^+}$ & $\{ss\}_{1^+}$ &
 $[ud]_{0^-}$ & $[sd]_{0^-}$ &
 $[ud]_{1^-}$ & $[sd]_{1^-}$ & -- \\\hline
SCI & 0.78 & 0.93 &
1.06 & 1.16 & 1.26 &
1.15 & 1.26 &
1.33 & 1.44 & -- \\\hline\hline

%%%
\end{tabular*}
\end{center}
\end{table*}

\begin{figure*}[!tbh]
\hspace{-0.03\textwidth}%
\includegraphics[width=0.33\textwidth]{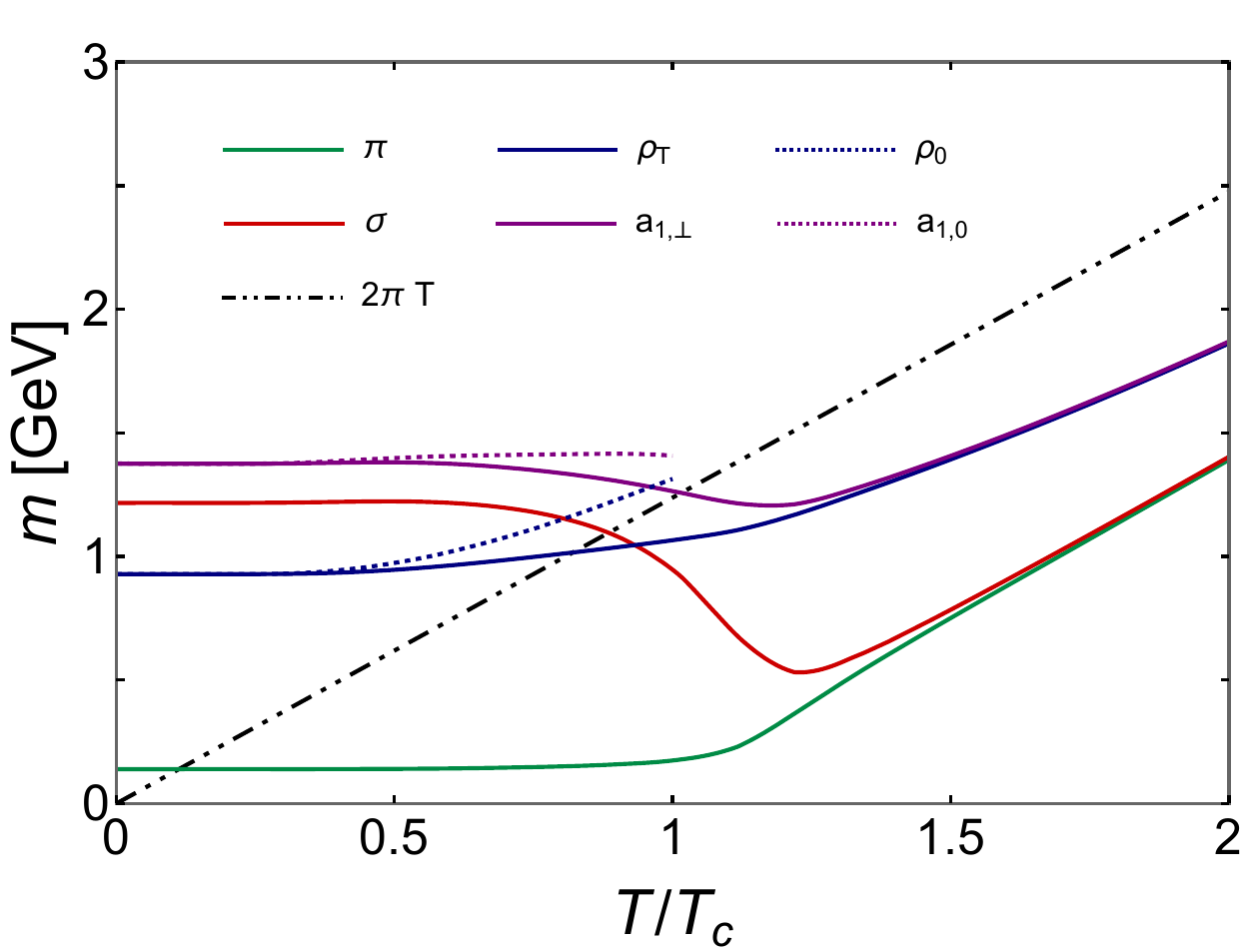}
\includegraphics[width=0.33\textwidth]{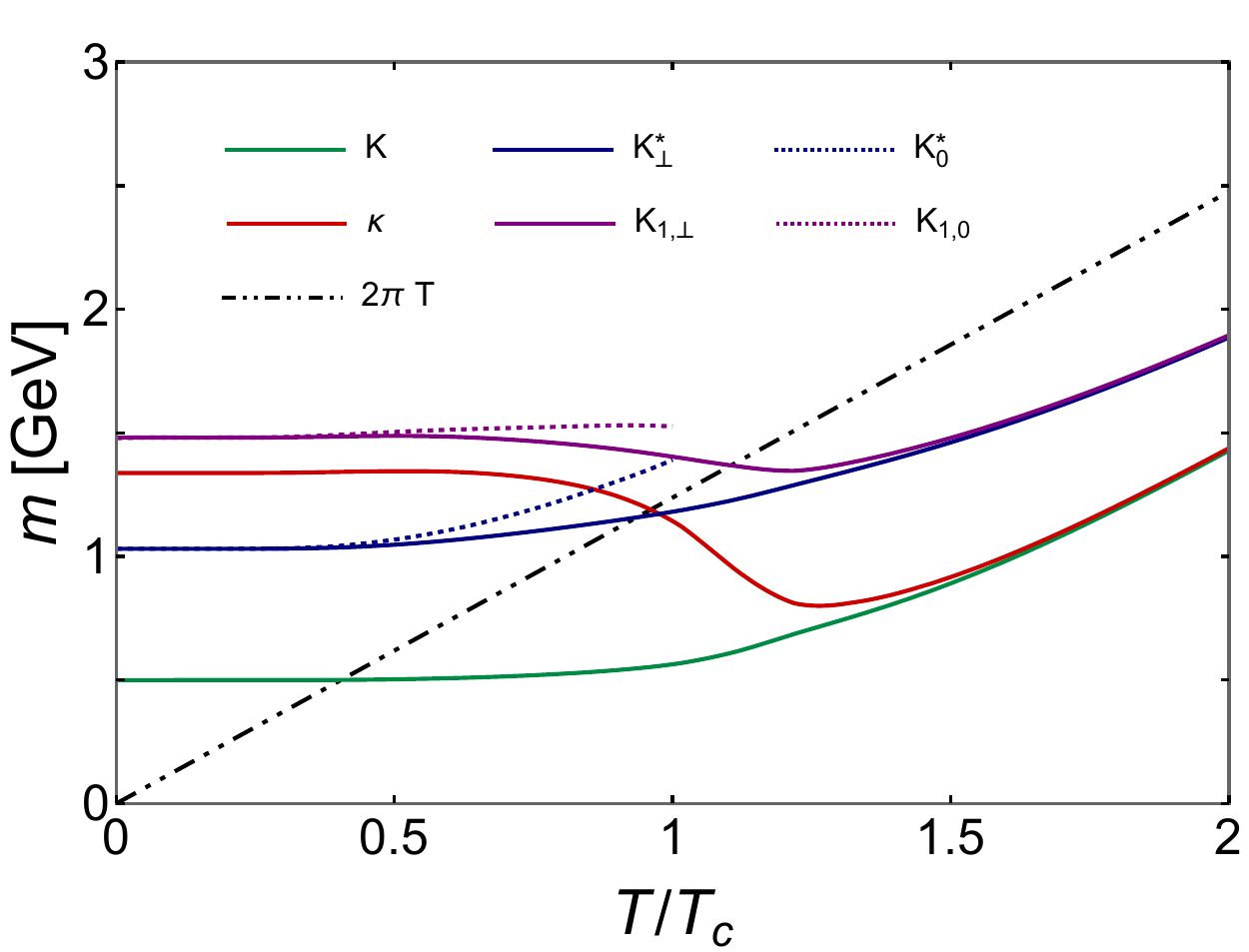}
\includegraphics[width=0.33\textwidth]{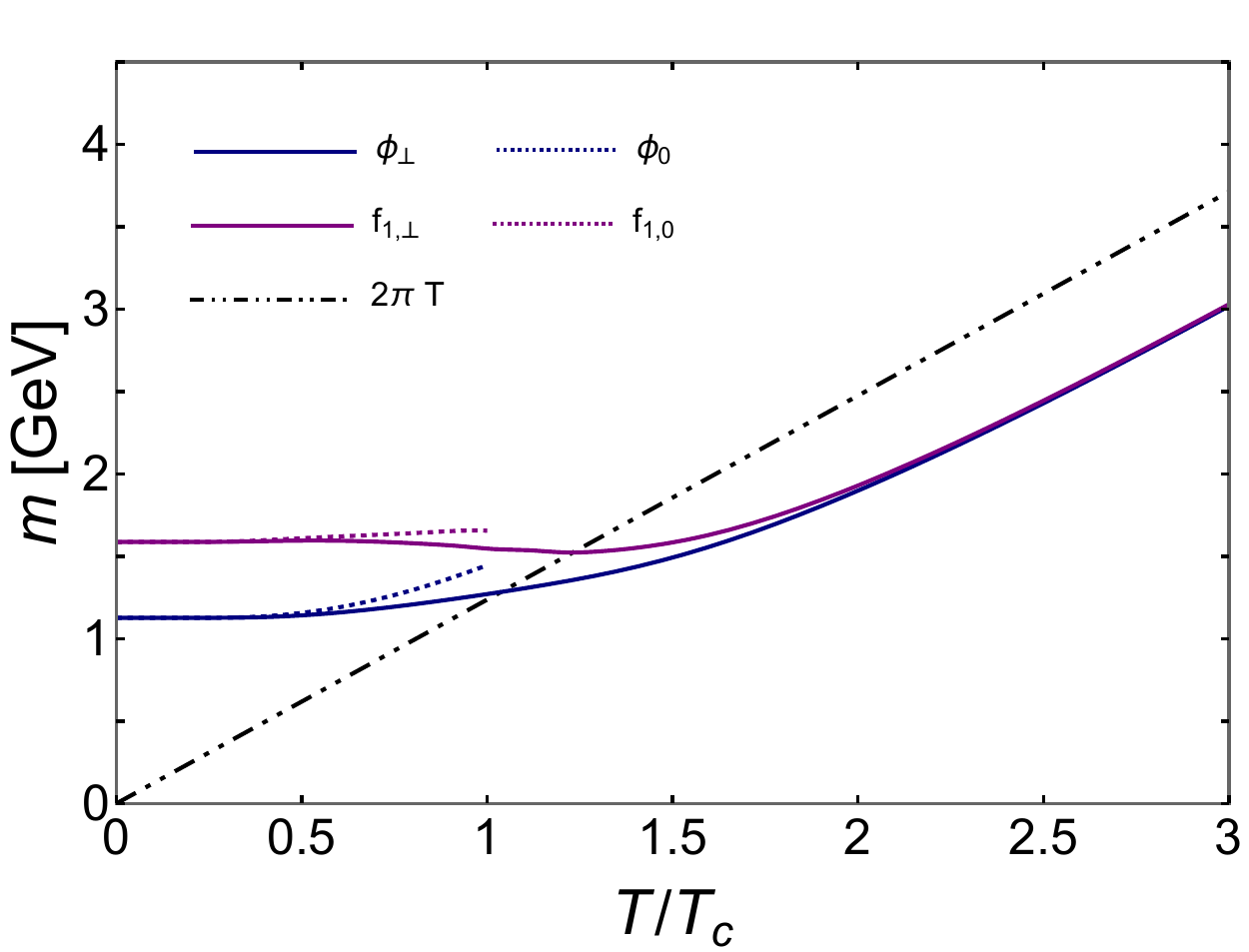}
\caption{(Left to right) Screening masses of mesons, for the light-light, light-strange, and strange-strange ($\bar{s}s$) sectors. In each figure: \emph{green curve}: pseudoscalar meson; \emph{red curve}: scalar meson; \emph{blue curves}: vector meson; \emph{purple curves}: axial-vector meson; and \emph{black dot-dot-dashed curve}: free theory limit of $m=2\pi T$. For the $J=1$ vector and axial-vector mesons, transverse modes are traced with solid lines and longitudinal modes with dotted lines.
\label{fig:mesons}}
\end{figure*}

Inserting Eqs.\,\eqref{bsamps}-\eqref{bsamax} into \eqref{bsem} and using Eqs.\,\eqref{gsom} for the associated channels, one can get the explicit expressions of the meson Bethe-Salpeter equations; we list them in Appendix\,\ref{appbse}. Their solutions, \emph{i.e.}, the screening masses of the ground-state mesons, are depicted in Fig.\,\ref{fig:mesons}.

To start with, we discuss the values at $T=0$, which are listed in Table\,\ref{tab:mdsmasses}. The results are the same as those given in Ref.\,\cite{Lu:2017cln}, and the results were discussed in great detail therein. Here, we would like to make two remarks. First, one should notice that the $T=0$ masses of the vector and axial-vector mesons are all larger than the corresponding empirical values\footnote{By fitting the data, we decide the current masses of $u$- and $s$-quarks. Therefore, the masses of the pion and kaon are equal to the empirical values.}; the reason is that the meson Bethe-Salpeter equation, Eq.\,\eqref{bsem}, does not contain the effects of the meson cloud, therefore, only produces solutions for dressed-quark-core masses, which are typically larger than the empirical results. Second, the scalar mesons, $\sigma$ and $\kappa$, may not be considered the Nature's lightest $0^+$ mesons which might have more complicated structures, \emph{e.g.}, tetraquarks\,\cite{Heupel:2012ua,Eichmann:2015cra}. Instead, the calculated masses in Table\,\ref{tab:mdsmasses} should be the artefacts of RL truncation. However, together with the pseudoscalar mesons, their screening masses are particularly useful for studying the chiral symmetry restoration transition, see below. 

In Fig.\,\ref{fig:mesons}, we first focus on the longitudinal modes of the vector and axial-vector mesons.  Our computation shows that the screening mass for each longitudinal mode disappears at phase transition temperature. This can be understood via their Bethe-Salpeter equations, Eqs.\,\eqref{bsevc} and \eqref{bseax}, in which the ${\mathcal R}^{\rm iu}$ term provides additional repulsion. At $T=0$, the mass of longitudinal mode is equal to that of its transverse counterparts under the current scheme ; it is natural, because the $Q(4)$ symmetry is not broken in vacuum. As temperature increases, their difference is marked. The screening masses of the longitudinal modes of vector mesons increase at a faster rate than those of their transverse counterparts. While for axial-vector mesons, the screening masses of their longitudinal modes are almost insensitive to changes in temperature.

For the remaining states, on the domain depicted, one can see that the ones with the same parity behave similarly. To expand on the elaboration, pseudoscalar and the transverse modes of vector mesons, which have negative-parity, their screening masses increase monotonously. Besides, before the temperature reaches $T_c$, the increase progresses slowly, especially for pseudoscalar mesons, whose screening masses are almost insensitive to $T$, while when $T>T_c$, the rate of increase significantly accelerates. But, for the scalar and the transverse modes of axial-vector mesons, the story is different: their screening masses are almost invariant for $T\lesssim T_c/2$. When the temperature continues to increase, the screening masses decrease (but are always larger than those of the corresponding parity partners) until $T\gtrsim 1.25T_c$, and then  rise again and finally degenerate with parity partners. Therefore, chiral symmetry restoration transitions happen. These observations are consistent with the contemporary lQCD result in Ref.\,\cite{Bazavov:2019www}.

To study the screening mass degeneracy quantitatively, we define a \emph{critical degenerate temperature} $T^{\delta m}_c$ for each parity-partner-pair, at which the relative difference of the two screening masses is equal to $0.5\%$, and when $T$ gets larger, their screening masses can be regarded as degenerated. For example, for $\pi$- and $\sigma$-mesons, this critical temperature is defined via
\begin{align}
\label{cdtpisig}
T_{c}^{\delta m_{[\sigma\!\pi]}} := \bigg\{T>0 \,\bigg| \,  \frac{m_\sigma(T) - m_\pi(T)}{m_\sigma(T) + m_\pi(T)}=0.5\% \bigg\}\,,
\end{align}
and for the transverse modes of $K^\ast$- and $K_1$-mesons, one has
\begin{align}
\label{cdtrhoa1}
T_{c}^{\delta m^{\perp}_{[K_1\!K^\ast]}} := \bigg\{T>0  \,\bigg| \,  \frac{m^{\perp}_{K_1}(T) - m^{\perp}_{K^\ast}(T)}{m^{\perp}_{K_1}(T) + m^{\perp}_{K^\ast}(T)}=0.5\% \bigg\}\,.
\end{align}
For other states, their $T^{\delta m}_c$ can be defined similarly. Our results are
\begin{subequations}
\label{cdtmj0}
\begin{align}
T_{c}^{\delta m_{[\sigma\!\pi]}} &= 1.95\,T_c\,,\\
T_{c}^{\delta m_{[\kappa\!K]}} &= 1.82\,T_c\,,
\end{align}
\end{subequations}
and
\begin{subequations}
\label{cdtmj1}
\begin{align}
T_{c}^{\delta m^{\perp}_{[a_1\!\rho]}} &= 1.51\,T_c\,,\\
T_{c}^{\delta m^{\perp}_{[K_1\!K^\ast]}} &= 1.58\,T_c\,,\\
T_{c}^{\delta m^{\perp}_{[f_1\!\phi]}} &= 2.51\,T_c\,.
\end{align}
\end{subequations}
In each of the light-light, light-strange, and strange-strange ($\bar{s}s$) sectors, $T^{\delta m}_c$ of the $J=1$ pair is obviously smaller than that of the $J=0$ pair, \emph{e.g.}, $T_{c}^{\delta m_{[\sigma\!\pi]}}>T_{c}^{\delta m^{\perp}_{[a_1\!\rho]}}$. One can also see that for $J=1$ pairs,  $T^{\delta m}_c$ increases, from the light-light to the strange-strange sector. However, for $J=0$ pairs, $T_{c}^{\delta m_{[\sigma\!\pi]}}>T_{c}^{\delta m_{[\kappa\!K]}}$. One reason is that the pion is the Nambu-Goldstone boson with an extremely small mass $m_\pi=0.14$\,GeV at $T=0$. Using the results in Table\,\ref{tab:mdsmasses}, it is easy to calculate the differences: $m_\sigma-m_\pi=1.08\,$GeV, which is about $32\%$ larger than $m_\kappa-m_K=0.82\,$GeV, so there could be $T_{c}^{\delta m_{[\sigma\!\pi]}}>T_{c}^{\delta m_{[K\!\kappa]}}$. These observations are consistent with the lQCD result in Ref.\,\cite{Bazavov:2019www}.

One also notices that in each sector, after the parity partners have degenerated, the screening mass of the $J=1$ pair is uniformly larger than that of the $J=0$ pair; this, again, agrees with the lQCD prediction in Ref.\,\cite{Bazavov:2019www}. However, the screening masses have never been larger than the free theory value $2\pi T$. This is due to the oversimplification of our SCI-RL kernel, Eq.\,\eqref{sciker}, which does not depend on $T$. At larger temperatures, it was found that there is a positive correction to the free theory limit\,\cite{Braaten:1995jr,Laine:2003bd}. Therefore, meson screening masses are expected to overshoot the free theory value first, confirmed by lQCD\,\cite{Bazavov:2019www}, and approach $2\pi T$ at extremely higher temperatures. In the CSMs framework, the screening mass calculation at large $T$ can only be performed using a realistic QCD-connected kernel, see, \emph{e.g.}, Ref.\,\cite{Fischer:2018sdj}. The relevant calculation is beyond the scope of this article and will be performed in the future.

%%%%%%%%%%%%%%%%%%%%%%%%%%%%%%%%%%%%%%%%%%%%%%%%%%%%%%%%%%%%%%%%%%%%%%%%%%%%%%
%%%%%%%%%%%%%%%%%%%%%%%%%%%%%%%%%%%%%%%%%%%%%%%%%%%%%%%%%%%%%%%%%%%%%%%%%%%%%%

\section{Screening masses: diquark correlations}
\label{secdiquark}

\begin{figure}[t]
\centerline{%
\includegraphics[clip, height=0.14\textwidth, width=0.45\textwidth]{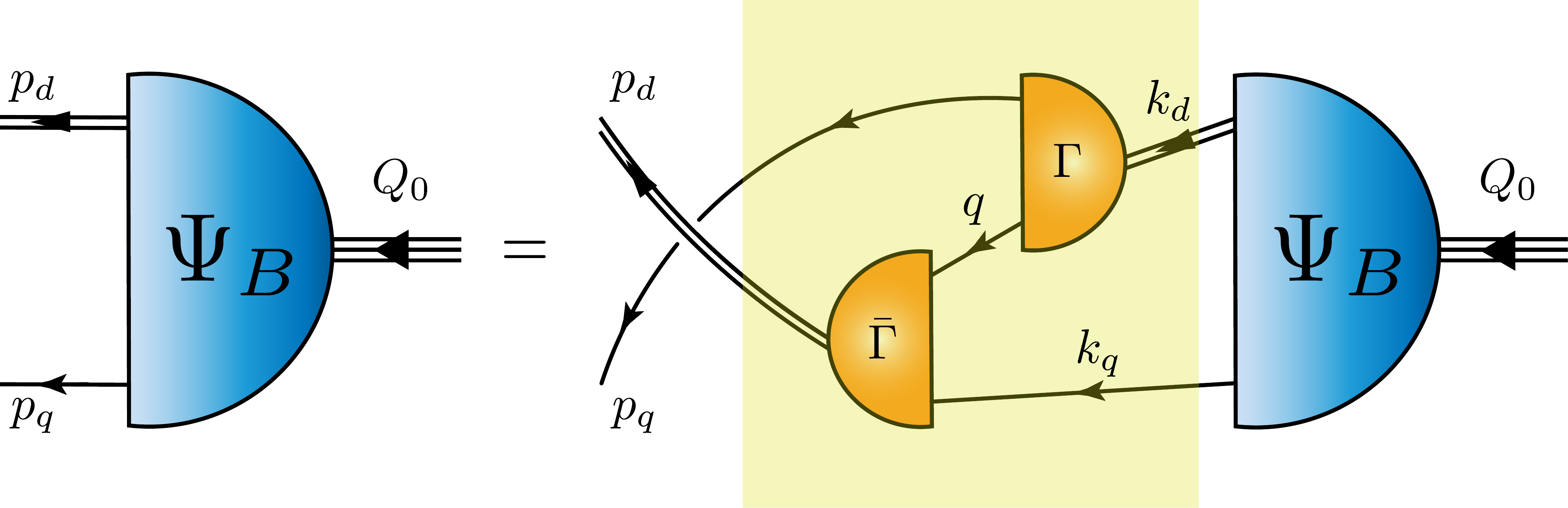}}
\caption{\label{figFaddeev}
Quark+diquark Faddeev equation, a linear integral equation for the Poincar\'e-covariant matrix-valued function $\Psi_B$, the Faddeev amplitude for a baryon  $B$ with total momentum $Q_0=p_q+p_d=k_q+k_d$. Legend. \emph{Shaded rectangle} -- Faddeev kernel; \emph{single line} -- dressed-quark propagator; $\Gamma$ -- diquark correlation amplitude; and \emph{double line} -- diquark propagator.}
\end{figure}

To compute the screening masses of baryons, herein we use a quark+diquark approximation to the three-body equations for baryon bound-states. Consequently, one only needs to solve a two-body Poincar\'e-covariant Faddeev equation depicted in Fig.\,\ref{figFaddeev}, first derived in Ref.\,\cite{Cahill:1988dx}. 

The Faddeev equation represents an eigenvalue problem. To solve it, the required ingredients include: the propagators of quarks and diquark correlations, and the canonically normalised Bethe-Salpeter amplitudes of diquarks. In principle, all these elements should depend on temperature. The quark propagators are the solutions to the gap equations in Eq.\,\eqref{gapequ}, and we have obtained the SCI solutions in Section.\,\ref{subsecpht}. The diquark propagators will be discussed in the next section when we solve the Faddeev equations. In this section, we focus on the Bethe-Salpeter equation of diquark correlations, from which one can obtain the masses and amplitudes of diquarks.

\begin{figure*}[!tbh]
\hspace{-0.03\textwidth}%
\includegraphics[width=0.33\textwidth]{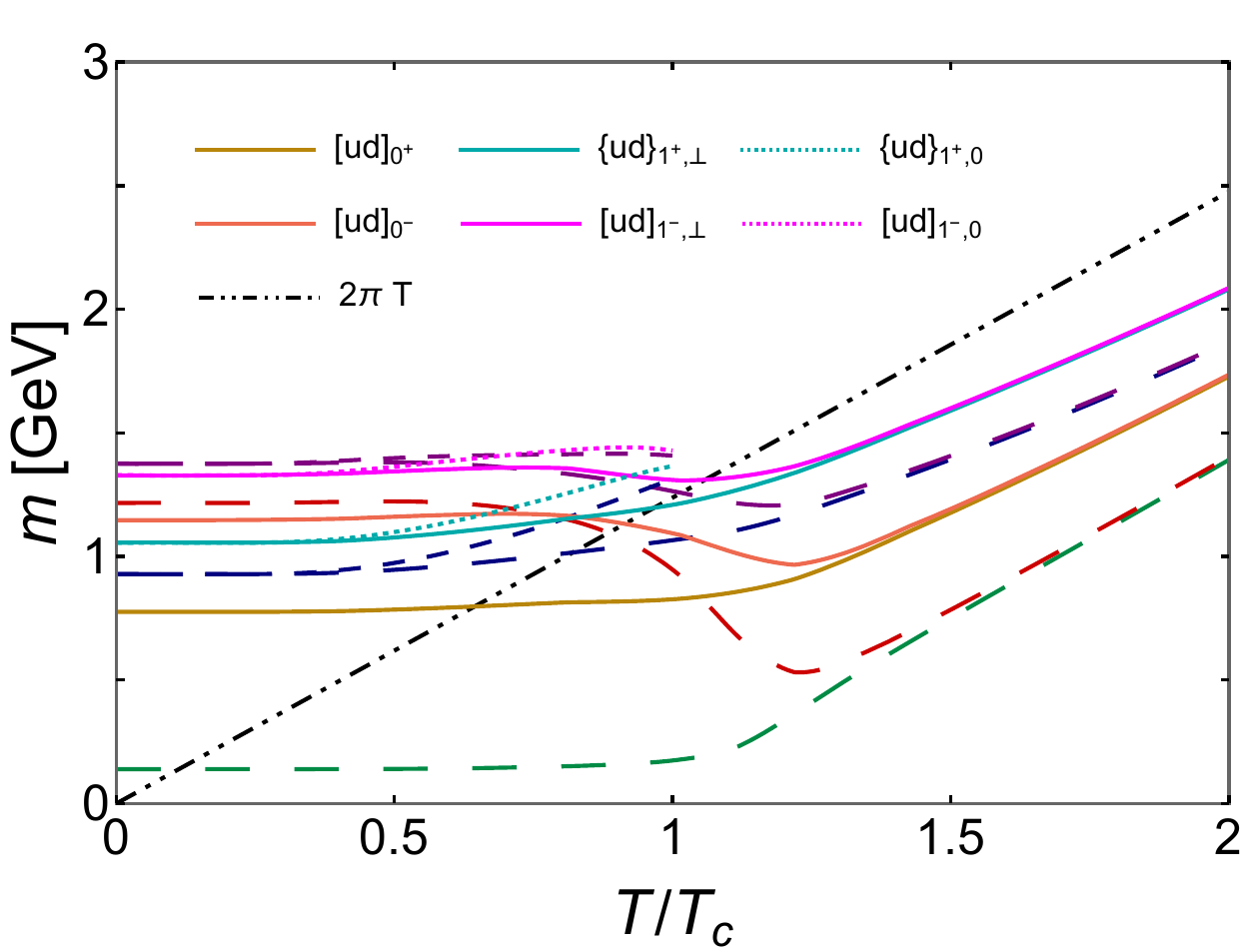}
\includegraphics[width=0.33\textwidth]{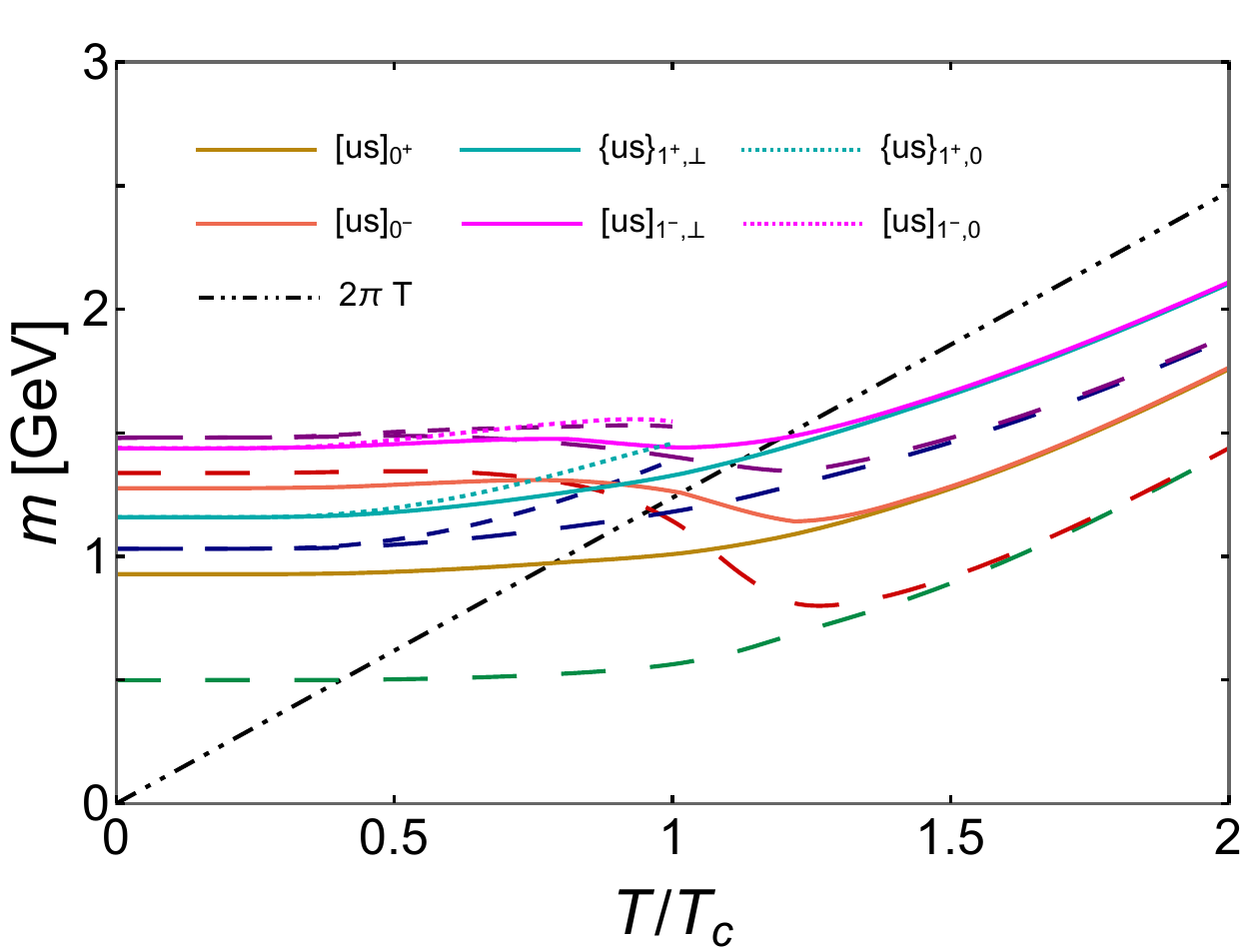}
\includegraphics[width=0.33\textwidth]{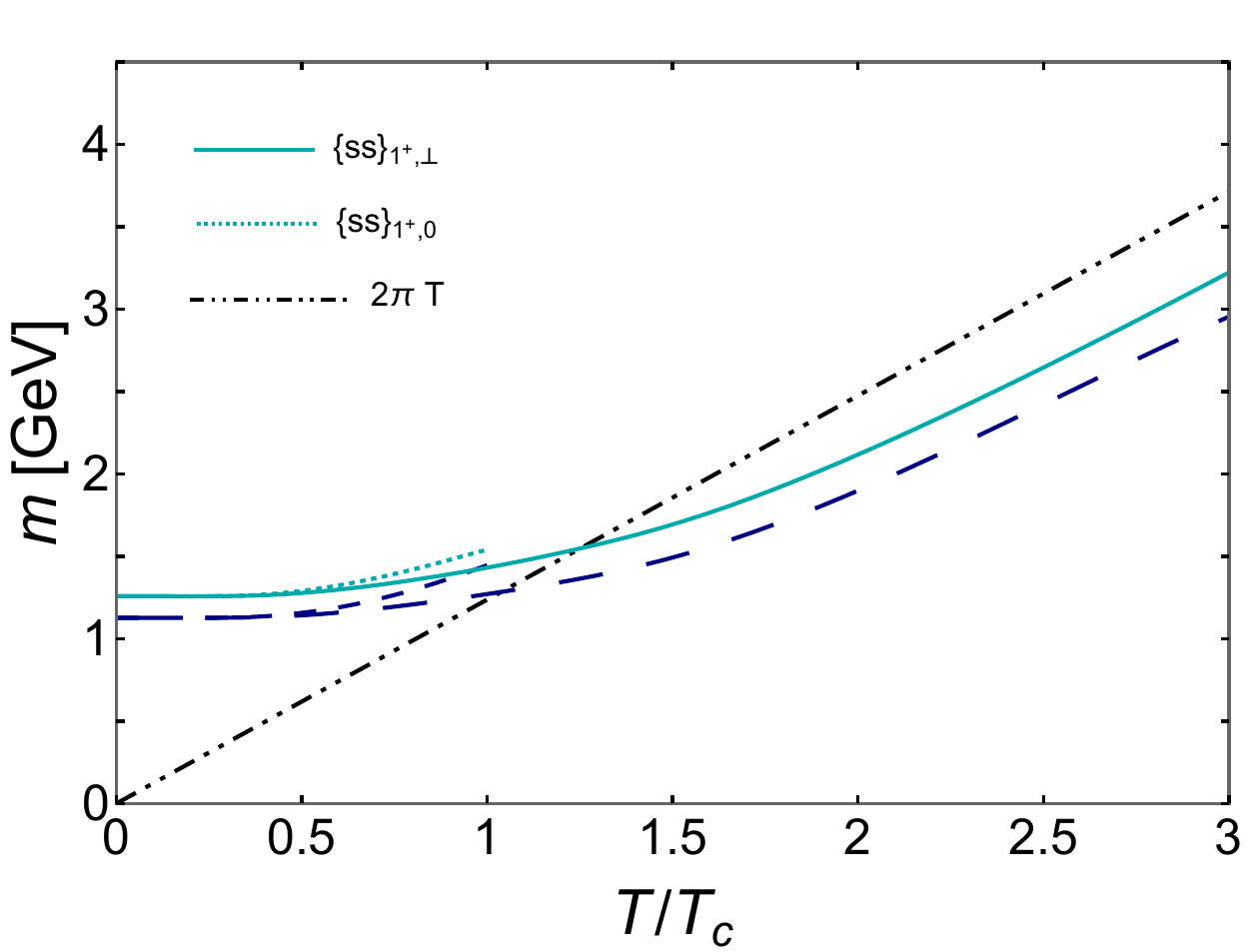}
\caption{(Left to right) Screening masses of diquarks with strangeness $S=0,1,2$, in comparison with the results of their meson partners from Fig.\,\ref{fig:mesons}. In each figure: \emph{gold curve}: scalar diquark; \emph{coral curve}: pseudoscalar diquark; \emph{cyan curves}: axial-vector diquark; \emph{magenta curves}: vector diquark; and \emph{black dot-dot-dashed curve}: free theory limit of $m=2\pi T$. For the $J=1$ axial-vector and vector diquarks, transverse modes are traced with solid lines and longitudinal modes with dotted lines. For the legends of mesons: the colours are the same as those in Fig.\,\ref{fig:mesons}, except that the \emph{long dashed curves} represent the $J=0$ and the transverse modes of the $J=1$ mesons, and the \emph{short dashed curves} are for the longitudinal modes of the $J=1$ mesons.
\label{fig:diquarks}}
\end{figure*}

It is straightforward to show that, using RL truncation, the Bethe-Salpeter equation of a $J^P$ diquark can be derived from that for its $J^{-P}$ meson partner directly, \emph{viz.}, simply multiplying the meson kernel by a factor 1/2\,\cite{Cahill:1987qr}. As a result, using Eqs.\,\eqref{sciker} and \eqref{bsem}, the Bethe-Salpeter equation for a $J^P$ diquark constituted from valence $f$- and $g$-quarks is expressed in the following form
\begin{align}
\label{bsedq}
\nonumber
\Gamma^{C,J^P}_{[fg]}(Q_0;T)
= &-\frac{1}{2}\frac{16\pi}{3}\frac{\alpha_{\rm IR}}{m_G^2}\int_{l,dq}\gamma_\mu S_f(q+Q_0;T)\\
&\times \Gamma^{C,J^P}_{[fg]}(Q_0;T)S_g(q;T)\gamma_\mu\,,
\end{align}
where 
\begin{align}
\Gamma^{C,J^P}_{[fg]}(Q_0;T) := 	\Gamma^{J^P}_{[fg]}(Q_0;T)C^\dagger\,,
\end{align}
with $\Gamma^{J^P}_{[fg]}$ being the diquark amplitude, and $C=\gamma_2\gamma_4$  the charge-conjugation matrix.

Because of the similarity, the Bethe-Salpeter amplitude for a $J^P$ diquark takes the same form as its $J^{-P}$ meson partner. For example, for a scalar ($0^+$) diquark, the amplitude is
\begin{align}
\label{bsadqsc}
\Gamma^{C,0^+}_{[fg]}(Q_0;T) &= i\gamma_5E^{0^+}_{[fg]}(T)+\frac{1}{2M_{fg}}\gamma_5\gamma\cdot Q_0F^{0^+}_{[fg]}(T)\,.
\end{align}	
And the Bethe-Salpeter amplitude for its parity partner, the pseudoscalar ($0^-$) diquark, is as follows:
\begin{align}
\label{bsadqps}
\Gamma^{C,0^-}_{[fg]}(Q_0;T) = {\mathbb I}_D E^{0^-}_{[fg]}(T)\,.
\end{align}	
Clearly, Eqs.\,\eqref{bsadqsc} and \eqref{bsadqps} are precisely the same as those of their meson partners, Eqs.\,\eqref{bsamps} and \eqref{bsamsc}.

For $J=1$ diquarks, their Bethe-Salpeter amplitudes also separate into longitudinal and transverse parts, namely,
\begin{align}
\label{bsadqax}
\Gamma^{C,1^+}_{[fg]}(Q_0;T) =
\left\{\begin{array}{l}
\gamma_4 \, E^{1^+,\parallel}_{[fg]}(T) \\
\vec{\gamma}_\perp E^{1^+,\perp}_{[fg]}(T)
\end{array}\right.\,,
\end{align}
for the axial-vector ($1^+$) diquark, and 
\begin{align}
\label{bsadqvc}
\Gamma^{C,1^-}_{[fg]}(Q_0;T) =
\left\{\begin{array}{l}
\gamma_5\gamma_4 \, E^{1^-,\parallel}_{[fg]}(T) \\
\gamma_5\vec{\gamma}_\perp E^{1^-,\perp}_{[fg]} (T)
\end{array}\right.\,,
\end{align}
for the vector ($1^-$) diquark. And, once again, they are the same as Eqs.\,\eqref{bsamvc} and \eqref{bsamax} separately.

In Eqs.\,\eqref{bsadqsc}-\eqref{bsadqvc}, we only wrote out the momentum-Dirac parts. The full Bethe-Salpeter amplitude of a diquark should also contain the colour and flavour parts. The colour part of each diquark amplitude is an antitriplet, and by combining it with the bystander colour-triplet quark, the whole baryon is a colour-singlet. Regarding the flavour part, the situation is more complicated. To make an amplitude antisymmetric under the exchange between two valence quarks, in the flavour-SU(3) sector, the flavour matrices of the scalar, pseudoscalar, and vector diquark correlations are antitriplets, while sextet for the axial-vector diquarks\footnote{Using the SCI-RL kernel, a Bethe-Salpeter amplitude does not depend on the relative momentum, therefore, a flavour-sextet vector diquark does not exsit, which, however, do exist by using a more complicated  kernel\,\cite{Chen:2017pse,Liu:2022ndb,Liu:2022nku,Eichmann:2016hgl}.}. The explicit expressions of the diquarks' flavour matrices are listed in Appendix\,\ref{appdqf}. In the diquark Bethe-Salpeter equation, Eq.\,\eqref{bsedq}, the colour and flavour matrices have been traced out and included in the coefficient in front of the integral, one only needs to consider the momentum-Dirac parts given in Eqs.\,\eqref{bsadqsc}-\eqref{bsadqvc}.

In the last section, for scalar and axial-vector mesons, we introduced the factors  in Eqs.\,\eqref{gsom} to simulate the SO effects, therefore, the flaw of RL-truncation can be overcome partly. It was found that for their diquark partners, \emph{i.e.}, the pseudoscalar and vector diquarks, the similar factors are also necessary\,\cite{Chen:2012qr,Lu:2017cln,Yin:2021uom}. We define
\begin{subequations}
\label{gsodq}
\begin{align}
\label{gsopsdq}
{\mathfrak g}_{\rm SO}^{qq,0^-}(T) &= 1-\frac{M_u(T)}{M_u(0)}(1-{0.58}^2)\,,\\
\label{gsovcdq}
{\mathfrak g}_{\rm SO}^{qq,1^-}(T) &= 1-\frac{M_u(T)}{M_u(0)}(1-{0.45}^2)\,,
\end{align}
\end{subequations}
for pseudoscalar ($0^-$) and vector ($1^-$) diquarks separately. At $T=0$, one has
\begin{subequations}
\begin{align}
{\mathfrak g}_{\rm SO}^{qq,0^-}(0) &= {0.58}^2\,,\\
{\mathfrak g}_{\rm SO}^{qq,1^-}(0) &= {0.45}^2\,.
\end{align}
\end{subequations}
These values, used in Refs.\,\cite{Lu:2017cln,Yin:2021uom}, were selected to ensure that the calculated baryon masses are realistic. For $T\gtrsim T_c$, ${\mathfrak g}_{\rm SO}^{qq,0^-}$ and ${\mathfrak g}_{\rm SO}^{qq,1^-}$ trend to 1. 

The explicit expression of the Bethe-Salpeter equation for each diquark correlation is given in Appendix\,\ref{appbse}. The calculated screening masses of diquarks are depicted in Fig.\,\ref{fig:diquarks}. Table\,\ref{tab:mdsmasses} lists the corresponding $T=0$ values, which equal the diquark masses obtained in Refs.\,\cite{Lu:2017cln}. These values have been discussed in full detail previously, so we will not reiterate those arguments here. For nonzero temperatures, the $T$-dependent behaviour of the screening mass of each diquark resembles that of its corresponding meson partner. When $T>T_c$, each parity-partner-pair trends to degenerate. Analogous to mesons, \emph{e.g.}, Eqs.\,\eqref{cdtpisig} and \eqref{cdtrhoa1}, we can also define the critical degenerate temperatures for diquark parity-partner-pairs. For example, we define
\begin{align}
\nonumber
&T_{c}^{\delta m_{0^\mp}^{S=0}}\\
:= &\bigg\{T>0 \,\bigg| \, \frac{m_{[ud]_{0^-}}(T) - m_{[ud]_{0^+}}(T)}{m_{[ud]_{0^-}}(T) + m_{[ud]_{0^+}}(T)}=0.5\% \bigg\}\,,
\end{align}
for the light-light $J=0$ diquark correlations; and define
\begin{align}
\nonumber
&T_{c}^{\delta m_{1^\mp}^{\perp,S=1}}\\
:= &\bigg\{T>0  \,\bigg| \, \frac{m_{\{us\}_{1^-}^\perp}(T) - m_{\{us\}_{1^+}^\perp}(T)}{m_{\{us\}_{1^-}^\perp}(T) + m_{\{us\}_{1^+}^\perp}(T)}=0.5\% \bigg\}\,,
\end{align}
for the transverse modes of the light-strange $J=1$ diquarks. The results are
\begin{subequations}
\label{tmcdq0}
\begin{align}
T_{c}^{\delta m_{0^\mp}^{S=0}} &= 1.59\,T_c\,,\\
T_{c}^{\delta m_{0^\mp}^{S=1}} &= 1.43\,T_c\,,
\end{align}
\end{subequations}
and
\begin{subequations}
\label{tmcdq1}
\begin{align}
T_{c}^{\delta m_{1^\mp}^{\perp,S=0}} &= 1.34\,T_c\,,\\
T_{c}^{\delta m_{1^\mp}^{\perp,S=1}} &= 1.40\,T_c\,.
\end{align}
\end{subequations}
Comparing these values with Eqs.\,\eqref{cdtmj0} and \eqref{cdtmj1}, it is clear that each value is smaller than that of the meson partners. This indicates that the diquarks are becoming degenerate more quickly. Besides, the qualitative features are the same as those of mesons, namely, for each $S$, the critical degenerate temperature of the $J=1$ pair is lower than that of the $J=0$ pair. And for the $J=1$ pairs, $T_c^{\delta m}$ increases with $S$; but it decreases for the $J=0$ pairs.

%%%%%%%%%%%%%%%%%%%%%%%%%%%%%%%%%%%%%%%%%%%%%%%%%%%%%%%%%%%%%%%%%%%%%%%%%%%%%%
%%%%%%%%%%%%%%%%%%%%%%%%%%%%%%%%%%%%%%%%%%%%%%%%%%%%%%%%%%%%%%%%%%%%%%%%%%%%%%

\section{Screening masses: baryons}
\label{secbaryon}

The Faddeev equation for baryons, depicted in Fig.\,\ref{figFaddeev}, describes the decomposition and reformation of diquarks through the exchange of a dressed quark within a baryon. $\Psi_B$ represents the Faddeev amplitude for baryon $B$, is the analogy of the Bethe-Salpeter amplitude for a meson or diquark, and it can be decomposed into the components associated with different diquark correlations. For example, for a spin-1/2 baryon, one has 
\begin{align}
\Psi_B = \Psi^1_B + \Psi^2_B + \Psi^3_B\,,
\end{align}
where the superscript refers the bystander quark, and the three components are related by each other via a cyclic permutation. In the following we will use $\Psi^3_B$ for illustration. In principle, a spin-1/2 baryon should include all four kinds of diquarks mentioned in Section\,\ref{secdiquark}, therefore, its Faddeev amplitude can be expressed as the sum of the contributions from each diquark:
\begin{align}
\Psi^3_B(p_j,\alpha_j,f_j;T) = {\mathcal N}^{0^+}_{\Psi^3_B}+{\mathcal N}^{1^+}_{\Psi^3_B}+{\mathcal N}^{0^-}_{\Psi^3_B}+{\mathcal N}^{1^-}_{\Psi^3_B}\,,
\end{align}
with $(p_j,\alpha_j,f_j)$ the momentum, spin, and flavour labels of the valence quarks of the baryon, and $Q_0=p_1+p_2+p_3=p_d+p_q=k_d+k_q$ the baryon's total momentum.

As we discussed in Section\,\ref{secdiquark}, the scalar ($0^+$), pseudoscalar  ($0^-$), and vector ($1^-$) diquarks are all flavour-antitriplets, with the flavour matrices ${\tt t}^{[f_1f_2]}\equiv{\tt t}^{1,2,3}$ given in Eq.\,\eqref{flavourarrays1}, and the Faddeev amplitudes take the form
\begin{align}
\label{octba3}
\nonumber
&{\mathcal N}^{J^P\in\{0^+,0^-,1^-\}}_{\Psi^3_B}(p_j,\alpha_j,f_j;T)\\
\nonumber
=&\sum_{[f_1f_2]f_3\in\Psi_B}\bigg[{\tt t}^{[f_1f_2]}\Gamma^{J^P}_{[f_1f_2]}(k_d;T)\bigg]^{f_1f_2}_{\alpha_1\alpha_2}\\
&\times\Delta_{[f_1f_2]}^{J^P}(k_d;T)\big[{\mathcal F}^B_{[f_1f_2]}(\ell,Q_0;T)u^B(Q_0;T)\big]^{f_3}_{\alpha_3}\,,
\end{align}
where $k_d=p_1+p_2$ is the momentum of the diquark correlation, $\ell=(2p_3-p_1-p_2)/3$ is the relative momentum between diquark and the bystander quark; $\Delta^{J^P}_{[f_1f_2]}$ is the propagator of the diquark with quantum number $J^P$; $\Gamma^{J^P}_{[f_1f_2]}$ is the canonically normalised Bethe-Salpeter amplitude of the diquark correlation, which is the solution of the corresponding Bethe-Salpeter equation \eqref{bsedq}, and detailed discussion is given in in Section\,\ref{secdiquark}; ${\mathcal F}^B$ is the quark-diquark correlation matrix, it is a linear combination of all allowed Dirac structures, with the coefficients the associated amplitudes; and $u^B$ is the Dirac spinor of the baryon. 

On the other hand, the axial-vector ($1^+$) diquarks are flavour-sextet, their flavour matrices are listed in Eq.\,\eqref{flavourarrays2}, and the Faddeev amplitude can be written as
\begin{align}
\label{octb6}
\nonumber
&{\mathcal N}^{1^+}_{\Psi^3_B}(p_j,\alpha_j,f_j;T)\\
\nonumber
=&\sum_{\{f_1f_2\}f_3\in\Psi_B}\bigg[{\tt t}^{\{f_1f_2\}}\Gamma^{1^+}_{\{f_1f_2\}}(k_d;T)\bigg]^{f_1f_2}_{\alpha_1\alpha_2}\\
&\times\Delta_{\{f_1f_2\}}^{1^+}(k_d;T)\big[{\mathcal F}^B_{\{f_1f_2\}}(\ell,Q_0;T)u^B(Q_0;T)\big]^{f_3}_{\alpha_3}\,.
\end{align}

As for the spin-3/2 baryons, they can only have axial-vector diquark correlations because of the constraint from spin and isospin. Therefore, their Faddeev amplitudes have the form
\begin{align}
\widetilde{\Psi}^3_B(p_j,\alpha_j,f_j;T) = \widetilde{\mathcal N}^{1^+}_{\widetilde{\Psi}^3_B}\,,
\end{align}
with
\begin{align}
\label{decpb6}
\nonumber
&\widetilde{\mathcal N}^{1^+}_{\widetilde{\Psi}^3_B}(p_j,\alpha_j,f_j;T)\\
\nonumber
=&\sum_{\{f_1f_2\}f_3\in\widetilde{\Psi}_B}\bigg[{\tt t}^{\{f_1f_2\}}\Gamma^{1^+}_{\{f_1f_2\}}(k_d;T)\bigg]^{f_1f_2}_{\alpha_1\alpha_2}\\
&\times\Delta_{\{f_1f_2\}}^{1^+}(k_d;T)\big[{\mathcal D}_{{\{f_1f_2\}}\nu\rho}^B(\ell,Q_0;T)u_\rho^B(Q_0;T)\big]^{f_3}_{\alpha_3}\,,
\end{align}
where $u_\rho$ is a Rarita-Schwinger spinor. 

Previous CSMs studies for baryon spectrum at $T=0$ found that for the negative-parity spin-1/2 baryons, the computed masses are too light. The reason is also from the oversimplification of the RL truncation, which is inappropriate to describe those systems that strongly depend on the SO effects, such as the negative-parity spin-1/2 baryons. To remedy this, a multiplying factor ${\mathfrak g}_{\rm DB}^{P_BP_d}$ for the diquark amplitudes was used in the channels where the diquark parities $P_d$ are different from the baryon parity $P_B$, see Refs.\,\cite{Chen:2017pse,Lu:2017cln,Yin:2021uom}. In this article, we still use this strategy, however, it should be modified to express the effect from nonzero temperatures. Analogous to Eqs.\,\eqref{gsom} and \eqref{gsodq}, we define
\begin{align}
\label{gsomsc}
{\mathfrak g}_{\rm DB}^{P_BP_d}(T) = 
\left\{\begin{array}{l}
1:\,{\rm if}\,P_B=P_d  \\
1-\frac{M_u(T)}{M_u(0)}(1-{0.1}^2):\,{\rm if}\,P_B=-P_d \\	
\end{array}\right.\,.
\end{align}
As a result, in Faddeev equation calculations, a diquark's Bethe-Salpeter amplitude $\Gamma^{J^{P_d}}_{[fg]}(Q_0;T)$ should be modified as
\begin{align}
{\mathfrak g}_{\rm DB}^{P_BP_d}(T)\times\Gamma^{J^{P_d}}_{[fg]}(Q_0;T).	
\end{align}
One should also notice that at $T=0$, ${\mathfrak g}_{\rm DB}^{P_BP_d}$ takes the value given in Refs.\,\cite{Lu:2017cln,Yin:2021uom}.

At $T\neq0$, except for the quark propagators and the diquark Bethe-Salpeter amplitudes, the diquark propagators, the Faddeev amplitudes and the spinors of baryons, should also depend on $T$, as one can see in Eqs.\,\eqref{octba3}, \eqref{octb6}, and \eqref{decpb6}. In this study, we follow the method used in Ref.\,\cite{Wang:2013wk}, \emph{i.e.}, using the $T$-dependent quark masses, the diquarks' screening masses and the normalised Bethe-Salpleter amplitudes\,\footnote{For the $J^P=1^\pm$ diquark correlations, we use the results from the transverse modes in the Faddeev equations.}. Consequently, the diquark propagators are also $T$-dependent and take the form
\begin{subequations}
\begin{align}
\Delta^{0^\pm}(k;T) &= \frac{1}{k^2 + (m_{0^\pm}(T))^2}\,,\\
\Delta^{1^\pm}_{\mu\nu}(k;T) &= \frac{1}{k^2 + (m_{1^\pm}(T))^2}\bigg(\delta_{\mu\nu}+\frac{k_\mu k_\nu}{(m_{1^\pm}(T))^2}\bigg)\,,
\end{align}
\end{subequations}
where $m_{0^\pm}(T)$ and $m_{1^\pm}(T)$ are the diquark screening masses. On the other hand, for simplicity, we use the baryon spinors at $T=0$, as outlined in Appendix\,D of Ref.\,\cite{Wang:2013wk}.

It is noticeable that the quark-diquark correlation matrices, ${\mathcal F}^B$ and ${\mathcal D}^B_{\nu\rho}$ in Eqs.\,\eqref{octba3}, \eqref{octb6}, and \eqref{decpb6}, should depend on the relative momentum $\ell$, even using the SCI-RL kernel. In this study, we utilise the method used in Ref.\,\cite{Wang:2013wk}, the key point is selecting a representative value for the momentum ratio of a diquark in relation to that of the baryon\footnote{Two step-by-step examples are provided in Appendix\,D of Ref.\,\cite{Wang:2013wk}.}. As a result, the dependence of the relative momentum can be removed, and then the correlation matrices ${\mathcal F}^B$ in Eq.\,\eqref{octba3} of a spin-1/2 baryon should include the following Dirac structures: 
\begin{subequations}
\label{octetdiraca3}
\begin{align}
{\mathcal S}_a^{B}(Q_0;T) &= {\mathfrak s}_a^{B}(T){\mathbb I}_{\rm D}{\mathcal G}^\pm\,,\\
{\mathcal P}_a^{B}(Q_0;T) &= {\mathfrak p}_a^{B}(T)i\gamma_5{\mathcal G}^\pm\,,\\
\label{octetdiraca3ax}
{\mathcal V}^{B}_{a,\mu}(Q_0;T) &= \big[{\mathfrak v}^{B}_{a,1}(T)i\gamma_\mu + {\mathfrak v}^{B}_{a,2}(T){\mathbb I}_{\rm D}(\hat{Q}_0)_\mu\big]{\mathcal G}^\pm\,.
\end{align}	
\end{subequations}
And the expression of ${\mathcal F}^B$ in Eq.\,\eqref{octb6} is
\begin{align}
\label{octetdirac6}
{\mathcal A}^{B}_{a,\mu}(Q_0;T) &= \big[{\mathfrak a}^{B}_{a,1}(T)i\gamma_5\gamma_\mu + {\mathfrak a}^{B}_{a,2}(T)\gamma_5(\hat{Q}_0)_\mu\big]{\mathcal G}^\pm\,,
\end{align}
In Eqs.\,\eqref{octetdiraca3} and \eqref{octetdirac6}, ${\mathcal G}^{+(-)}={\mathbb I}_{\rm D}(\gamma_5)$, $(\hat{Q}_0)_\mu=(Q_0)_\mu/(im_B(T))$ with $m_B(T)$ the baryon screening mass; the index $a$ indicates the specific diquark content for different baryons, see Appendices\,\ref{appdqf} and \ref{appfad}; and ${\mathfrak s}_a^{B}$, ${\mathfrak p}_a^{B}$, ${\mathfrak v}^{B}_{a,1(2)}$, ${\mathfrak a}^{B}_{a,1(2)}$ are scalar coefficients, they are all $T$-dependent, and determined via solving the Faddeev equation.

As for a spin-3/2 baryon, using the SCI-RL kernel and removing the $\ell$-dependence, its correlation matrix in Eq.\,\eqref{decpb6} is
\begin{align}
{\mathcal D}^{B}_{a,\mu\nu}(Q_0;T) = {\mathfrak d}_a^{B}(T){\mathbb I}_{\rm D}\delta_{\mu\nu}{\mathcal G}^\pm\,,
\end{align}
with $ {\mathfrak d}_a^{B}$ the $T$-dependent coefficient.

%%%%%%%%%%%%%%%%%%%%%%%%%%%%%%%%%%%%%%%%%%%%%%%%%%%%%%%%%%%%%%%%%%%%%%%%%%%%%%

\subsection{$J^P=1/2^\pm$}

\begin{table*}[t]
\caption{\label{OctetDecupletMasses}
Computed mass for each ground-state $J^P=1/2^\pm$ baryon at $T=0$.
Row~1: Baryon ground-states. 
Row~3: Parity partners of the baryon ground-states.
Masses in the rows labelled ``expt.'' are empirical mass values, taken from Ref.\,\cite{ParticleDataGroup:2024cfk}. ``--'' indicates no empirical results available for comparison.
(All Masses are listed in GeV.)
}
\begin{center}
\begin{tabular*}%{|c|c|c|c|c|c|c|}\hline
{\hsize}
{
l@{\extracolsep{0ptplus1fil}}
l@{\extracolsep{0ptplus1fil}}
|l@{\extracolsep{0ptplus1fil}}
l@{\extracolsep{0ptplus1fil}}
l@{\extracolsep{0ptplus1fil}}
l@{\extracolsep{0ptplus1fil}}
|l@{\extracolsep{0ptplus1fil}}
l@{\extracolsep{0ptplus1fil}}
l@{\extracolsep{0ptplus1fil}}
l@{\extracolsep{0ptplus1fil}}}\hline\hline
& & \rule{0ex}{2.5ex}
$N$ & $\Lambda$ & $\Sigma$ & $\Xi$
   & $\Delta$ & ${\Sigma^\ast}$ & ${\Xi^\ast}$ & $\Omega$ \\\hline
$P=+$ & \rule{0ex}{2.5ex}
SCI & 1.09 & 1.20 & 1.26 & 1.36\,\, & 1.42 & 1.50 & 1.60 & 1.70\rule{0em}{2.5ex} \\
%
%& Ref.\,\cite{Lu:2017cln}\,\, & 1.14 & 1.26 & 1.36 & 1.43 & 1.39 & 1.51 & 1.63 & 1.76\rule{0em}{2.5ex} \\
%
& \rule{0ex}{2.5ex}
expt. & 0.94 & 1.12 & 1.19 &  1.31 & $1.23$ & $1.39$ & $1.53$ & 1.67\\\hline
$P=-$ & \rule{0ex}{2.5ex}
SCI & 1.85 & 1.95 & 1.97 & 2.05 & 1.78 & 1.89 & 2.00 & 2.11\rule{0em}{2.5ex} \\
%
%& Ref.\,\cite{Lu:2017cln}\,\, & 1.82 & 1.92 & 1.96 & 2.04 & 2.07 & 2.16 & 2.26 & 2.36\rule{0em}{2.5ex} \\
%
& \rule{0ex}{2.5ex}
expt. & $1.54$ & $1.67$ & $1.75$ & -- & $1.65$ & $1.67$ & $1.82$ & -- \\\hline\hline
\end{tabular*}
\end{center}
\end{table*}

\begin{figure*}[!t]
\begin{center}
\begin{tabular}{lr}
\includegraphics[clip,width=0.4\linewidth]{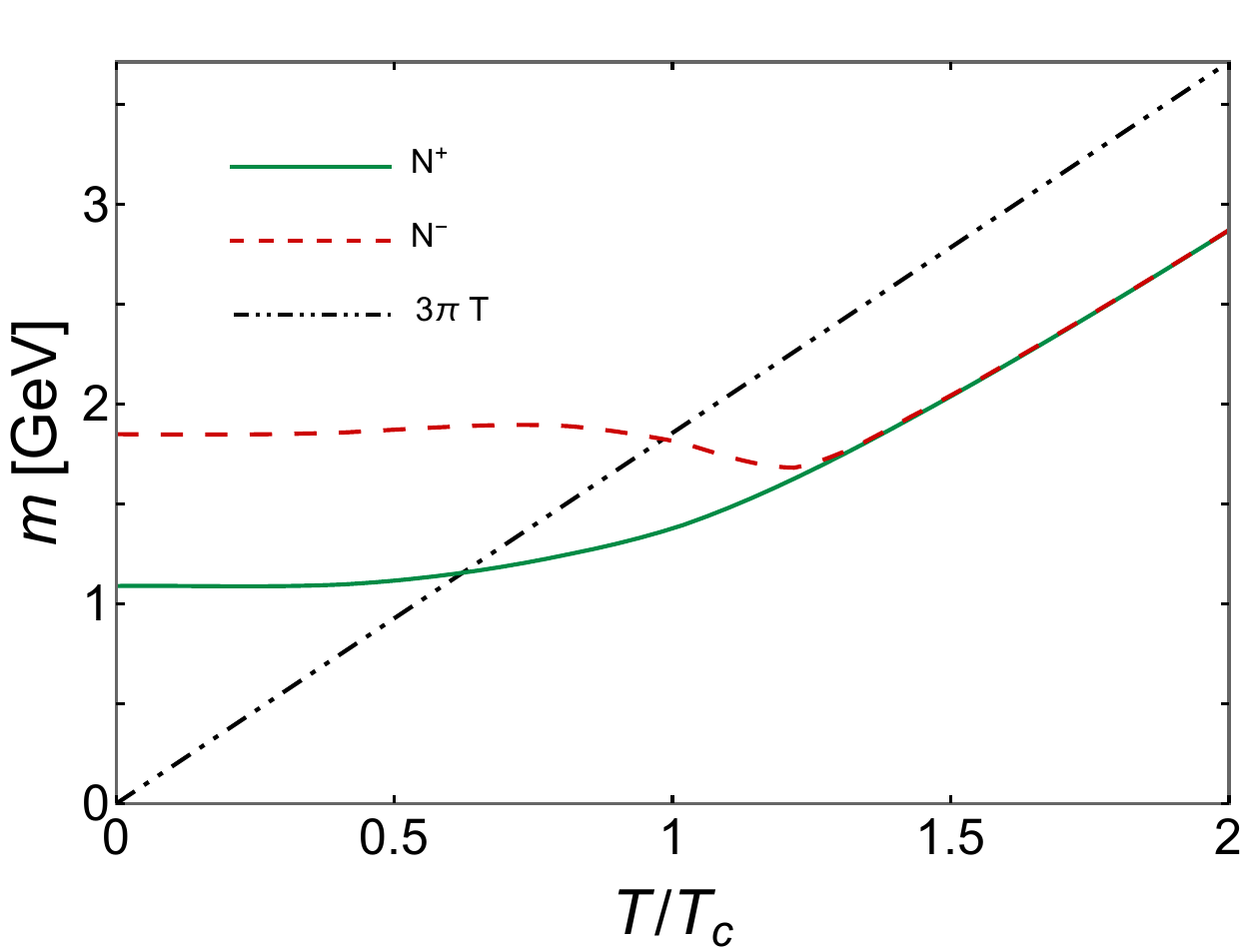}\hspace*{2ex } &
\includegraphics[clip,width=0.4\linewidth]{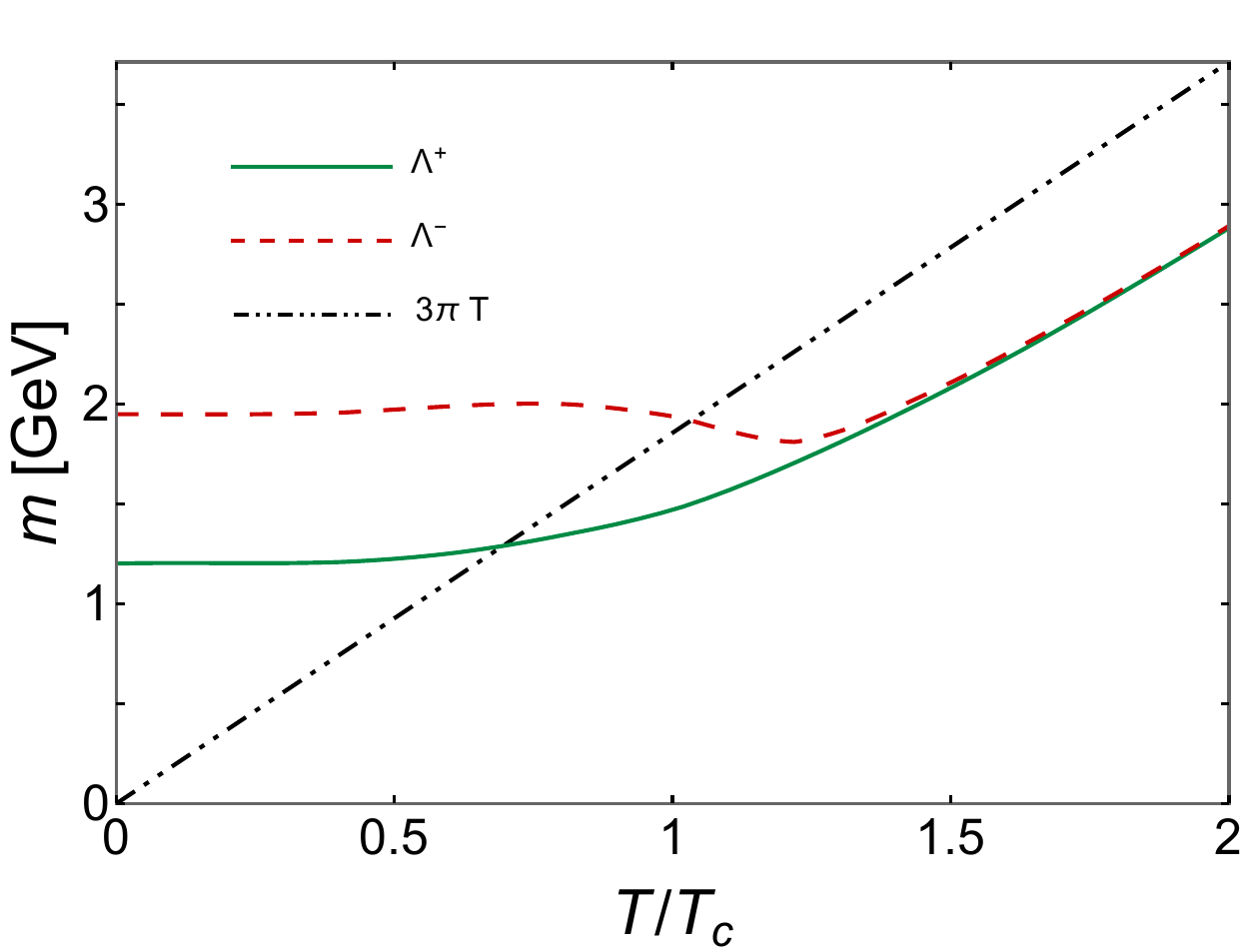}\vspace*{-0ex}
\end{tabular}
\begin{tabular}{lr}
\includegraphics[clip,width=0.4\linewidth]{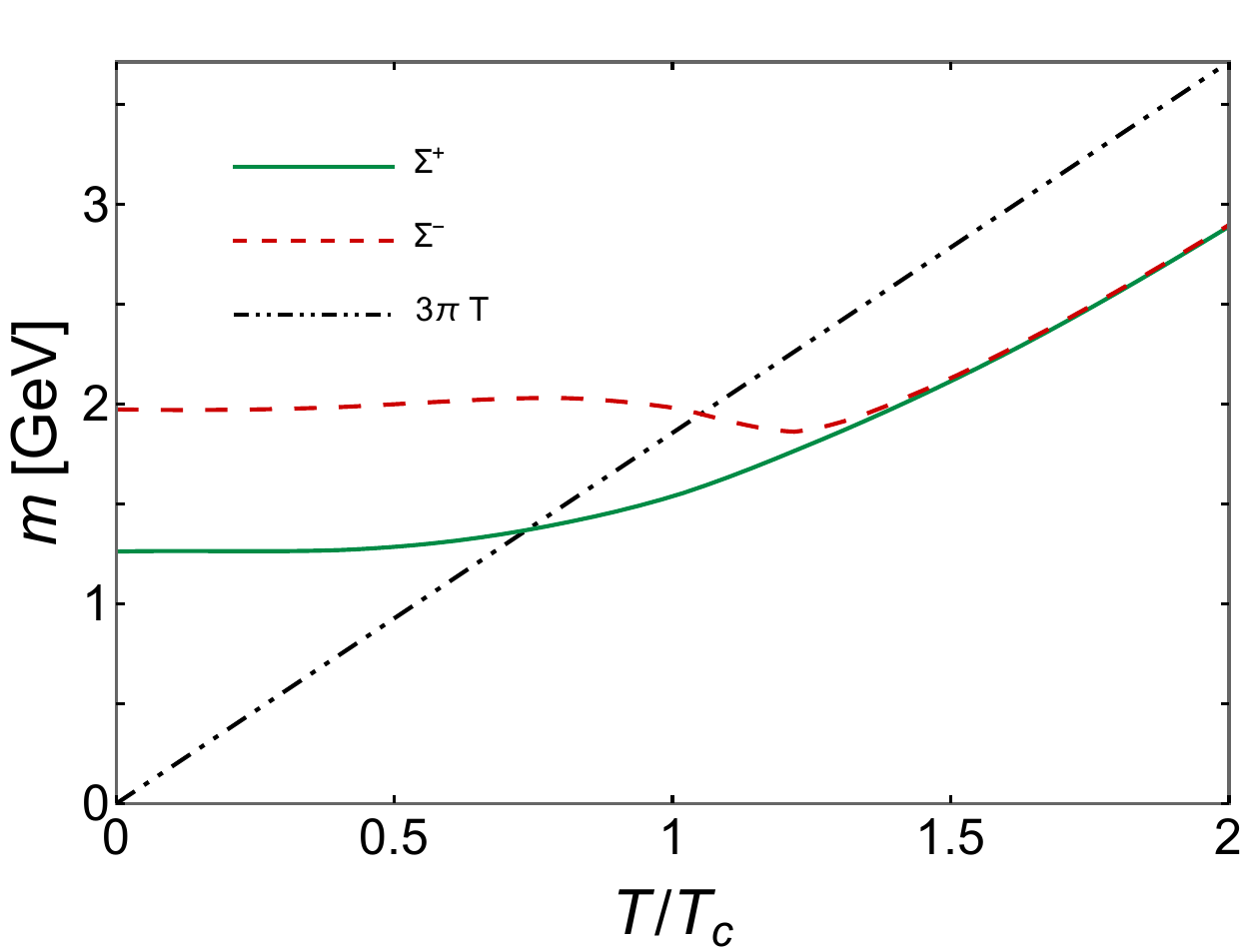}\hspace*{2ex } &
\includegraphics[clip,width=0.4\linewidth]{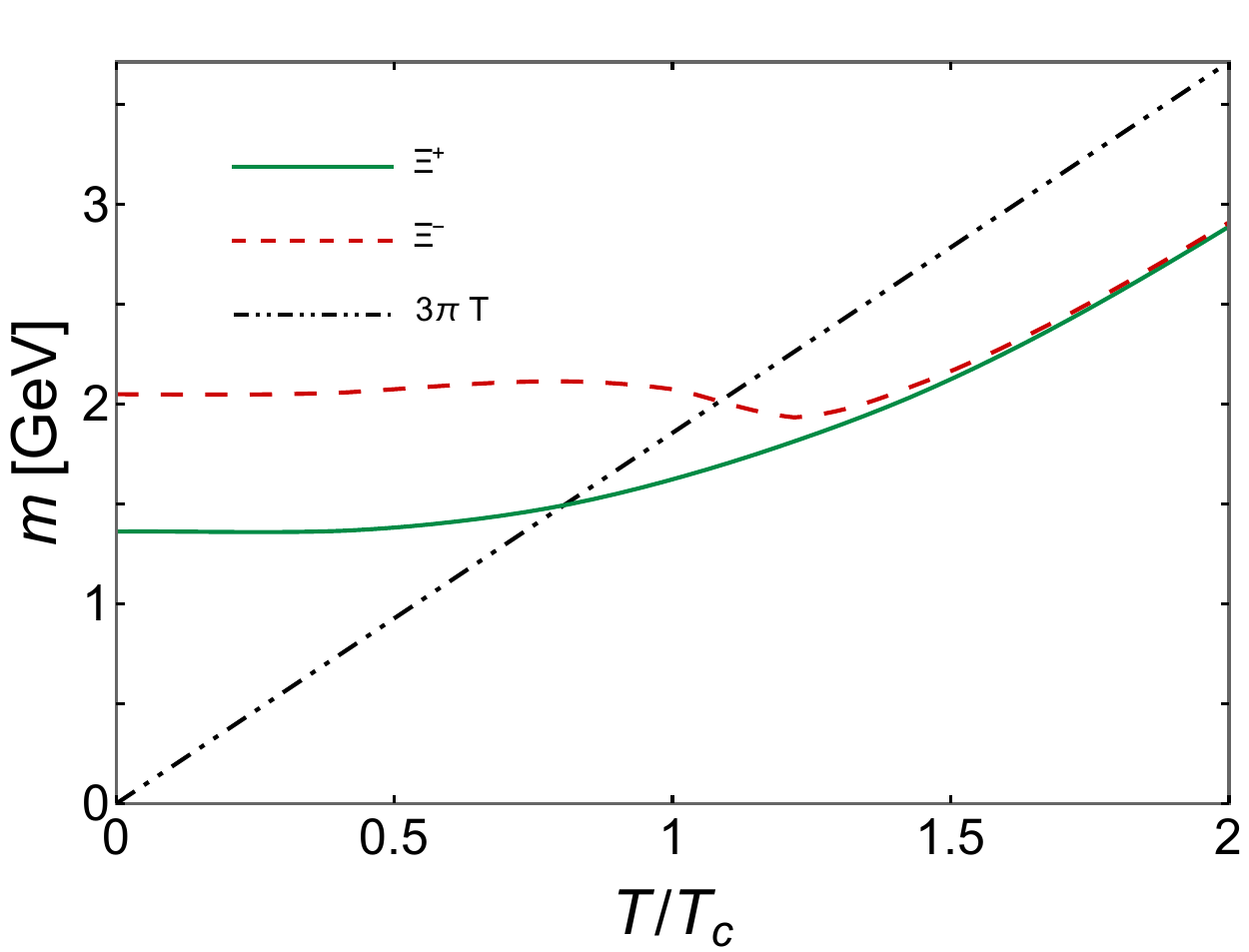}\vspace*{-1ex}
\end{tabular}
\end{center}
\caption{\label{fig:octet}
Screening masses of $J^P=1/2^\pm$ baryon ground states. In each figure: \emph{solid green curve}: positive-parity baryon; \emph{dashed red curve}: negative-parity baryon; and \emph{black dot-dot-dashed curve}: free theory limit of $m=3\pi T$.
}
\end{figure*}

\begin{table*}[t]
\caption{\label{FaddeevAmps}
Computed Faddeev amplitudes at $T=0$ (unit normalised), for each ground-state $J^P=1/2^\pm$ baryon.}
\begin{center}
\begin{tabular*}%{|c|c|c|c|c|c|c|}\hline
{\hsize}
{
l@{\extracolsep{0ptplus1fil}}
|c@{\extracolsep{0ptplus1fil}}
c@{\extracolsep{0ptplus1fil}}
c@{\extracolsep{0ptplus1fil}}
|c@{\extracolsep{0ptplus1fil}}
c@{\extracolsep{0ptplus1fil}}
c@{\extracolsep{0ptplus1fil}}
c@{\extracolsep{0ptplus1fil}}
c@{\extracolsep{0ptplus1fil}}
c@{\extracolsep{0ptplus1fil}}
c@{\extracolsep{0ptplus1fil}}
c@{\extracolsep{0ptplus1fil}}
c@{\extracolsep{0ptplus1fil}}
c@{\extracolsep{0ptplus1fil}}
|c@{\extracolsep{0ptplus1fil}}
c@{\extracolsep{0ptplus1fil}}
c@{\extracolsep{0ptplus1fil}}
|c@{\extracolsep{0ptplus1fil}}
c@{\extracolsep{0ptplus1fil}}
c@{\extracolsep{0ptplus1fil}}
c@{\extracolsep{0ptplus1fil}}
c@{\extracolsep{0ptplus1fil}}
c@{\extracolsep{0ptplus1fil}}
}\hline\hline

& $\mathfrak{s}_1^B$ & $\mathfrak{s}_2^B$ & $\mathfrak{s}_{\{23\}}^B$ &
    $\mathfrak{a}_{4,1}^B$ & $\mathfrak{a}_{4,2}^B$ & $\mathfrak{a}_{\{45\},1}^B$ & $\mathfrak{a}_{\{45\},2}^B$ &
    $\mathfrak{a}_{6,1}^B$ & $\mathfrak{a}_{6,2}^B$ & $\mathfrak{a}_{\{68\},1}^B$ & $\mathfrak{a}_{\{68\},2}^B$ &
   $\mathfrak{a}_{9,1}^B$ & $\mathfrak{a}_{9,2}^B$ & $\mathfrak{p}_{1}^B$ & $\mathfrak{p}_{2}^B$ & $\mathfrak{p}_{\{23\}}^B$
   & $\mathfrak{v}_{1,1}^B$ & $\mathfrak{v}_{1,2}^B$ & $\mathfrak{v}_{2,1}^B$ & $\mathfrak{v}_{2,2}^B$ & $\mathfrak{v}_{\{23\},1}^B$ & $\mathfrak{v}_{\{23\},2}^B$ 
\\\hline
$N^+$ & 0.86 & & & & & 0.36 & -0.35 & & & & & & & 0.01 & & & 0.04 & -0.05 & & & \\
$\Lambda^+$ & 0.52 & & 0.43 & & & & & & & -0.53 & 0.51 & & & 0.02 & & 0.01 & 0.01 & -0.02 & & & 0.02 & -0.03\\
$\Sigma^+$ & & 0.78 & & -0.37 & 0.34 & & & 0.28 & -0.26 & & & & & & 0.00 & & & & 0.03 & -0.03 & &\\
$\Xi^+$ & & 0.78 & & & & & & 0.24 & -0.33 & & & -0.29 & 0.28 & & 0.01 & & & & 0.02 & -0.03 & & \\\hline
%%%%
$N^-$ & 0.20 & & & & & 0.07 & -0.08 & & & & & & & 0.93 & & & 0.23 & -0.18 & & & \\
$\Lambda^-$ & 0.16 & & 0.12 & & & & & & & -0.13 & 0.13 & & & 0.66 & & 0.61 & 0.20 & -0.17 & & & 0.18 & -0.16\\
$\Sigma^-$ & & 0.20 & & -0.10 & 0.11 & & & 0.09 & -0.10 & & & & & & 0.94 & & & & 0.14 & -0.08 & &\\
$\Xi^-$ & & 0.17 & & & & & & 0.07 & -0.06 & & & -0.06 & -0.07 & & 0.90 & & & & 0.29 & -0.26 & & \\\hline\hline
%%%%
%
\end{tabular*}
\end{center}
\end{table*}

Our computed screening masses of the $J^P=1/2^\pm$ baryons are depicted in Fig.\,\ref{fig:octet}. First, we look at their values at $T=0$, listed in Table\,\ref{OctetDecupletMasses}. Clearly, each SCI result is uniformly larger than the corresponding empirical value. This is natural, just like the meson case. The Faddeev equation in Fig.\,\ref{figFaddeev} describes a baryon's dressed-quark-core, which excludes the effects of the meson cloud that can lower the observed masses in measurements. Our values are also in agreement with other SCI results obtained by using the so-called static approximation, which introduces one more parameter, see Refs.\,\cite{Lu:2017cln,Yin:2021uom}. Compared to the results therein, a simple calculation shows that the relative differences for the $J^P=1/2^+$ baryons are all smaller than $5\%$, except the $\Sigma^+$-baryon, which is $8\%$; and for the $J^P=1/2^-$ ones, the relative differences are all smaller than $2\%$. 

In Fig.\,\ref{fig:octet}, the screening masses of the four parity-partner-pairs have almost the same feature as those of mesons and diquarks described in the previous two sections: on the domain depicted, the screening masses of the positive parity baryons all increase slowly with temperature for $T\lesssim T_c$, and for $T>T_c$, they grow more rapidly, but always smaller than the free theory value $3\pi T$; whereas their parity partners are all almost invariant for $T\lesssim T_c$, when $T$ gets close to $T_c$, the screening masses decrease, and at one temperature around $1.25\,T_c$, they increase once again, and quickly merge with every positive parity partner; therefore, chiral symmetry restores.  

The critical degeneracy temperature $T^{\delta m}_c$ can be helpful in studying the screening mass degeneracy quantitatively. For example, for the nucleon and its parity partner, we define
\begin{align}
\label{cdtba}
\nonumber
&T_{c}^{\delta m_{N^\mp}^{S=0}}\\
:= &\bigg\{T>0  \,\bigg| \, \frac{m_{N^-}(T) - m_{N^+}(T)}{m_{N^-}(T) + m_{N^+}(T)}=0.5\% \bigg\}\,.
\end{align}
And that for the other three parity-partner-pairs can be defined similarly. The results are
\begin{subequations}
\begin{align}
T_c^{\delta m^{S=0}_{N^\pm}} &= 1.29\,T_c\,,\\
T_c^{\delta m^{S=1}_{{\Lambda}^{\pm}}} &= 1.63\,T_c\,,\\
T_c^{\delta m^{S=1}_{{\Sigma}^{\pm}}} &= 1.45\,T_c\,,\\
T_c^{\delta m^{S=2}_{{\Xi}^{\pm}}} &= 1.78\,T_c\,.
\end{align}
\end{subequations}
We find that $T^{\delta m}_c$ also increases with strangeness, just like the results of $J=1$ mesons and diquarks. On the other side, although both have $S=1$, $T^{\delta m}_c$ for $\Lambda^\pm$ is larger than that of $\Sigma^\pm$. Checking their flavour contents, Eqs.\,\eqref{lambdadq} and \eqref{sigmadq}, one possible reason is that $\Lambda^\pm$ contain the scalar and pseudoscalar diquark correlations $[ud]_{0^+}$ and $[ud]_{0^-}$, they have the largest $T^{\delta m}_c$ in diquark parity-partner-pairs, see Eqs.\,\eqref{tmcdq0} and \eqref{tmcdq1}.

\begin{figure*}[!t]
\begin{center}
\begin{tabular}{lr}
\includegraphics[clip,width=0.4\linewidth]{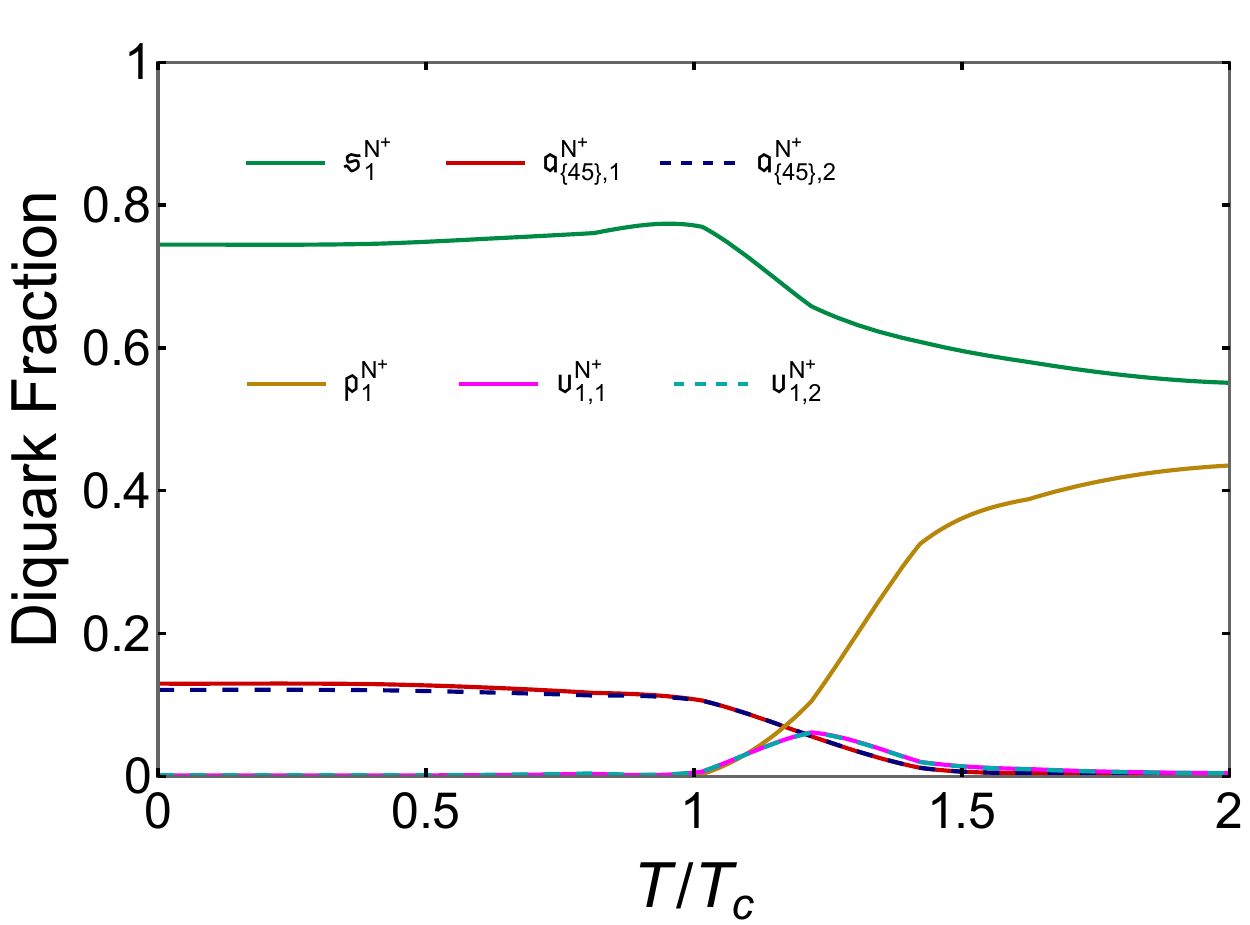}\hspace*{2ex } &
\includegraphics[clip,width=0.4\linewidth]{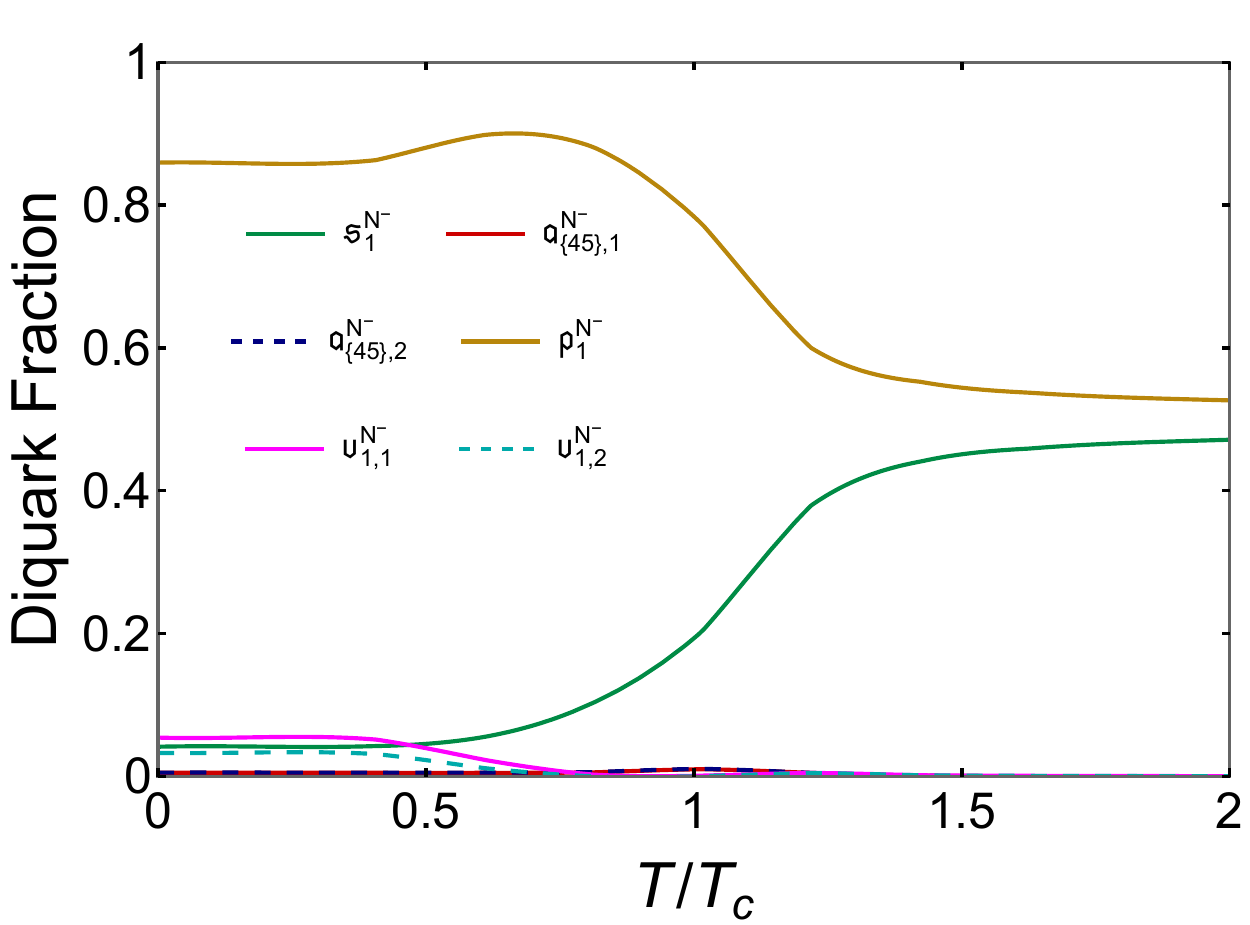}\vspace*{-0ex}
\end{tabular}
\begin{tabular}{lr}
\includegraphics[clip,width=0.4\linewidth]{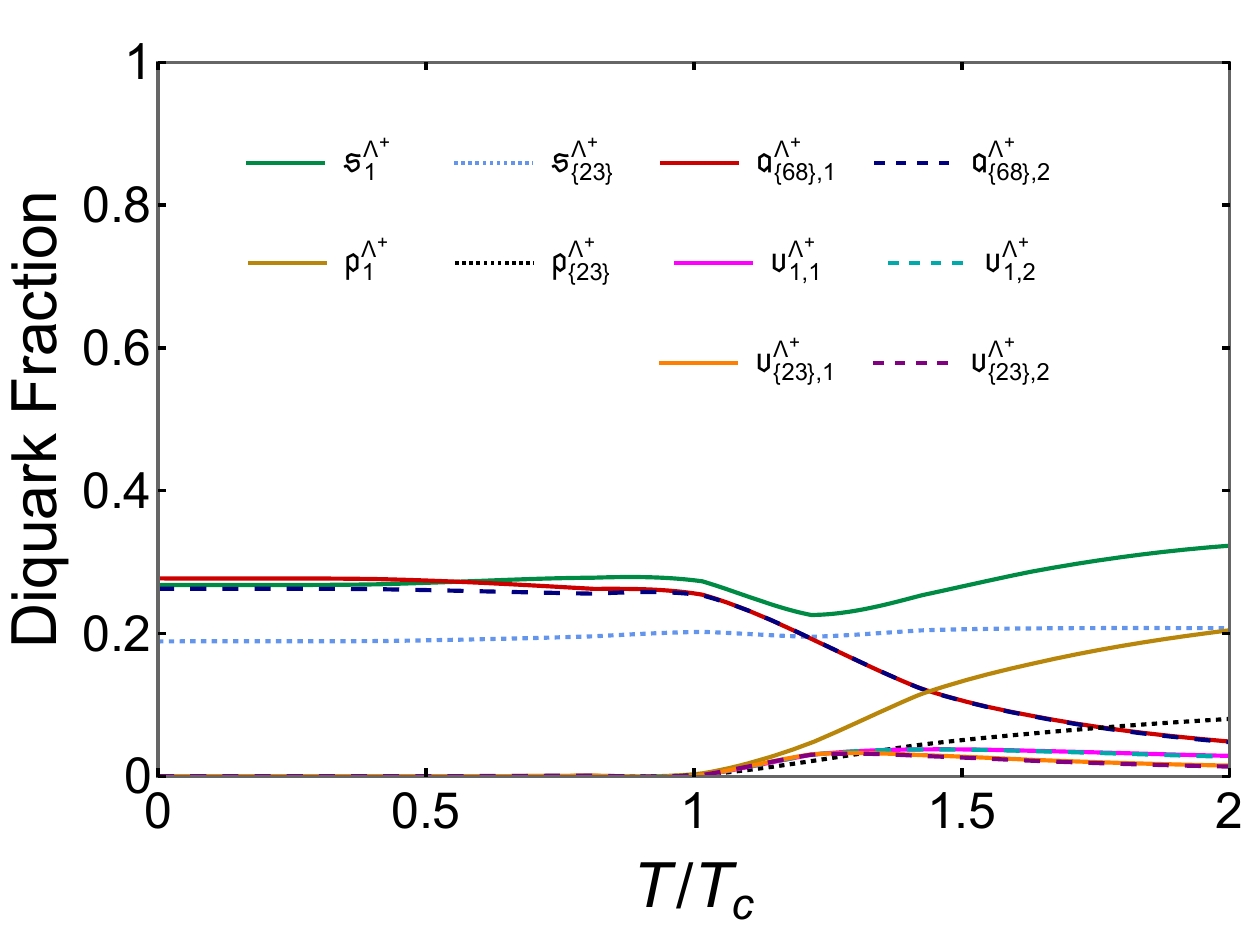}\hspace*{2ex } &
\includegraphics[clip,width=0.4\linewidth]{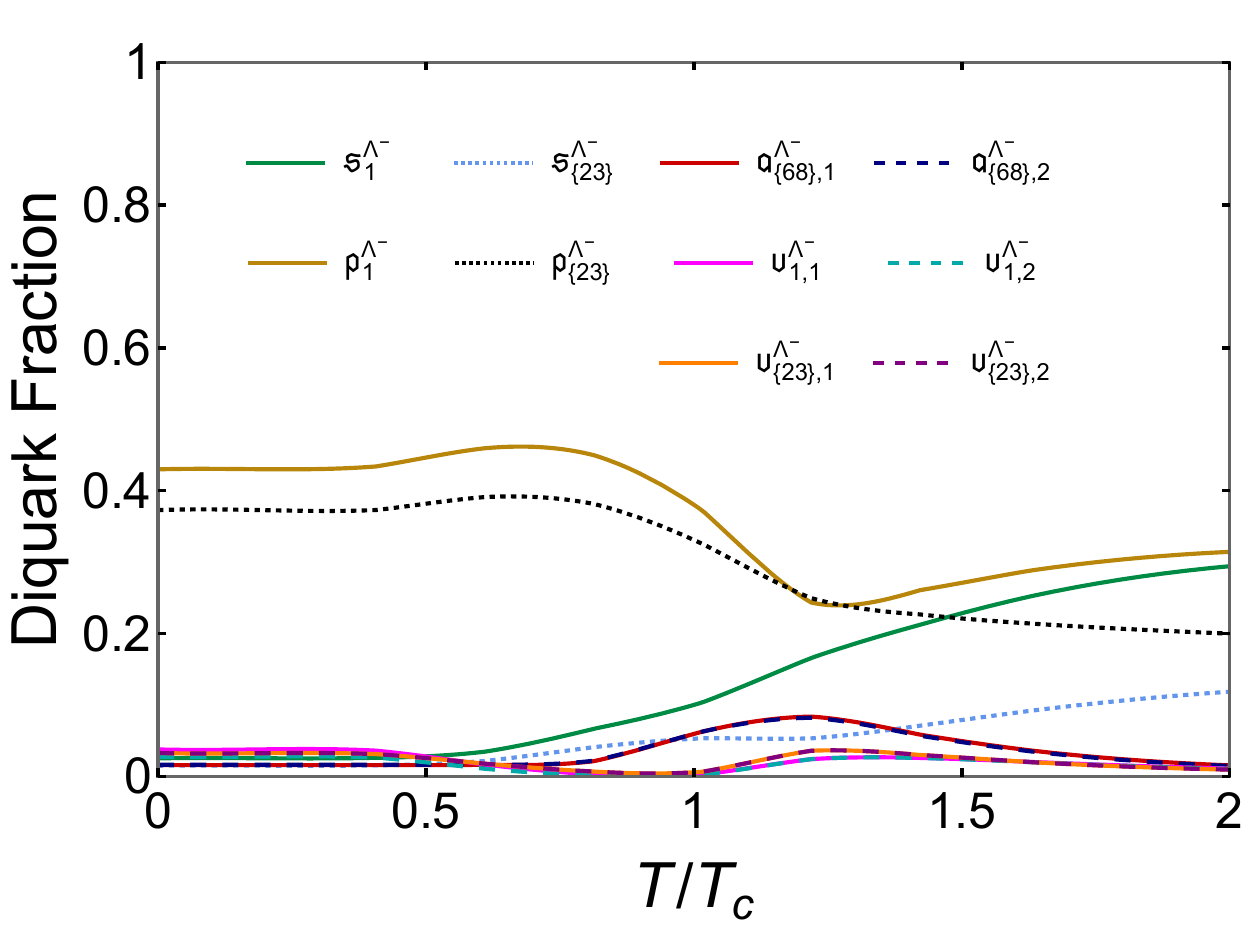}\vspace*{-1ex}
\end{tabular}
\begin{tabular}{lr}
\includegraphics[clip,width=0.4\linewidth]{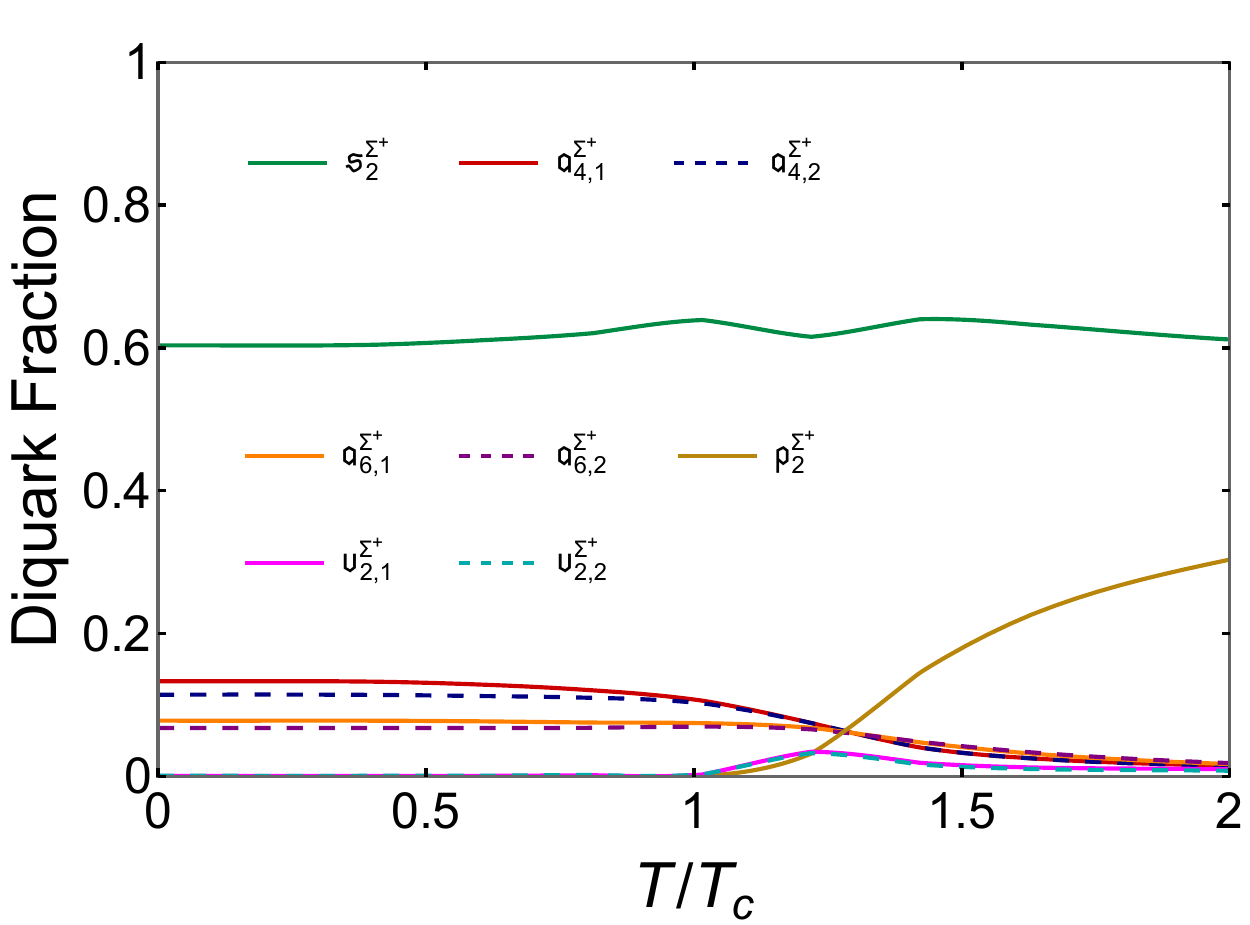}\hspace*{2ex } &
\includegraphics[clip,width=0.4\linewidth]{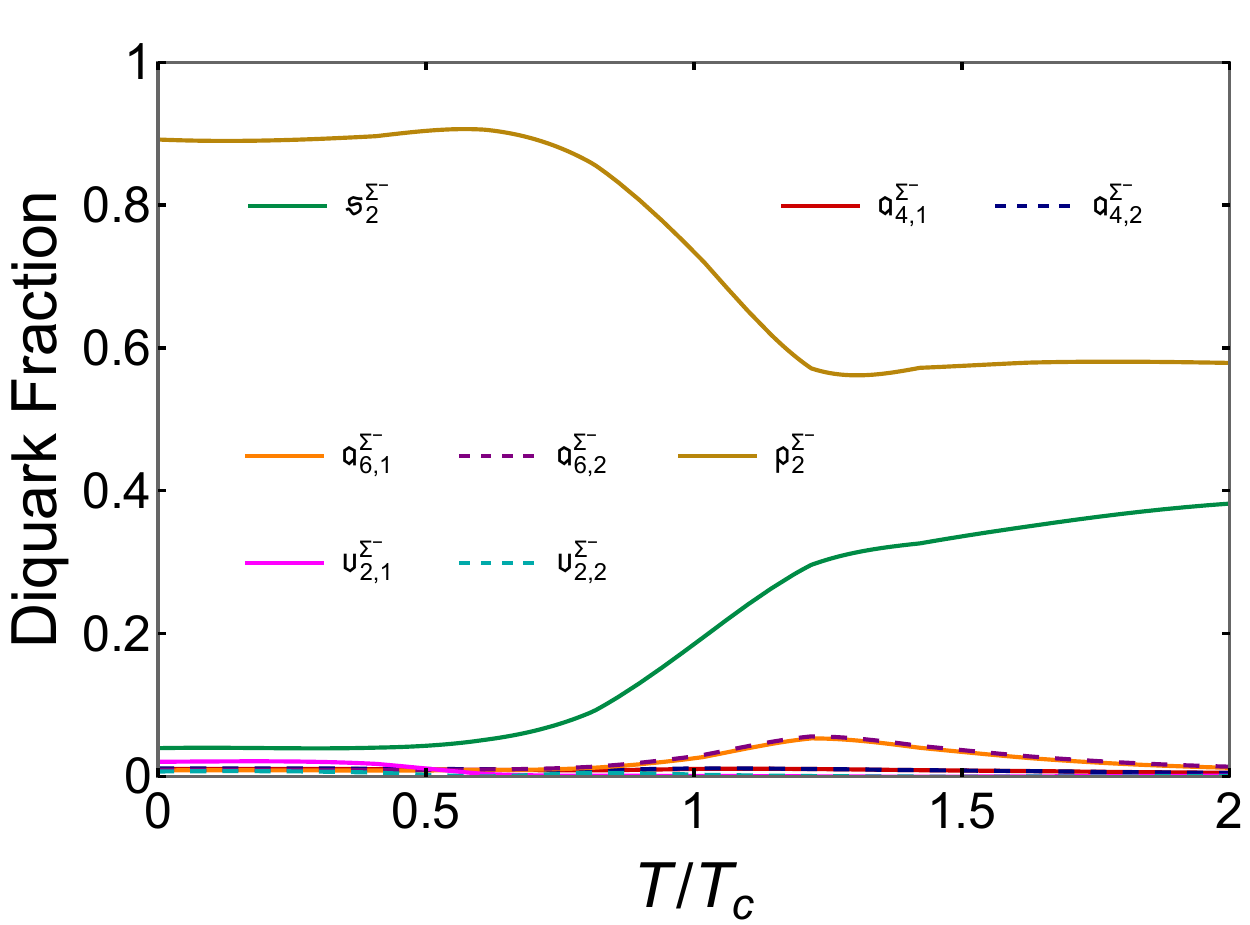}\vspace*{-1ex}
\end{tabular}
\begin{tabular}{lr}
\includegraphics[clip,width=0.4\linewidth]{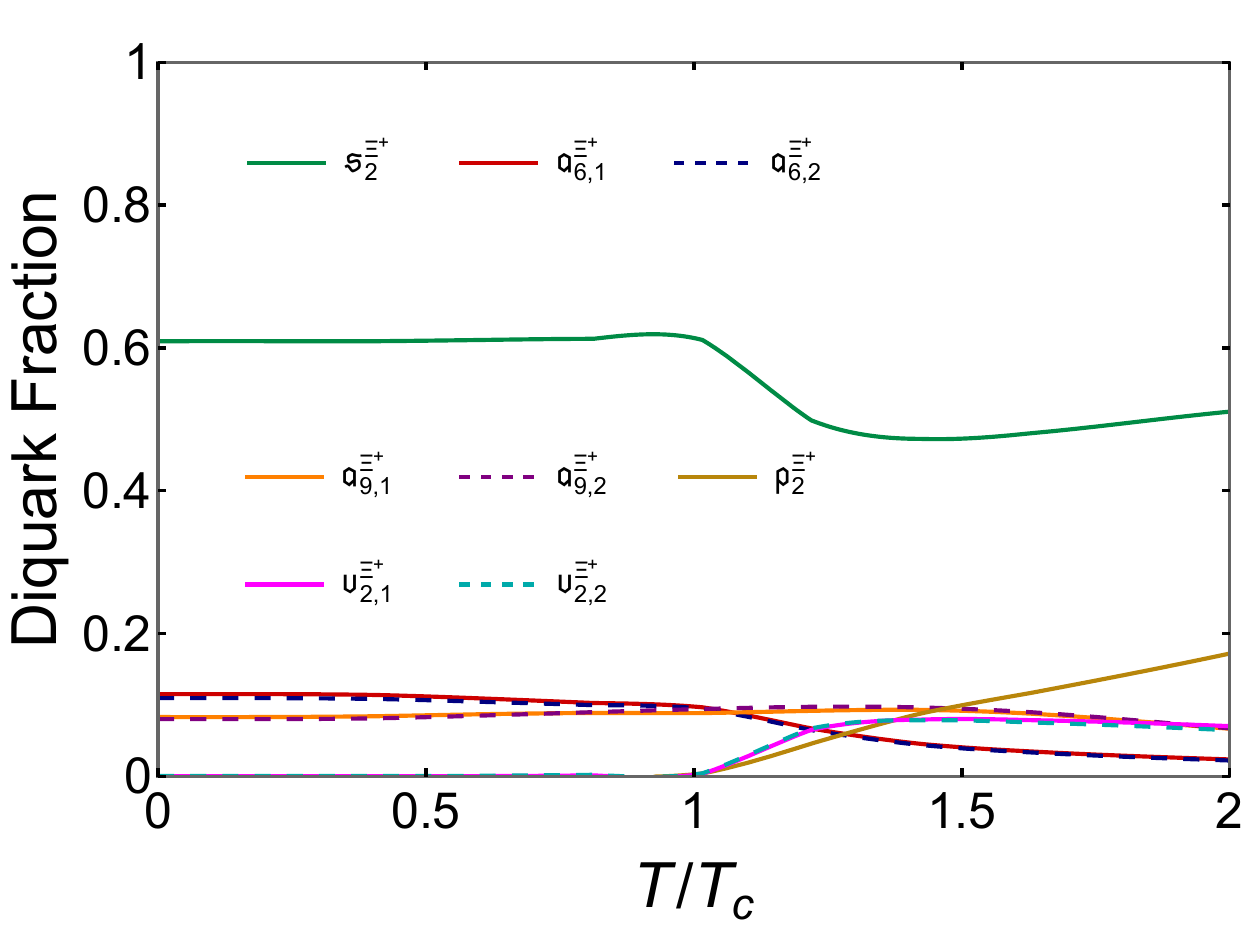}\hspace*{2ex } &
\includegraphics[clip,width=0.4\linewidth]{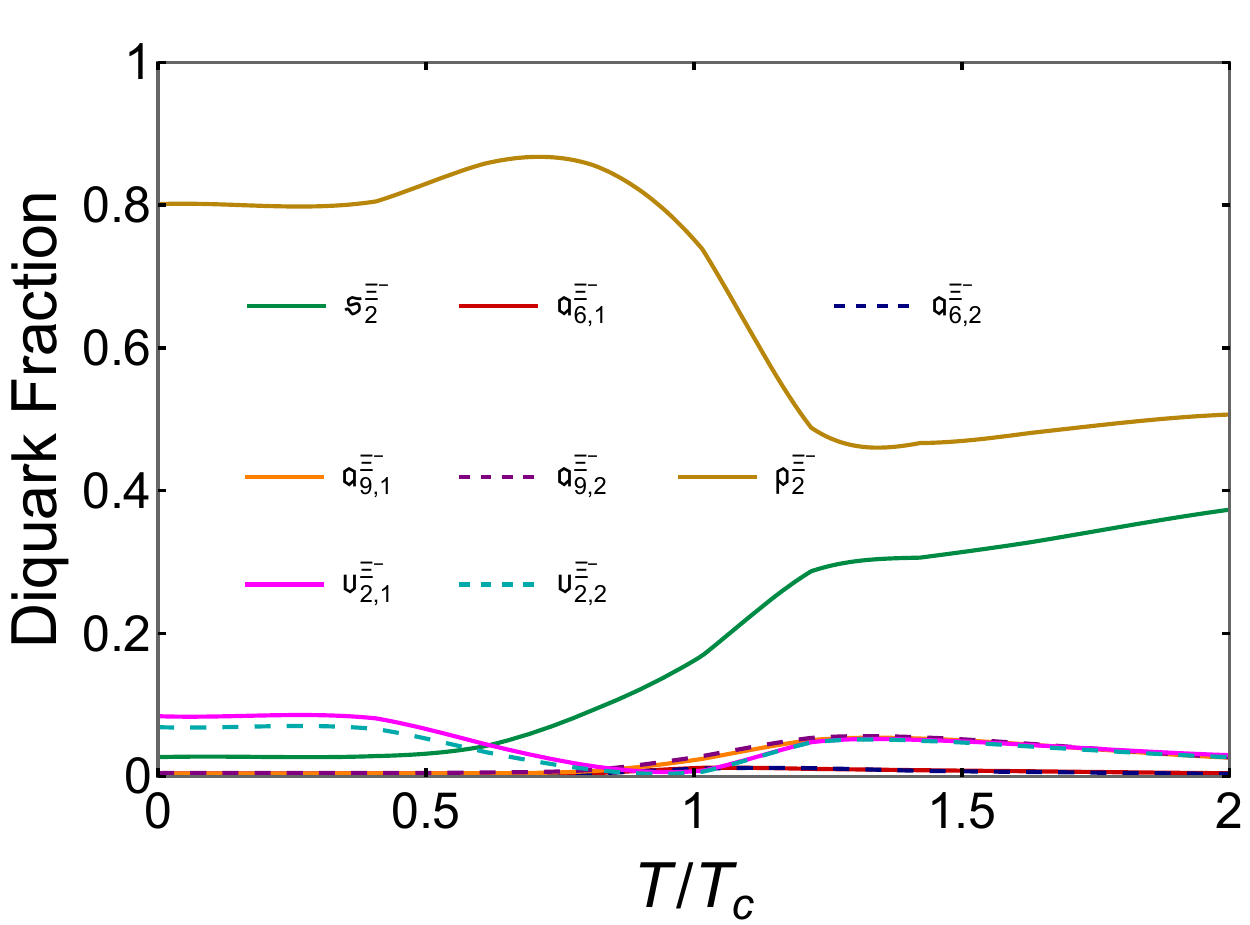}\vspace*{-1ex}
\end{tabular}
\end{center}
\caption{\label{fig:octetfrac}
Temperature evolutions of the fractions of various diquark components of the $J^P=1/2^\pm$ baryons. For each curve, the associated diquark content is listed in Eqs.\,\eqref{octetdiquarkcs}. \emph{Left panels}: positive parity baryons; and \emph{Right panels}: negative parity baryons, the parity partners. From top to bottom: $N^\pm$-, $\Lambda^\pm$-, $\Sigma^\pm$-, and $\Xi^\pm$-baryons. For instance, $N^+$ is the top-left panel, the top-right is $N^-$, and so on, with the $\Xi^-$-baryon being the bottom-right panel.
\newline
}
\end{figure*}

It is very interesting to study how different diquark fractions evolve with temperature, and the results are depicted in Fig.\,\ref{fig:octetfrac}. To this end, we use \emph{unit normalisation} for the Faddeev amplitudes. As a result, each square of the normalised Faddeev amplitude is the fraction of the associated diquark component inside a baryon. In Table\,\ref{FaddeevAmps}, we list the unit normalised amplitudes at $T=0$. The results agree well with those obtained using the static approximation in Refs.\,\cite{Lu:2017cln,Yin:2021uom}. To name a few, the dominant diquark contents in $N^+$-, $\Sigma^+$-, and $\Xi^+$-baryons are the scalar correlations, and the axial-vector diquarks also play material roles; for $\Lambda^+$, its scalar and axial-vector diquark fractions are almost equal, however, this is exactly the feature obtained from the calculation using a realistic QCD-connected interaction\,\cite{Chen:2019fzn}; and in these positive parity baryons, the contributions from negative parity diquarks are negligible. Instead, for those negative parity baryons, the pseudoscalar diquarks are dominant; and the fractions of the axial-vector diquarks are the smallest.

The $T$-evolutions of the diquark fractions given in Fig.\,\ref{fig:octetfrac} are very interesting. On the domain depicted, for the positive parity baryons, for $T\lesssim T_c$, each fraction is almost invariant, especially, the contributions from the negative parity diquarks are practically zero. 

When the temperature becomes larger than $T_c$, a novel phenomenon occurs: the fractions of scalar diquarks are still the largest. However, the pseudoscalar diquark fractions increase strongly and approach those of their parity partners. The result of the $\Lambda^+$-baryon is clearer: the fractions of the ${\mathfrak s}_1^{N^+}$-${\mathfrak p}_1^{N^+}$ and ${\mathfrak s}_{\{23\}}^{N^+}$-${\mathfrak p}_{\{23\}}^{N^+}$ pairs trend to close seperately. The axial-vector diquark fractions decrease and approach zero. And the fractions of vector diquarks initially increase but subsequently decrease, also trending toward zero. We judge that at enough large temperatures, only $J=0$ scalar and pseudoscalar diquarks can survive, and their contributions should be equal. The $J=1$ diquarks will disappear.

The observation above is remarkably different from the result in the previous study in Ref.\,\cite{Wang:2013wk}. The authors therein assumed that the nucleon only has scalar and axial-vector diquarks and predicted that the fraction of each diquark is steady. In conclusion, we judge that in any realistic computation for positive parity baryons, at large temperatures, one should take into account of all kinds of diquark correlations, and at very large $T$, perhaps using $J=0$ scalar and pseudoscalar diquarks is enough.  

Turning to the predictions for the negative parity baryons. Apparently, the dominant diquark contents are the pseudoscalar correlations. The fractions of scalar diquarks start with very small values, increase quickly as the temperature rises, and finally come close to those of the pseudoscalar partners. The fractions of axial-vector and vector diquarks are always very small, and trend to vanish at large $T$. Again, when $T$ gets large enough, we judge that only scalar and pseudoscalar correlations can survive, and their contributions could be equal.

%%%%%%%%%%%%%%%%%%%%%%%%%%%%%%%%%%%%%%%%%%%%%%%%%%%%%%%%%%%%%%%%%%%%%%%%%%%%%%

\subsection{$J^P=3/2^\pm$}

\begin{figure*}[!t]
\begin{center}
\begin{tabular}{lr}
\includegraphics[clip,width=0.4\linewidth]{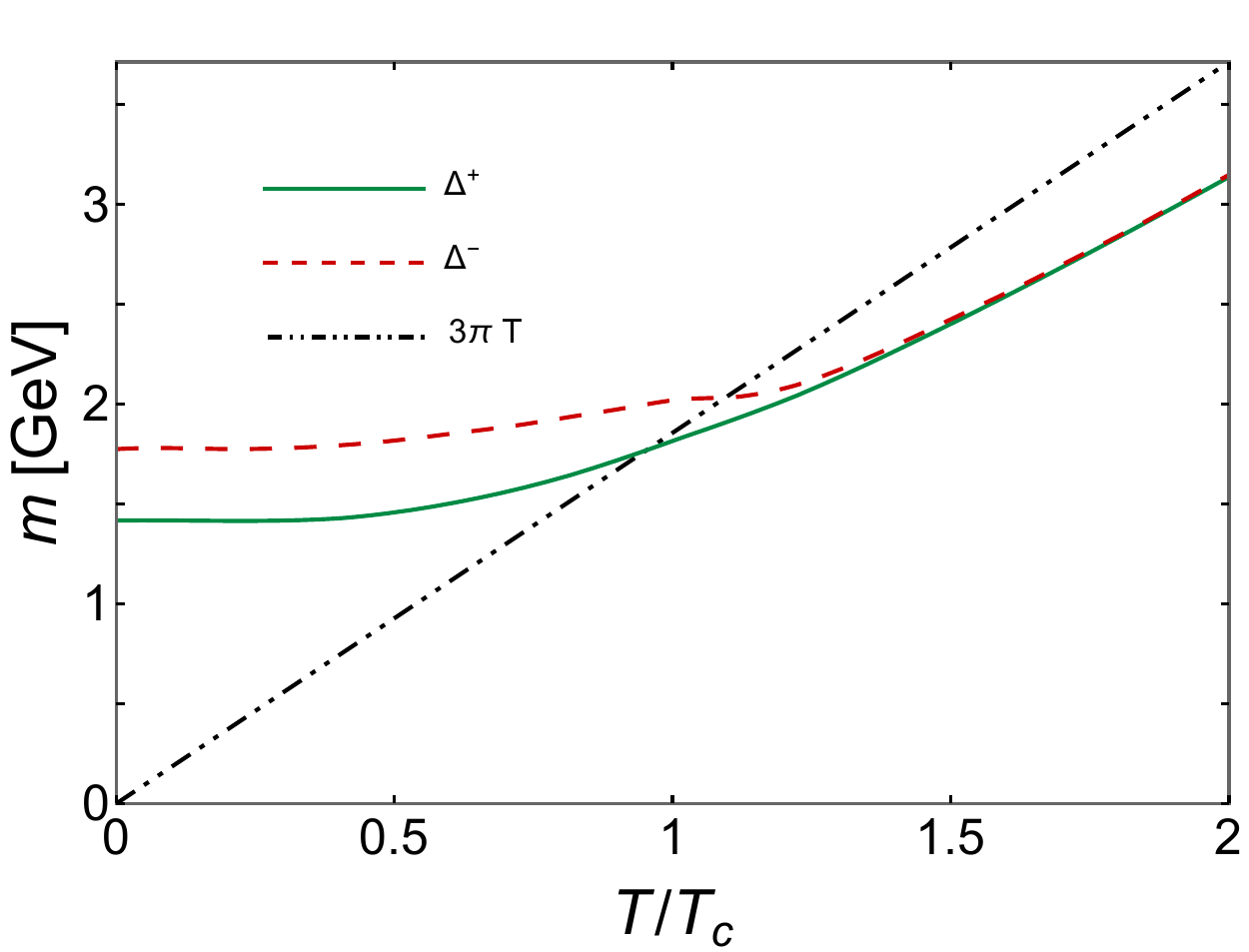}\hspace*{2ex } &
\includegraphics[clip,width=0.4\linewidth]{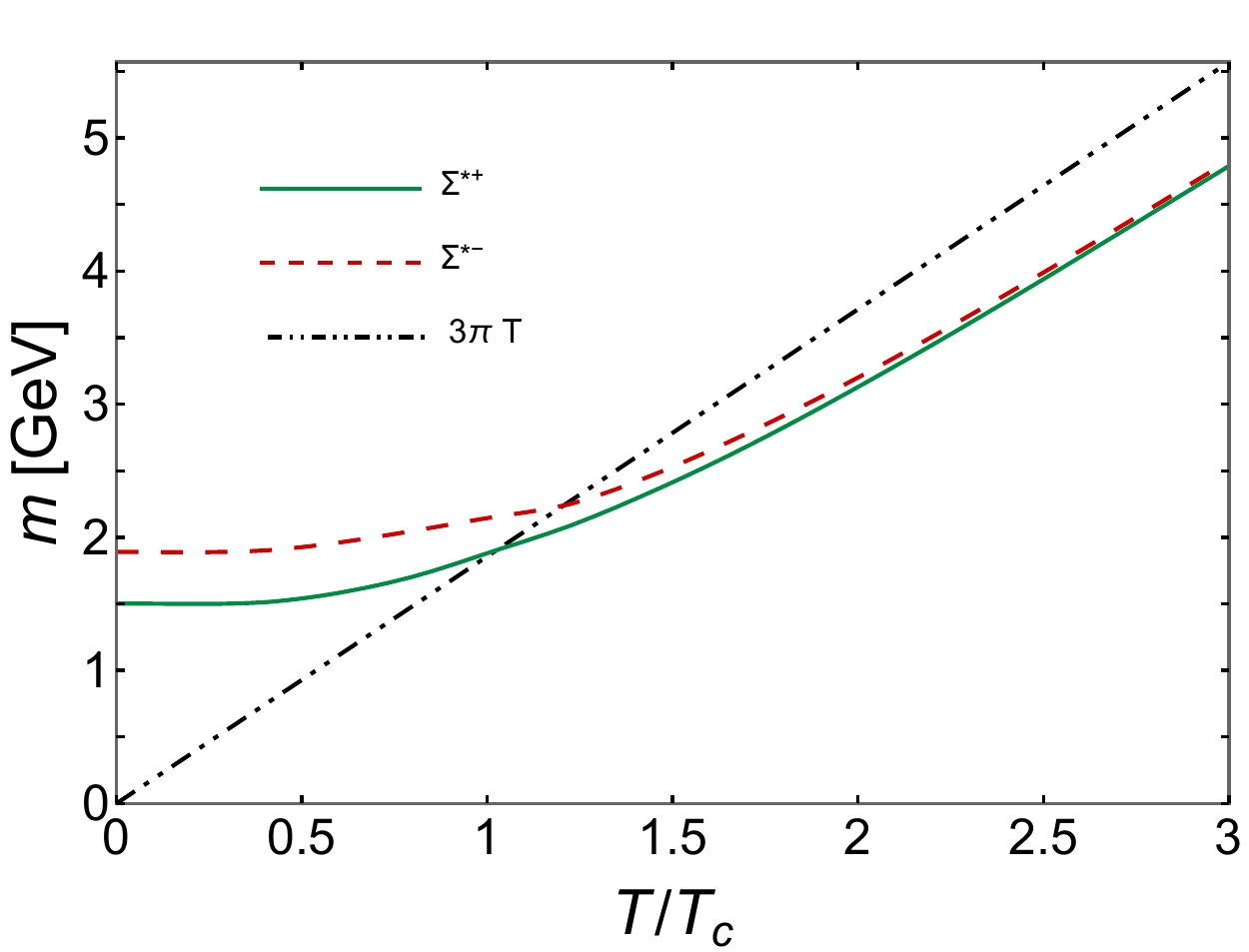}\vspace*{-0ex}
\end{tabular}
\begin{tabular}{lr}
\includegraphics[clip,width=0.4\linewidth]{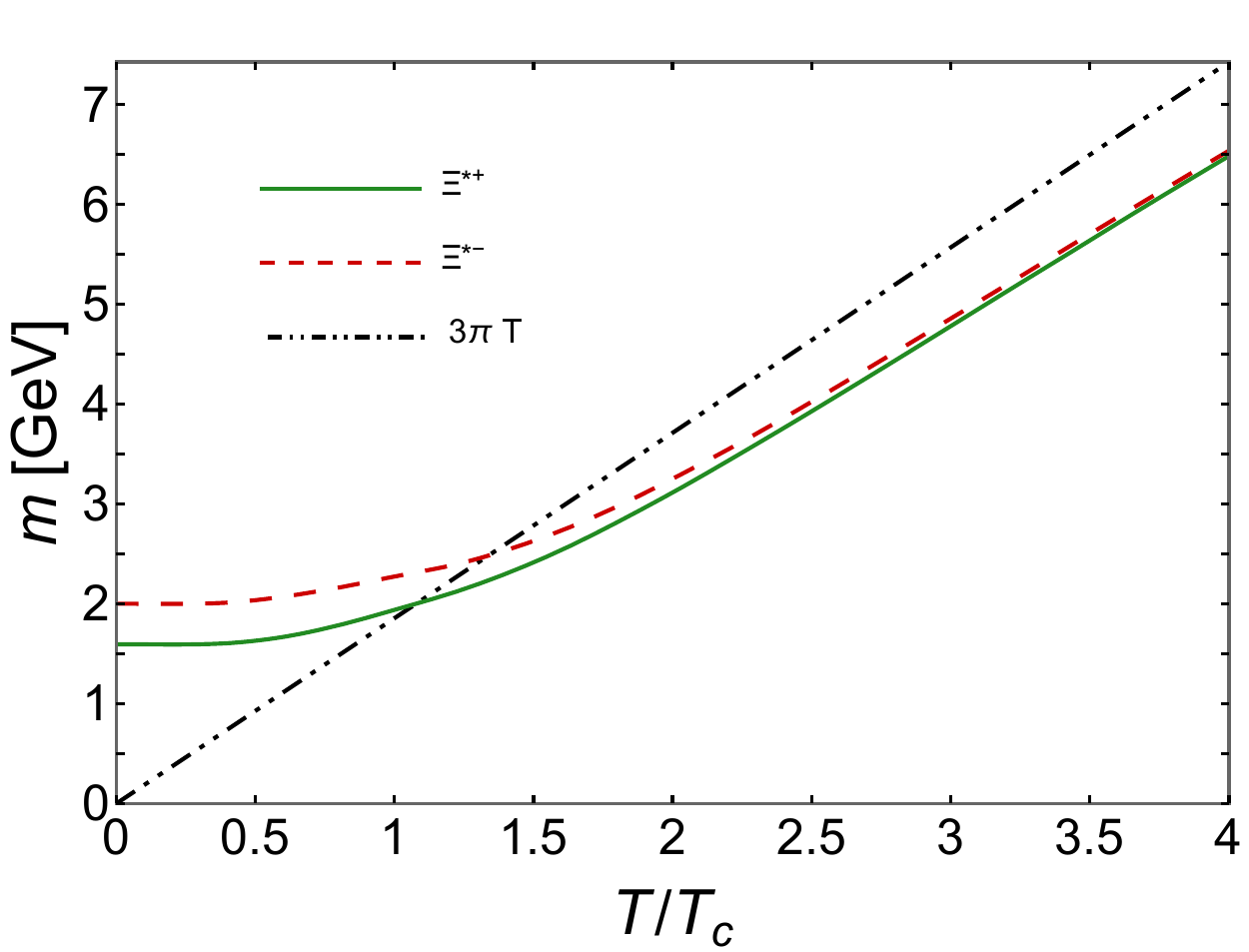}\hspace*{2ex } &
\includegraphics[clip,width=0.4\linewidth]{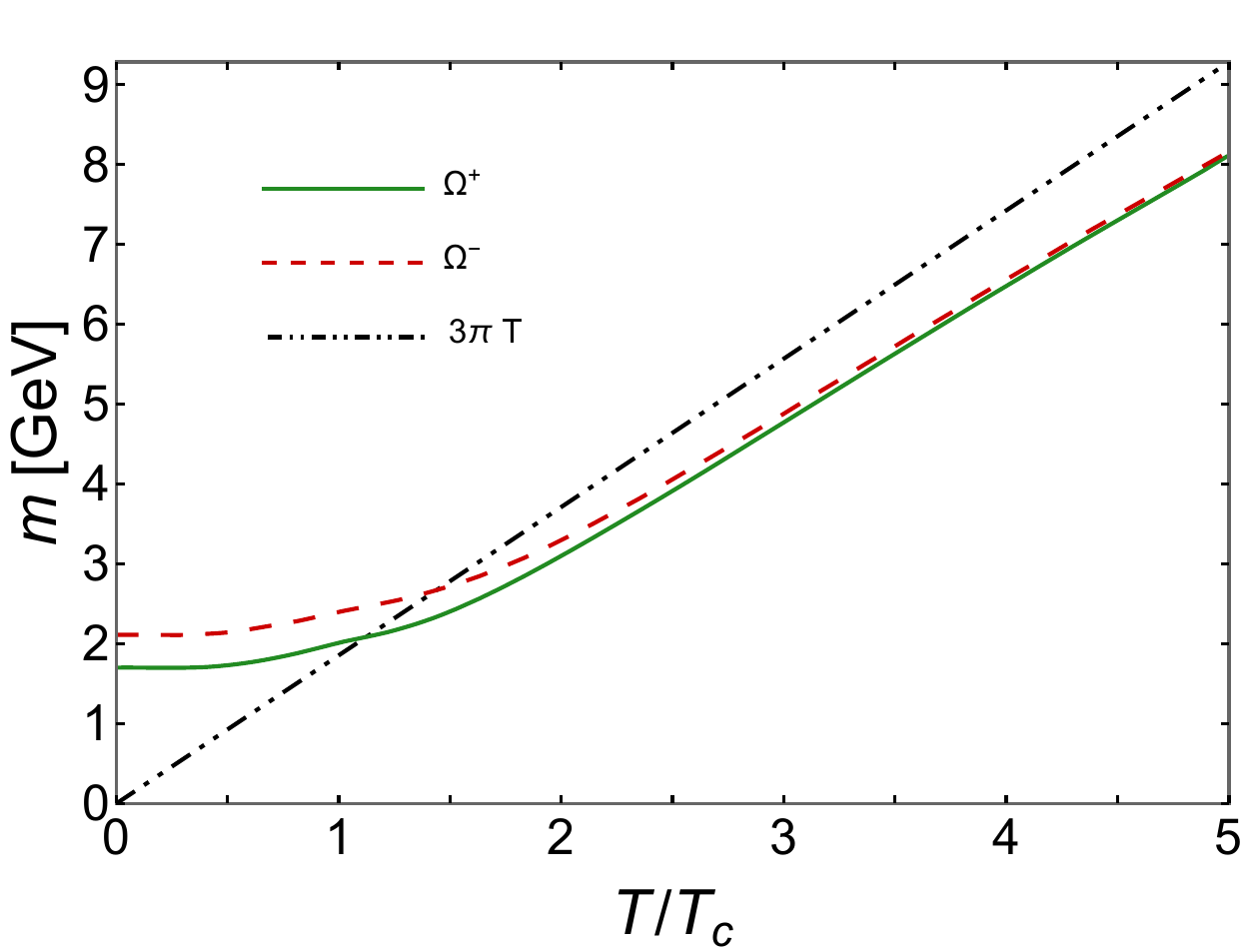}\vspace*{-1ex}
\end{tabular}
\end{center}
\caption{\label{fig:decuplet}
Screening masses of $J^P=3/2^\pm$ baryon ground states. In each figure: \emph{solid green curve}: positive-parity baryon; \emph{dashed red curve}: negative-parity baryon; and \emph{black dot-dot-dashed curve}: free theory limit of $m=3\pi T$.
}
\end{figure*}

The screening masses of the $J^P=3/2^\pm$ baryons are depicted in Fig.\,\ref{fig:decuplet}, and their values at $T=0$ are listed in Table\,\ref{OctetDecupletMasses}. Compared to the SCI results obtained via the static approximation in Ref.\,\cite{Yin:2021uom}, the relative differences are all smaller than $4\%$. In Fig.\,\ref{fig:decuplet}, on the domain depicted, each parity-partner-pair degenerates when $T$ gets large, and all the curves of positive parity baryons monotonously increase with temperature. As for the negative parity baryons, only the result of the $\Delta^-$-baryon exhibits a similar (but inconspicuous) behaviour with the $J^P=1/2^-$ counterparts, \emph{i.e.}, at $T\sim T_c$, the screening mass first decreases and soon turns to increase, finally merges with the curve of $\Delta^+$. The other three negative parity baryons, $\Sigma^-$, $\Xi^-$, and $\Omega^-$, all monotonously increase with temperature without any clear turning point; therefore, do not exhibit the pattern found previously for mesons, diquark correlations, and $J^P=1/2^\pm$ baryons. However, this might be an artefact due to the oversimplification of our treatment herein. The $J^P=3/2^\pm$ baryons have higher spin, so the formulae at nonzero temperatures become much more complicated, see, \emph{e.g.}, Ref.\,\cite{Korpa:2004sh}. Using a realistic QCD-connected interaction and  careful construction of the relevant ingredients, we expect that the screening masses of the $J^P=3/2^-$ baryons should behave similarly to those of $J^P=1/2^-$ baryons. This work will be conducted in the future.

The critical degenerate temperatures can be calculated analogously. Our calculation yields
\begin{subequations}
\begin{align}
T_c^{\delta m^{S=0}_{\Delta^\pm}} &= 1.47\,T_c\,,\\
T_c^{\delta m^{S=1}_{{\Sigma}^{\ast\pm}}} &= 2.71\,T_c\,,\\
T_c^{\delta m^{S=2}_{{\Xi}^{\ast\pm}}} &= 3.64\,T_c\,,\\
T_c^{\delta m^{S=3}_{\Omega^\pm}} &= 4.42\,T_c\,,
\end{align}
\end{subequations}
apparently, they increase with the strangeness. Unlike the $J=1/2$ baryons, some of the results are already within the domain $T\gtrsim 3T_c$, namely, close to the perturbative domain, on which our SCI kernel can only provide a qualitative guide. A quantitative calculation can only be obtained using a realistic QCD-connected treatment for the interaction kernel and the formulae of the $J=3/2$ baryons.

\begin{figure*}[!t]
\begin{center}
\begin{tabular}{lr}
\includegraphics[clip,width=0.4\linewidth]{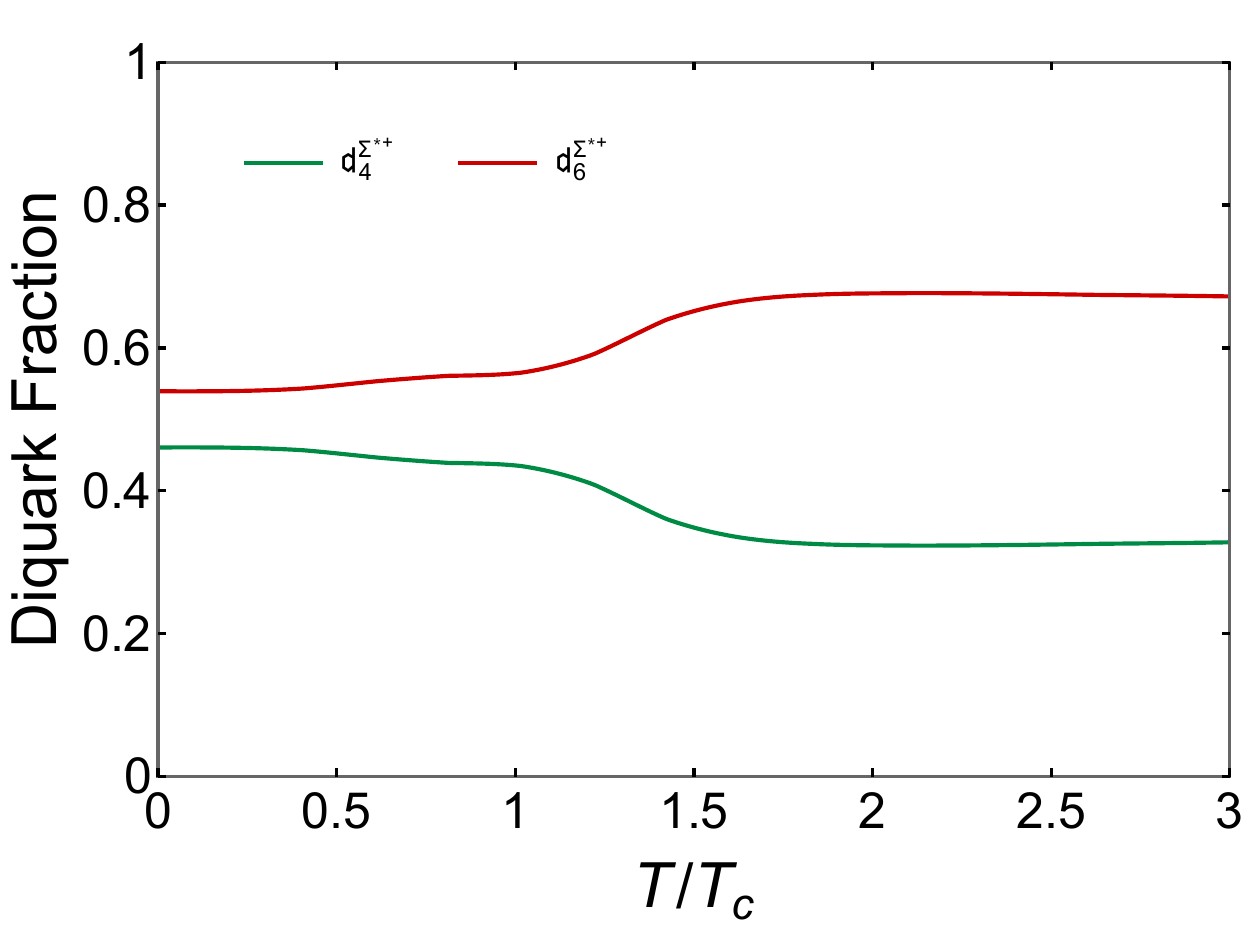}\hspace*{2ex } &
\includegraphics[clip,width=0.4\linewidth]{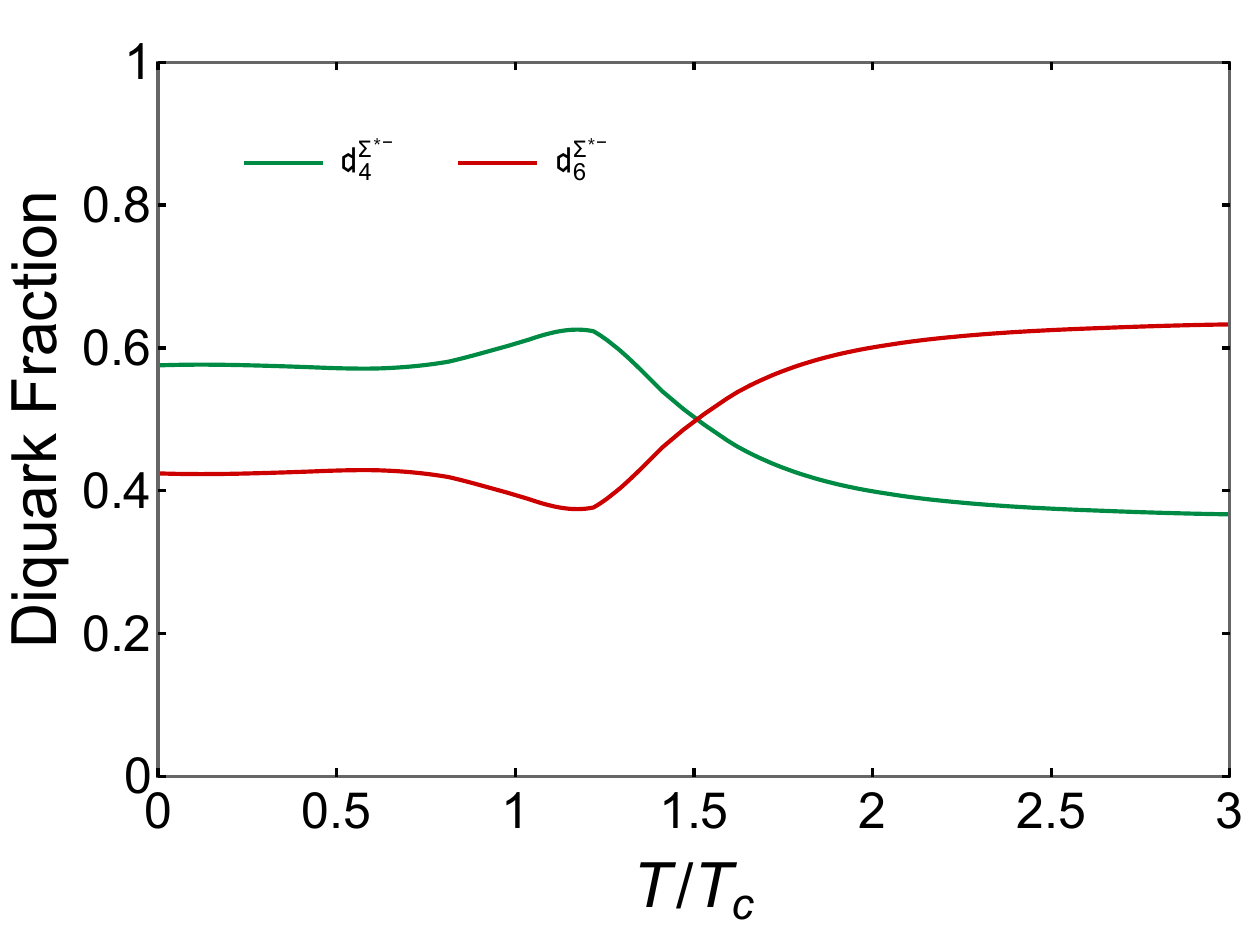}\vspace*{-0ex}
\end{tabular}
\begin{tabular}{lr}
\includegraphics[clip,width=0.4\linewidth]{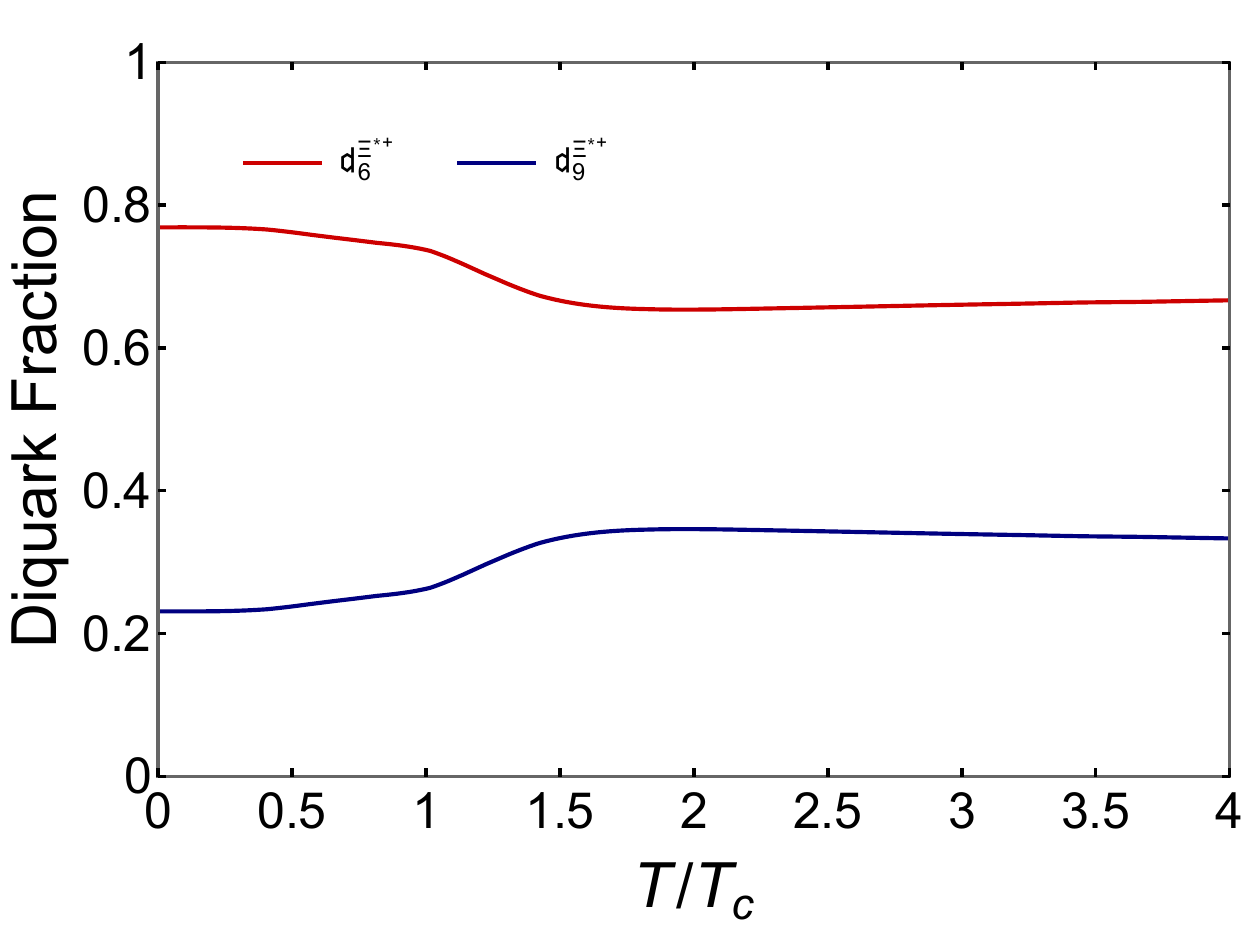}\hspace*{2ex } &
\includegraphics[clip,width=0.4\linewidth]{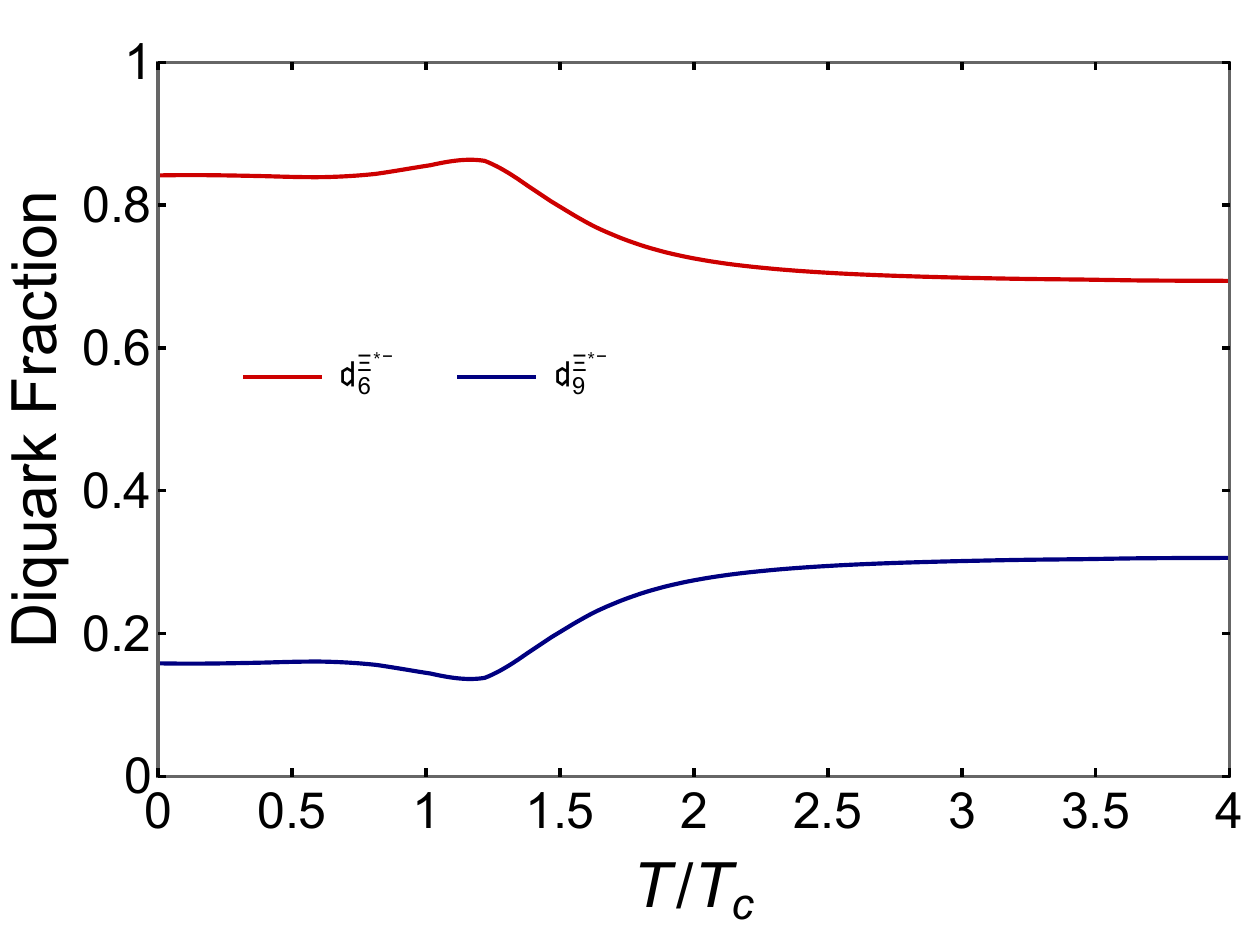}\vspace*{-1ex}
\end{tabular}
\end{center}
\caption{\label{fig:decupletfrac}
Temperature evolutions of the fractions of various diquark components of the $J^P=3/2^\pm$ baryons. For each curve, the associated diquark content is listed in Eqs.\,\eqref{decupletdiquarkcs}. \emph{Top-left panel}: $\Sigma^{\ast+}$; \emph{Top-right panel}: $\Sigma^{\ast-}$; \emph{Bottom-left panel}: $\Xi^{\ast+}$; and \emph{Bottom-right panel}: $\Xi^{\ast-}$.
} 
\end{figure*}

Lastly, we discuss the diquark fractions. As previously explained, using the SCI-RL kernel, one only needs to consider the axial-vector diquarks for $J=3/2$ baryons. According to Eq.\,\eqref{decupletdiquarkcs}, only the $\Sigma^{\ast\pm}$- and $\Xi^{\ast\pm}$-baryons have nontrivial flavour structures, and their unit-normalised Faddeev amplitudes at $T=0$ are
\begin{align}
\label{sigmaxifadv}
\begin{array} {l|rrr}\hline\hline
 & {\mathfrak d}_4 & {\mathfrak d}_6 & {\mathfrak d}_9 \\\hline
\Sigma^{\ast +} & 0.68 & 0.73 & \\
\Xi^{\ast +} & & 0.88 & 0.48 \\\hline
%%%%
\Sigma^{\ast -} & 0.76 & 0.65 & \\
\Xi^{\ast -} & & 0.92 & 0.40 \\\hline\hline
\end{array}\,,
\end{align}
again, these predictions agree well with the results from the static approximation in Ref.\,\cite{Yin:2021uom}. 

The $T$-evolutions of the diquark fractions are depicted in Fig.\,\ref{fig:decupletfrac}. At large temperatures, the axial-vector diquark $\{us\}_{1^+}$ is the dominant correlation, and its fraction lies within the domain $[0.6,0.7]$. This result comes as no surprise. Imagine a $\Delta^0$-baryon, which processes two flavour components: $d\{ud\}_{1^+}$ and $u\{dd\}_{1^+}$, according to isospin symmetry, at any temperature, the fraction of the latter ($=1/3$) should always be half of that of the former ($=2/3$). On the other hand, at extremely high $T$, the result should trend to the prediction in the free theory limit, for which case the $u$- and $s$-quarks become indistinguishable. As a result, for both the $\Sigma^{\ast \pm}$- and $\Xi^{\ast \pm}$-baryons, the fractions of the $\{us\}_{1^+}$ diquark correlation should trend to $2/3\simeq0.67$, this is just our prediction here; and the remaining one, $\{uu\}_{1^+}$ for $\Sigma^{\ast \pm}$, and $\{ss\}_{1^+}$ for $\Xi^{\ast \pm}$, should get close to $1/3\simeq0.33$.

%%%%%%%%%%%%%%%%%%%%%%%%%%%%%%%%%%%%%%%%%%%%%%%%%%%%%%%%%%%%%%%%%%%%%%%%%%%%%%
%%%%%%%%%%%%%%%%%%%%%%%%%%%%%%%%%%%%%%%%%%%%%%%%%%%%%%%%%%%%%%%%%%%%%%%%%%%%%%

\section{Summary and perspective}
\label{secsum}

A symmetry-preserving treatment of a vector $\times$ vector contact interaction (SCI) at $T\ge0$ was used to compute the screening masses of flavour-SU(3) hadron ground-states. Upon solving the gap equation of $u$-quark, the Nambu and Wigner solutions were obtained, from which we fixed the critical temperature $T_c$ of the chiral symmetry restoration transition. Our calculation yielded $T_c/m_\rho=0.212$, which is close to the contemporary lQCD result $0.200(3)$. Assuming the deconfinement transition happens simultaneously, one has $T_d=T_c$, with $T_d$ the critical temperature of the deconfinement transition. Based on these results, we reconstructed the model's infrared regulator $\Lambda^m_{\rm ir}(T)$ so that it equals zero for $T>T_d$, which implements the deconfinement transition.

After that, we solved the Bethe-Salpeter equations and obtained the screening masses of $J^P=0^\pm$, $1^\pm$ mesons, including those with strangeness. It is worth emphasizing that the masses at $T=0$ are uniformly larger than the empirical values because our Bethe-Salpeter kernel does not contain the meson-cloud contributions that should induce a material reduction in hadron masses. For nonzero temperatures, the $T$-dependent behaviours of the meson screening masses qualitatively agree with those obtained from the contemporary lQCD simulations. One of the most remarkable features is that each parity-partner-pair degenerates for $T>T_c$, with $T_c$ being the critical temperature. For each pair, the screening mass of the negative parity meson increases monotonously with temperature. In contrast, the screening mass of the meson with positive parity is almost invariant on the domain $T\lesssim T_c/2$; when $T$ gets close to $T_c$, it decreases but soon increases again and finally degenerates with its parity partner, which signals the restoration of chiral symmetry. Furthermore, in each of the light-light, light-strange, and strange-strange ($\bar{s}s$) sectors, the critical degenerate temperature $T_c^{\delta m}$ of the $J=1$ pair is lower than that of the $J=0$ pair. And, from the light-light to the strange-strange sector, $T_c^{\delta m}$ increases for the $J=1$ pairs, but decreases for the $J=0$ ones. One also notices that in each sector, after the parity partners degenerate, the screening mass of $J=1$ pair is uniformly larger than that of the $J=0$ pair.

We did not use the static approximation when dealing with the baryon Faddeev equations because it is not appropriate for $T\neq0$. Instead, we employed the method in Ref.\,\cite{Wang:2013wk}. The key point of this method is to select a representative value for the momentum ratio of a diquark in relation to that of the baryon, therefore, the dependence of the relative momentum can be removed. One feature of this method is that, compared to the static approximation, this method removes an additional parameter. Besides, for any baryon, we took account of all the diquark correlations that the SCI supports. At $T=0$, the baryon mass is generally larger than the empirical value; the reason is the same as the meson case, namely, the Faddeev equation kernel does not contain the meson-cloud effects. Compared to the results obtained via the static approximation, we found that the relative differences for the $J^P=1/2^+$ baryons are all smaller than $5\%$, except the $\Sigma^+$-baryon, which is $8\%$; and for the $J^P=1/2^-$ baryons, the relative differences are all smaller than $2\%$. Furthermore, the relative differences are all smaller than $4\%$ for the $J^P=3/2^\pm$ baryons.

As the temperature increases, we found that the $T$-dependent behaviours of baryon screening masses are quite similar to those of the meson. For all the four parity-partner-pairs of the $J=1/2$ baryons and the $J=3/2$ $\Delta^\pm$-pair, the screening masses of the positive parity baryons all increase slowly with temperature when $T\lesssim T_c$, and for $T>T_c$, they grow more rapidly, but always smaller than the free theory value $3\pi T$; whereas for their parity partners, the screening masses   are all almost invariant for $T\lesssim T_c$;  they  decrease when $T$ gets close to $T_c$, and at the temperature around $1.25\,T_c$, they increase once again, and quickly merge with every positive parity partner. The $T$-evolutions of the $\Sigma^{\ast\pm}$-, $\Xi^{\ast\pm}$-, and $\Omega^{\ast\pm}$-pairs are somewhat different; however, this should be due to the oversimplification of our treatment of the $J=3/2$ baryons, and should be improved using a realistic QCD-connected interaction and the corresponding carefully constructed ingredients. Besides, for baryons with $J=1/2$ or $3/2$, the critical degenerate temperature increases with strangeness. 

After applying unit normalisation, we also calculated the Faddeev amplitudes. At $T=0$, we found that the results agree well with those obtained via the static approximation. We furthermore explored the $T$-evolution of the diquark fractions within baryons. For $J=3/2$ baryons, only the $\Sigma^{\ast\pm}$- and $\Xi^{\ast\pm}$-baryons have nontrivial flavour structures, for them, we found that at large temperatures, the axial-vector diquark $\{us\}_{1^+}$ is the dominant correlation, and its fraction lies on the domain $[0.6,0.7]$, which is consistent with the result in the free field limit. The $T$-dependent behaviours of the diquark fractions of the $J=1/2$ baryons are much more interesting and novel: at high temperatures, only $J=0$ scalar and pseudoscalar diquark correlations can survive and their fractions trend to close, while the $J=1$ axial-vector and vector diquarks become disappear. Therefore, it is necessary to consider all diquark correlations to study baryons at nonzero temperatures; and at very large $T$, perhaps, only using $J=0$ scalar and pseudoscalar diquarks is enough.

This work can provide a valuable qualitative and semiquantitative guide for studying the properties of baryons at $T\neq0$ in the future. For example, using a similar framework, the screening masses of hadrons with heavy quarks can be calculated. And one can also use the strategy in Ref.\,\cite{Xu:2015kta} to eliminate the approximation for the exchanged quark in Faddeev equation, upon which, the baryon Faddeev amplitudes will depend on relative momentum.  A long-term goal is to compute the hadron screening masses using a realistic QCD-connected interaction kernel. In doing so, employing both the quark+diquark approximation and the three-body Faddeev equation approach will be necessary for baryons.

%%%%%%%%%%%%%%%%%%%%%%%%%%%%%%%%%%%%%%%%%%%%%%%%%%%%%%%%%%%%%%%%%%%%%%%%%%%%%%
%%%%%%%%%%%%%%%%%%%%%%%%%%%%%%%%%%%%%%%%%%%%%%%%%%%%%%%%%%%%%%%%%%%%%%%%%%%%%%

\acknowledgments

We are grateful for constructive communications with C.~D.~Roberts and C.~S.~Fischer.
Work supported by:
National Natural Science Foundation of China (Grants No.~12247103, No.~12305134).
%

%%%%%%%%%%%%%%%%%%%%%%%%%%%%%%%%%%%%%%%%%%%%%%%%%%%%%%%%%%%%%%%%%%%%%%%%%%%%%%
%%%%%%%%%%%%%%%%%%%%%%%%%%%%%%%%%%%%%%%%%%%%%%%%%%%%%%%%%%%%%%%%%%%%%%%%%%%%%%
%%%%%%%%%%%%%%%%%%%%%%%%%%%%%%%%%%%%%%%%%%%%%%%%%%%%%%%%%%%%%%%%%%%%%%%%%%%%%%

\appendix
\setcounter{equation}{0}
\setcounter{figure}{0}
\setcounter{table}{0}
\renewcommand{\theequation}{\Alph{section}.\arabic{equation}}
\renewcommand{\thetable}{\Alph{section}.\arabic{table}}
\renewcommand{\thefigure}{\Alph{section}.\arabic{figure}}

%%%%%%%%%%%%%%%%%%%%%%%%%%%%%%%%%%%%%%%%%%%%%%%%%%%%%%%%%%%%%%%%%%%%%%%%%%%%%%

\section{Bethe-Salpeter equations}
\label{appbse}

\subsection{Symmetry}

For sensible calculations in hadron physics, accounting for relevant symmetries is crucial. In this work, we will utilise the axial-vector and vector Ward-Green-Takahashi identities (WGTIs) whenever necessary\,\cite{Ward:1950xp,Green:1953te,Takahashi:1957xn}, and in the SCI framework, they can be expressed as\,\cite{Wang:2013wk}
\begin{align}
\label{wtisci}
\int^1_0 d\alpha [{\mathcal C}^{\rm iu}(\varsigma_{fg};T) + {\mathcal C}^{\rm iu}_1(\varsigma_{fg};T) + {\mathcal R}^{\rm iu}(\varsigma_{fg};T)]=0\,,
\end{align}
where
\begin{align}
\varsigma_{fg} = \varsigma_{fg}(\alpha,Q_0^2) = \hat{\alpha}M_f^2+\alpha M_g^2+\alpha\hat{\alpha}Q_0^2\,,	
\end{align}
with $\hat{\alpha}\equiv1-\alpha$;
\begin{subequations}
\begin{align}
\overline{{\mathcal C}}^{\rm iu}_n(\varsigma_{fg};T) &= \frac{(-1)^n}{n!}\frac{d^n}{d\varsigma_{fg}^n}{\mathcal C}^{\rm iu}(\varsigma_{fg};T)\,,\\
{\mathcal C}^{\rm iu}_n(\varsigma_{fg};T) &= \varsigma_{fg}^n\overline{{\mathcal C}}^{\rm iu}_n(\varsigma_{fg};T)\,,
\end{align}	
\end{subequations}
and
\begin{align}
\nonumber
&{\mathcal R}^{\rm iu}(\varsigma_{fg};T)\\
 = &\int^{1/\Lambda_{\rm ir}^2}_{1/\Lambda_{\rm uv}^2}d\tau {\rm e}^{-\tau\varsigma_{fg}}\sqrt{\frac{\pi}{\tau}}\bigg[-\frac{d}{d\tau}-\frac{1}{2\tau}\bigg]2T\vartheta_2({\rm e}^{-\tau4\pi^2T^2})\,,
\end{align}
with ${\mathcal C}^{\rm iu}$ defined in Eqs.\,\eqref{cfun}, and $\vartheta_2(x)$ being the Jacobi theta function\,\cite{Gradshteyn:1943cpj}. 

Eq.\,\eqref{wtisci} is the generalisation of the WGTIs for $T\neq0$. Using Eq.\,\eqref{jacbi2}, it is easy to prove 
\begin{align}
{\mathcal R}^{\rm iu}(\varsigma_{fg};T\to 0)=0\,,
\end{align}
thus the WGTIs in the vacuum can be retrieved again\,\cite{Chen:2012qr}.

%%%%%%%%%%%%%%%%%%%%%%%%%%%%%%%%%%%%%%%%%%%%%%%%%%%%%%%%%%%%%%%%%%%%%%%%%%%%%%

\subsection{Pseudoscalar ($0^-$) mesons and scalar ($0^+$) diquarks}

Inserting Eq.\,\eqref{bsamps} into \eqref{bsem} produces the Bethe-Salpeter equation for a pseudoscalar meson:
\begin{equation}
\label{bsefinalE}
\left[
\begin{array}{c}
E^{0^-}_{[f\bar{g}]}\\
F^{0^-}_{[f\bar{g}]}
\end{array}
\right]
= \frac{4 \alpha_{\rm IR}}{3\pi m_G^2}
\left[
\begin{array}{cc}
{\mathcal K}_{{[f\bar{g}]}}^{0^-,EE} & {\mathcal K}_{{[f\bar{g}]}}^{0^-,EF} \\
{\mathcal K}_{{[f\bar{g}]}}^{0^-,FE} & {\mathcal K}_{{[f\bar{g}]}}^{0^-,FF}
\end{array}\right]
\left[\begin{array}{c}
E^{0^-}_{[f\bar{g}]}\\
F^{0^-}_{[f\bar{g}]}
\end{array}
\right],
\end{equation}
where
{\allowdisplaybreaks
\begin{subequations}
\begin{align}
\nonumber
{\mathcal K}_{[f\bar{g}]}^{0^-,EE} &=
\int_0^1d\alpha \bigg\{
{\cal C}^{\rm iu}(\varsigma_{f g}(\alpha, Q_0^2);T)  \\
&+ \bigg[ M_f M_{g}-\alpha \hat\alpha Q_0^2 - \varsigma_{f g}(\alpha, Q_0^2)\bigg]\nonumber \\
& \times 
\overline{\mathcal C}^{\rm iu}_1(\varsigma_{f g}(\alpha, Q_0^2);T)\bigg\}\,,\\
\nonumber
{\cal K}_{[f\bar{g}]}^{0^-,EF} &= \frac{Q_0^2}{2 M_{f g}} \int_0^1d\alpha\, \bigg[\hat \alpha M_f+\alpha M_{g}\bigg]\\
& \times \overline{\cal C}^{\rm iu}_1(\varsigma_{f g}(\alpha, Q_0^2);T)\,,\\
{\mathcal K}_{[f\bar{g}]}^{0^-,FE} &= M_{f g} {\mathcal K}_{[f\bar{g}]}^{0^-,EF} ,\\
\nonumber
{\mathcal K}_{[f\bar{g}]}^{0^-,FF} &= - \frac{1}{2} \int_0^1d\alpha\, \bigg[ M_f M_{g}+\hat\alpha M_f^2+\alpha M_{g}^2\bigg]\\
&  \times \overline{\mathcal C}^{\rm iu}_1(\varsigma_{f g}(\alpha, Q_0^2);T)\,.
\end{align}
\end{subequations}}
Solving Eq.\,\eqref{bsefinalE} yields the solution for $Q_0^2=-(m_{[f\bar{g}]}^{0^-})^2$, with $m_{[f\bar{g}]}^{0^-}=m_{[f\bar{g}]}^{0^-}(T)$ the screening mass. 

To compute any observables one must utilise the canonically normalised amplitude, therefore, the Bethe-Salpeter amplitude must satisfy 
\begin{align}
1= \frac{d}{d Q_0^2}\Pi_{[f\bar{g}]}^{0^-}(K,Q_0;T)\bigg|_{K=Q_0},
\end{align}
where
\begin{align}
\nonumber
\Pi_{[f\bar{g}]}^{0^-}(K,Q_0;T) &= 6 {\rm tr}_{\rm D} \int_{l,dq} \Gamma_{[f\bar{g}]}^{0^-}(-K;T)
 S_f(q+Q_0;T)\\
&\times \Gamma_{[f\bar{g}]}^{0^-}(K;T)S_g(q;T)\,.
\end{align}

Considering the discussion in Section\,\ref{secdiquark}, the Bethe-Salpeter equation for a scalar diquark with the amplitude given in Eq.\,\eqref{bsadqsc} takes the form
\begin{equation}
\left[
\begin{array}{c}
E^{0^+}_{[fg]}\\
F^{0^+}_{[fg]}
\end{array}
\right]
= \frac{2 \alpha_{\rm IR}}{3\pi m_G^2}
\left[
\begin{array}{cc}
{\mathcal K}_{{[f\bar{g}]}}^{0^-,EE} & {\mathcal K}_{{[f\bar{g}]}}^{0^-,EF} \\
{\mathcal K}_{{[f\bar{g}]}}^{0^-,FE} & {\mathcal K}_{{[f\bar{g}]}}^{0^-,FF}
\end{array}\right]
\left[\begin{array}{c}
E^{0^+}_{[fg]}\\
F^{0^+}_{[fg]}
\end{array}
\right],
\end{equation}
in this situation the canonical normalisation condition is
\begin{align}
1= \frac{d}{d Q_0^2}\Pi_{[fg]}^{0^+}(K,Q_0;T)\bigg|_{K=Q_0},
\end{align}
where
\begin{align}
\nonumber
\Pi_{[fg]}^{0^+}(K,Q_0;T) &= 4 {\rm tr}_{\rm D} \int_{l,dq} \Gamma_{[fg]}^{0^+}(-K;T)
 S_f(q+Q_0;T)\\
&\times \Gamma_{[fg]}^{0^+}(K;T)S_g(q;T)\,.
\end{align}

%%%%%%%%%%%%%%%%%%%%%%%%%%%%%%%%%%%%%%%%%%%%%%%%%%%%%%%%%%%%%%%%%%%%%%%%%%%%%%

\subsection{Vector ($1^-$) mesons and axial-vector($1^+$) diquarks}

Using Eq.\,\eqref{bsamvc}, it is straightforward to derive the Bethe-Salpeter equations for vector mesons. For a vector meson's longitudinal component, the Bethe-Salpeter equation is
\begin{align}
\label{bsevc}
1+{\mathcal K}_{[f\bar{g}]}^{1^-,\parallel}(Q_0^2=-(m_{[f\bar{g}]}^{1^-})^2;T) =0\,,
\end{align}
where
\begin{align}
\label{bsekervc}
\nonumber
{\mathcal K}_{[f\bar{g}]}^{1^-,\parallel} = &\frac{2\alpha_{\rm IR}}{3\pi m_G^2}\int^1_0d\alpha \bigg\{\bigg[M_fM_g-\hat{\alpha}M_f^2-\alpha M_g^2\\
\nonumber
&-2\alpha\hat{\alpha}Q_0^2\bigg]
\overline{\mathcal C}^{\rm iu}_1(\varsigma_{f g}(\alpha, Q_0^2);T)\\
&+2{\mathcal R}^{\rm iu}(\varsigma_{f g}(\alpha, Q_0^2);T)\bigg\}\,,
\end{align}
and the canonical normalisation condition is
\begin{align}
\label{canovc}
\frac{1}{(E_{[f\bar{g}]}^{1^-,\parallel})^2} = \frac{9m_G^2}{4\pi\alpha_{\rm IR}}\frac{d}{dz}{\mathcal K}_{[f\bar{g}]}^{1^-,\parallel}(z;T)\bigg|_{z=-(m_{[f\bar{g}]}^{1^-,\parallel})^2}\,.
\end{align}

Alternatively, the Bethe-Salpeter equation for the longitudinal component of an axial-vector diquark is  
\begin{align}
1+\frac{1}{2}{\mathcal K}_{[f\bar{g}]}^{1^-,\parallel}(Q_0^2=-(m_{[fg]}^{1^+})^2;T) =0\,,
\end{align}
and 
\begin{align}
\label{canovc}
\frac{1}{(E_{[fg]}^{1^+,\parallel})^2} = \frac{3m_G^2}{2\pi\alpha_{\rm IR}}\frac{d}{dz}{\mathcal K}_{[f\bar{g}]}^{1^-,\parallel}(z;T)\bigg|_{z=-(m_{[fg]}^{1^+,\parallel})^2}\,.
\end{align}
is the canonical normalisation condition.

The previous discussion is on the longitudinal components. Removing the ${\mathcal R}^{\rm iu}$-term in Eq.\,\eqref{bsekervc}, the Bethe-Salpeter equations and the canonical normalisation conditions for the corresponding transverse components can be obtained straightforwardly.

%%%%%%%%%%%%%%%%%%%%%%%%%%%%%%%%%%%%%%%%%%%%%%%%%%%%%%%%%%%%%%%%%%%%%%%%%%%%%%

\subsection{Scalar ($0^+$) mesons and pseudoscalar  ($1^+$) diquarks}

Inserting Eq.\,\eqref{bsamsc} into \eqref{bsem} and using Eq.\eqref{gsomsc}, yields the Bethe-Salpeter equation for a scalar meson
\begin{align}
1+{\mathfrak g}_{\rm SO}^{q\bar{q},0^+}(T)\cdot{\mathcal K}_{[f\bar{g}]}^{0^+}(Q_0^2=-(m_{[f\bar{g}]}^{0^+})^2;T) =0\,,	
\end{align}
where
\begin{align}
\nonumber
{\mathcal K}_{[f\bar{g}]}^{0^+} = &-\frac{4\alpha_{\rm IR}}{3\pi m_G^2}\int^1_0d\alpha \bigg\{\bigg[{\mathcal C}^{\rm iu}(\varsigma_{f g}(\alpha, Q_0^2);T)\\
\nonumber
&-{\mathcal C}^{\rm iu}_1(\varsigma_{f g}(\alpha, Q_0^2);T)\bigg]\\
&-(M_fM_g+\alpha\hat{\alpha}Q_0^2)\overline{\mathcal C}^{\rm iu}_1(\varsigma_{f g}(\alpha, Q_0^2);T)\bigg\}\,,
\end{align}
and the canonical normalisation condition is
\begin{align}
\frac{1}{(E_{[f\bar{g}]}^{0^+})^2} = \frac{9m_G^2}{8\pi\alpha_{\rm IR}}\frac{d}{dz}{\mathcal K}_{[f\bar{g}]}^{0^+}(z;T)\bigg|_{z=-(m_{[f\bar{g}]}^{0^+})^2}\,.
\end{align}

Employing Eq.\,\eqref{gsopsdq}, the Bethe-Salpeter equation for a pseudoscalar diquark is
\begin{align}
1+\frac{1}{2}{\mathfrak g}_{\rm SO}^{qq,0^-}(T)\cdot{\mathcal K}_{[f\bar{g}]}^{0^+}(Q_0^2=-(m_{[fg]}^{0^-})^2;T) =0\,,	
\end{align}
and the associated canonical normalisation condition is
\begin{align}
\frac{1}{(E_{[fg]}^{0^-})^2} = \frac{3m_G^2}{4\pi\alpha_{\rm IR}}\frac{d}{dz}{\mathcal K}_{[f\bar{g}]}^{0^+}(z;T)\bigg|_{z=-(m_{[fg]}^{0^-})^2}\,.
\end{align}

%%%%%%%%%%%%%%%%%%%%%%%%%%%%%%%%%%%%%%%%%%%%%%%%%%%%%%%%%%%%%%%%%%%%%%%%%%%%%%

\subsection{Axial-vector ($1^+$) mesons and vector ($1^-$) diquarks}

At nonzero temperatures, the axial-vector mesons also separate into longitudinal and transverse components. Using Eqs.\,\eqref{bsamax} and \eqref{gsomax}, one obtains the Bethe-Salpeter equation for the longitudinal part of an axial-vector meson 
\begin{align}
\label{bseax}
1+{\mathfrak g}_{\rm SO}^{q\bar{q},1^+}(T)\cdot{\mathcal K}_{[f\bar{g}]}^{1^+,\parallel}(Q_0^2=-(m_{[f\bar{g}]}^{1^+,\parallel})^2;T) =0\,,
\end{align}
where 
\begin{align}
\label{bsekerax}
\nonumber
{\mathcal K}_{[f\bar{g}]}^{1^+,\parallel} &= \frac{2\alpha_{\rm IR}}{3\pi m_G^2}\int^1_0d\alpha \bigg\{ {\mathcal C}^{\rm iu}_1(\varsigma_{f g}(\alpha, Q_0^2);T) \\
\nonumber
&+ (M_fM_g+\alpha\hat{\alpha}Q_0^2)\overline{\mathcal C}^{\rm iu}_1(\varsigma_{f g}(\alpha, Q_0^2);T)\\
&+2{\mathcal R}^{\rm iu}(\varsigma_{f g}(\alpha, Q_0^2);T)\bigg\}\,,
\end{align}
and
\begin{align}
\frac{1}{(E_{[f\bar{g}]}^{1^+,\parallel})^2} = -\frac{9m_G^2}{4\pi\alpha_{\rm IR}}\frac{d}{dz}{\mathcal K}_{[f\bar{g}]}^{1^+,\parallel}(z;T)\bigg|_{z=-(m_{[f\bar{g}]}^{1^+,\parallel})^2}\,,
\end{align}
is the related canonical normalisation condition.

Instead, for the longitudinal part of a vector diquark, the Bethe-Salpeter equation and the canonical normalisation condition are
\begin{align}
1+\frac{1}{2}{\mathfrak g}_{\rm SO}^{qq,1^-}(T)\cdot{\mathcal K}_{[f\bar{g}]}^{1^+,\parallel}(Q_0^2=-(m_{[fg]}^{1^-,\parallel})^2;T) =0\,,
\end{align}
and 
\begin{align}
\frac{1}{(E_{[fg]}^{1^-,\parallel})^2} = -\frac{3m_G^2}{2\pi\alpha_{\rm IR}}\frac{d}{dz}{\mathcal K}_{[f\bar{g}]}^{1^+,\parallel}(z;T)\bigg|_{z=-(m_{[fg]}^{1^-,\parallel})^2}\,,
\end{align}
separately, where ${\mathfrak g}_{\rm SO}^{qq,1^-}(T)$ is given in Eq.\,\eqref{gsovcdq}.

Again, the discussion above is on the longitudinal components. Omitting the ${\mathcal R}^{\rm iu}$-term in Eq.\,\eqref{bsekerax}, the equations for the corresponding transverse components can be derived.

%%%%%%%%%%%%%%%%%%%%%%%%%%%%%%%%%%%%%%%%%%%%%%%%%%%%%%%%%%%%%%%%%%%%%%%%%%%%%%
%%%%%%%%%%%%%%%%%%%%%%%%%%%%%%%%%%%%%%%%%%%%%%%%%%%%%%%%%%%%%%%%%%%%%%%%%%%%%%

\section{Diquark flavour matrices}
\label{appdqf}

In flavour-SU(3) sector, the flavour matrices of diquark correlations can be divided into the following two categories:
{\allowdisplaybreaks
%\begin{subequations}
\begin{align}
\label{flavourarrays1}
{\tt t}^{1=[ud]} = \left[\begin{array}{ccc}
                    0 & 1 & 0 \\
                    -1 & 0 & 0 \\
                    0 & 0 & 0                 
                    \end{array}\right], \nonumber\\
{\tt t}^{2=[us]} = \left[\begin{array}{ccc}
                    0 & 0 & 1 \\
                    0 & 0 & 0 \\
                    -1 & 0 & 0                    
                    \end{array}\right],\\
{\tt t}^{3=[ds]} = \left[\begin{array}{ccc}
                    0 & 0 & 0 \\
                    0 & 1 & 0 \\
                    -1 & 0 & 0                    
                    \end{array}\right], \nonumber
\end{align}
%\end{subequations}
}
for flavour-triplets; and 
%{\allowdisplaybreaks
%\begin{subequations}
\begin{align}
\label{flavourarrays2}
{\tt t}^{4=\{uu\}} &= \left[\begin{array}{ccc}
                    \sqrt{2} & 0 & 0 \\
                    0 & 0 & 0 \\
                    0 & 0 & 0                 
                    \end{array}\right], \nonumber\\
{\tt t}^{5=\{ud\}} &= \left[\begin{array}{ccc}
                    0 & 1 & 0 \\
                    1 & 0 & 0 \\
                    0 & 0 & 0                    
                    \end{array}\right], \nonumber\\
{\tt t}^{6=\{us\}} &= \left[\begin{array}{ccc}
                    0 & 0 & 1 \\
                    0 & 0 & 0 \\
                    1 & 0 & 0                    
                    \end{array}\right], \\
{\tt t}^{7=\{dd\}} &= \left[\begin{array}{ccc}
                    0 & 0 & 0 \\
                    0 & \sqrt{2} & 0 \\
                    0 & 0 & 0                 
                    \end{array}\right], \nonumber\\
{\tt t}^{8=\{ds\}} &= \left[\begin{array}{ccc}
                    0 & 0 & 0 \\
                    0 & 1 & 0 \\
                    1 & 0 & 0                    
                    \end{array}\right] , \nonumber\\
{\tt t}^{9=\{ss\}} &= \left[\begin{array}{ccc}
                    0 & 0 & 0 \\
                    0 & 0 & 0 \\
                    0 & 0 & \sqrt{2}                    
                    \end{array}\right], \nonumber
\end{align}
%\end{subequations}
%}
for flavour-sextets. In Eqs.\,\eqref{flavourarrays1} and \eqref{flavourarrays2}, the superscripts indicate the numbered diquarks' valence quark contents. For example, ${\tt t}^{1=[ud]}$ means the diquark is numbered by ``1'', constituted from valence $u$- and $d$-quarks, and is antisymmetric; while ${\tt t}^{8=\{ds\}}$ indicates the diquark is labeled  ``8'', made of valence $d$- and $s$-quarks, and is symmetric. In this article, we utilise isospin symmetry, which means that all diquarks and baryons within an isospin multiplet are degenerate. Therefore, we focus on solving for the baryons within a specific isospin multiplet with the simplest flavour structure.

%%%%%%%%%%%%%%%%%%%%%%%%%%%%%%%%%%%%%%%%%%%%%%%%%%%%%%%%%%%%%%%%%%%%%%%%%%%%%%
%%%%%%%%%%%%%%%%%%%%%%%%%%%%%%%%%%%%%%%%%%%%%%%%%%%%%%%%%%%%%%%%%%%%%%%%%%%%%%

\section{Flavour structure of baryons}
\label{appfad}

In the flavour-SU(3) sector, the SCI supports the following flavour-combinations for the $J^P=1/2^\pm$ baryons:
\vspace{1pt}
%{\allowdisplaybreaks
\begin{subequations}
\label{octetdiquarkcs}
\begin{align}
\Psi_{N^\pm} & =
\frac{1}{\sqrt{3}}\left[
\begin{array}{l}
\sqrt{3}u[ud]_{0^{+}}   \\
\sqrt{2}d\{uu\}_{1^{+}}-u\{ud\}_{1^{+}}\\
\sqrt{3}u[ud]_{0^{-}} \\
\sqrt{3}u[ud]_{1^{-}} \\
\end{array} \right]
\leftrightarrow
\left[
\begin{array}{l}
{\mathfrak s}^{N^\pm}_1 \\
{\mathfrak a}^{N^\pm}_{\{45\},i} \\
{\mathfrak p}^{N^\pm}_1 \\
{\mathfrak v}^{N^\pm}_{1,i} \\
\end{array} \right]\,, \\
\label{lambdadq}
\Psi_{\Lambda^\pm} & =
\frac{1}{\sqrt{2}}\left[
\begin{array}{l}
\sqrt{2}s[ud]_{0^{+}} \\
d[us]_{0^{+}}-u[ds]_{0^{+}} \\
d\{us\}_{1^{+}}-u\{ds\}_{1^{+}} \\
\sqrt{2}s[ud]_{0^{-}} \\
d[us]_{0^{-}}-u[ds]_{0^{-}} \\
\sqrt{2}s[ud]_{1^{-}} \\
d[us]_{1^{-}}-u[ds]_{1^{-}} \\
\end{array}
\right]
\leftrightarrow
\left[
\begin{array}{l}
{\mathfrak s}^{\Lambda^\pm}_1 \\
{\mathfrak s}^{\Lambda^\pm}_{\{23\}} \\
{\mathfrak a}^{\Lambda^\pm}_{\{68\},i} \\
{\mathfrak p}^{\Lambda^\pm}_1 \\
{\mathfrak p}^{\Lambda^\pm}_{\{23\}} \\
{\mathfrak v}^{\Lambda^\pm}_{1,i} \\
{\mathfrak v}^{\Lambda^\pm}_{\{23\},i} \\
\end{array}
\right]\,, \\
\label{sigmadq}
\Psi_{\Sigma^\pm} & =
\left[
\begin{array}{l}
u[us]_{0^{+}}   \\
s\{uu\}_{1^{+}} \\
u\{us\}_{1^{+}} \\
u[us]_{0^{-}}   \\
u[us]_{1^{-}}   \\
\end{array}
\right]
\leftrightarrow
\left[
\begin{array}{l}
{\mathfrak s}^{\Sigma^\pm}_{2} \\
{\mathfrak a}^{\Sigma^\pm}_{4,i} \\
{\mathfrak a}^{\Sigma^\pm}_{6,i} \\
{\mathfrak p}^{\Sigma^\pm}_{2} \\
{\mathfrak v}^{\Sigma^\pm}_{2,i} \\
\end{array}
\right]\,,\\
\Psi_{\Xi^\pm} & =
\left[
\begin{array}{l}
s[us]_{0^{+}}   \\
s\{us\}_{1^{+}} \\
u\{ss\}_{1^{+}} \\
s[us]_{0^{-}}   \\
s[us]_{1^{-}}   \\
\end{array}
\right]
\leftrightarrow
\left[
\begin{array}{l}
{\mathfrak s}^{\Xi^\pm}_{2} \\
{\mathfrak a}^{\Xi^\pm}_{6,i} \\
{\mathfrak a}^{\Xi^\pm}_{9,i} \\
{\mathfrak p}^{\Xi^\pm}_{2} \\
{\mathfrak v}^{\Xi^\pm}_{2,i} \\
\end{array}
\right]\,,
\end{align}
\end{subequations}
%}
where $[fg]_{J^P}$ and $\{fg\}_{J^P}$ indicate the flavour-antitriplet and -sextet diquarks with spin $J$ and parity $P$, respectively. Each rightmost column in Eqs.\,\eqref{octetdiquarkcs} represents the associated coefficients in Eqs.\,\eqref{octetdiraca3} and \eqref{octetdirac6}, with the subscript indicating the flavour content provided in Appendix\,\ref{appdqf}; and $i=1$, $2$ refers to different Dirac structures of the axial-vector or vector diquark, \emph{cf.}, Eqs.\,\eqref{octetdiraca3ax} and \eqref{octetdirac6}. 

The analogous vectors for the $J^P=3/2^\pm$ baryons are
\begin{subequations}
\label{decupletdiquarkcs}
\begin{align}
\Psi_{\Delta^\pm} & =
\left[
\begin{array}{l}
u\{uu\}_{1^{+}} \\
\end{array}
\right]
\leftrightarrow
\left[
\begin{array}{l}
{\mathfrak d}^{\Delta^\pm}_4 \\
\end{array}
\right]\,, \\
\Psi_{\Sigma^{\ast\pm}} & =
\left[
\begin{array}{l}
s\{uu\}_{1^{+}} \\
u\{us\}_{1^{+}} \\
\end{array}
\right]
\leftrightarrow
\left[
\begin{array}{l}
{\mathfrak d}^{\Sigma^{\ast\pm}}_4 \\
{\mathfrak d}^{\Sigma^{\ast\pm}}_6 \\
\end{array}
\right]\,, \\
\Psi_{\Xi^{\ast\pm}} & =
\left[
\begin{array}{l}
s\{us\}_{1^{+}} \\
u\{ss\}_{1^{+}} \\
\end{array}
\right]
\leftrightarrow
\left[
\begin{array}{l}
{\mathfrak d}^{\Xi^{\ast\pm}}_6 \\
{\mathfrak d}^{\Xi^{\ast\pm}}_9 \\
\end{array}
\right]\,, \\
\Psi_{\Omega^\pm} & = \left[
\begin{array}{l}
s\{ss\}_{1^{+}} \\
\end{array}
\right]
\leftrightarrow
\left[
\begin{array}{l}
{\mathfrak d}^{\Omega^{\pm}}_9\\
\end{array}
\right].
\end{align}
\end{subequations}

%%%%%%%%%%%%%%%%%%%%%%%%%%%%%%%%%%%%%%%%%%%%%%%%%%%%%%%%%%%%%%%%%%%%%%%%%%%%%%
%%%%%%%%%%%%%%%%%%%%%%%%%%%%%%%%%%%%%%%%%%%%%%%%%%%%%%%%%%%%%%%%%%%%%%%%%%%%%%

%%\bibliographystyle{../../../../zProc/z10/z10KITPC/h-physrev4}
%%\bibliography{NucleonAxial}
%%\bibliographystyle{../../apsrev4-2}
%%\bibliography{../../../../../CollectedBiB}


\end{document}